\documentclass[a4paper,11pt]{book}

\usepackage[utf8x]{inputenc}
\usepackage{amssymb}
\usepackage{amsmath}
\usepackage{nicefrac}
\usepackage{multirow}
\usepackage{bbold}

\usepackage{graphicx}% Include figure files
\usepackage{dcolumn}% Align table columns on decimal point
\usepackage{mathrsfs}
\usepackage{comment}
\usepackage{amsbsy}
\usepackage{units}

\newcommand{\Slash}[1]{\ooalign{\hfil/\hfil\crcr$#1$}}

\usepackage[width=0.8\textwidth,font=small,labelfont=bf]{caption}
\usepackage[sort&compress,numbers]{natbib}

\usepackage{color}
\definecolor{rltred}{rgb}{0.75,0,0}
\definecolor{rltgreen}{rgb}{0,0.5,0}
\definecolor{rltblue}{rgb}{0,0,1}
\definecolor{rltblack}{rgb}{0,0,0}

\author{{\bf {\LARGE H\`elios Sanchis Alepuz}}}
\title{{\bf {\huge Baryon properties and glueballs from Poincar\'e-covariant bound-state equations}}}
\date{}

\begin{document}

\thispagestyle{empty}
%\maketitle
\begin{center}
{\bf {\huge Baryon properties and glueballs from Poincar\'e-covariant bound-state equations}}
\end{center}

\vspace{2cm}
\begin{center}
{\bf {\LARGE H\`elios Sanchis Alepuz}}
\end{center}

\vspace{10cm}
%\vspace*{\fill} 
\begin{center}
%\begin{minipage}{0.8\textwidth}
\begin{center}
{\LARGE Dissertation}\\
zur Erlangung des Doktorgrades der Naturwissenschaften\\
verfasst am Institut fur Physik\\
an der Karl-Franzens-Universitat Graz\\
Betreuer: Univ.-Prof. Dr. R. Alkofer\\
Graz, 2012
\end{center}
%\end{minipage}
\end{center}

\newpage
\thispagestyle{empty}
~~
\clearpage
\thispagestyle{empty}
\vspace*{\fill} 
\begin{flushright}
\textit{per a Helena, Joana i Sergi}   
\end{flushright}
\vspace*{\fill} 
\clearpage
\thispagestyle{empty}

\tableofcontents

\chapter{Introduction}\label{ch:introduction}

Ordinary hadronic matter, i.e. matter made of protons and neutrons, is composed of elementary particles called quarks and has quantum numbers as dictated by the quark model \cite{GellMann:1964nj}. The quark model is a non-dynamical theory that classifies the different hadrons in terms of their quark content. This model allows two families of hadrons: mesons, formed by a quark and an antiquark and baryons, formed by three quarks. States found experimentally but not describable by the quark model are called exotic states.

Quantum Chromodynamics (QCD) is the theory that describes the strong interaction \cite{Fritzsch1973}, which is responsible for the formation of hadrons as bound-states of quarks. It is a quantum gauge-field theory that has quarks and gluons as the elementary degrees of freedom (the quantization process additionally introduces unphysical auxiliary fields, called ghosts). The charge corresponding to the strong interaction in QCD is called color charge. It is an yet unproven statement (and so far an extraordinarily well-established experimental fact) that all physical states must have no net color charge; this is the most intuitive definition of color confinement.

QCD was built upon the understanding of hadrons via the quark model and was intended to provide a dynamical description of the formation of hadrons as states composed of quarks and bound by gluons. However, the consensus that QCD provides a correct picture of strong interactions is due to its success at describing high-energy processes. The reason for this is that at high-energies QCD becomes a weakly-coupled theory and therefore perturbative methods can be applied in this regime. The calculation of hadron properties developed, instead, at a lower pace. The reason for this is two-fold: at low energies QCD becomes a strongly-interacting theory and, on the other hand, bound-state formation is an essentially non-perturbative phenomenon. In conclusion, perturbative techniques no longer apply in the low-energy regime and new methods had to be developed.

The challenge of low-energy QCD calculations is to understand how hadrons emerge as the physical degrees of freedom, out of the elementary degrees of freedom of  the theory, namely quarks and gluons (and ghosts). At the phenomenological level, this amounts to calculate hadron properties (mass, radius, shape, etc.) from the QCD Lagrangian
\begin{flalign}\label{eq:QCD_lagrangian}
 \mathcal{L}_{QCD}=&Z_2~\bar{q}\left(-\Slash{\partial}+Z_m~m\right)q+Z_3~\frac{1}{2}A_\mu^a\left(-\partial^2\delta_{\mu\nu}-\left(\frac{1}{Z_3\xi}-1\right)\partial_\mu\partial_\nu\right)A_\nu^a \nonumber\\
  &+\widetilde{Z}_3~\bar{c}^a\partial^2~c^a-Z_{1F}ig~\bar{q}\gamma_\mu\frac{\lambda^a}{2}qA_\mu^a-Z_1g~f^{abc}(\partial_\mu A_\nu^a)A_\mu^bA_\nu^c \nonumber \\ 
  &+Z_4\frac{1}{4}g^2~f^{abe}f^{cde}A_\mu^aA_\nu^bA_\mu^cA_\nu^d+\widetilde{Z}_1g~f^{abc}\bar{c}^a\partial_\mu(A_\mu^cc^b)~,
\end{flalign}
where $q$, $A$ and $c$ are the quark, gluon and ghost fields, respectively. Latin indices represent color (in the adjoint representation; quarks carry a color index in the fundamental representation, which is ommitted here for simplicity) and $\lambda$ are the Gell-Mann matrices. This expression defines the renormalization constants, which are not completely independent but have to fulfill the following relations
\begin{equation}\label{eq:Z_relations}
 Z_{1F}=Z_gZ_2Z_3^{\nicefrac{1}{2}}~,\qquad Z_1=Z_gZ_3^{\nicefrac{3}{2}}~,\qquad\widetilde{Z}_1=Z_g\widetilde{Z}_3Z_3^{\nicefrac{1}{2}}~,\qquad Z_4=Z_g^2Z_3^2
\end{equation}
which are consequence of the Slavnov-Taylor identities.

The spatial distribution of some of the baryon properties, such as mass or electric charge, is of especial relevance to understand low-energy QCD dynamics, since they probe the details of the the quark-quark and quark-gluon interactions. In particular the question arises whether the shape of baryons deviate from sphericity. For this reason, the electromagnetic properties of the nucleon and the Delta resonance have been subject to extensive experimental research. The evolution of the nucleon electromagnetic form factors with the photon momenta is now very well known experimentally (see, e.g. \cite{Arrington2011,Arrington2007} for recent reviews). On the other hand, the experimental information for the Delta electromagnetic form factors is very limited due to the short lifetime of this resonance. Most of this information comes indirectly from the study of the $\gamma N\rightarrow\Delta$ transition \cite{Beck2000a,Blanpied2001,Pospischil2001,Tiator2003,Sparveris2005,Schmieden2006,Elsner2006,Stave2006}. The only direct information on the Delta electromagnetic properties is limited to the $\Delta^{++}$ and $\Delta^{+}$ magnetic moments \cite{Kotulla2003,Nakamura2010o}, but with large errors. A more precise measurement of the $\Delta^{+}$ magnetic moment is expected to be performed at MAMI \cite{Kotulla2007}. With this state of affairs, any theoretical calculation of the Delta electromagnetic form factors which can lead to model-independent statements constitute a prediction. We will address this problem in Chapter \ref{ch:formfactors}.

Traditionally, hadron properties have been studied by modeling QCD with effective degrees of freedom. For instance, constituent quark models (see e.g \cite{Glozman1996,Glozman1998e} and references therein) describe hadrons as bound-states of effective quarks, whose mass includes dynamical effects (the typical constituent quark mass is one-third the nucleon mass), and use a wide variety of, more or less complicated, interaction potentials among the quarks.  They have been very successful in describing hadron spectra as well as other properties such as electromagnetic form factors. The main problem of these approaches is that it is not clear whether they capture all features, or any, of QCD dynamics.

Lattice QCD methods provide non-perturbative calculations of hadron properties using the fundamental degrees of freedom of QCD (see e.g. \cite{Fodor2012} and references therein). In many cases they almost have the status of \textit{theoretical experiments}, when no experimental data for the situation of interest is available (for example, in Chapter \ref{ch:baryons} we calculate the mass of triple-charm and triple-beauty baryons which have not been observed yet and, therefore, we can only compare to lattice predictions for their masses). Lattice approaches have, nevertheless, some drawbacks. First of all, most of the calculations are performed in the so-called quenched approximation, in which the quantum fluctuations involving quarks are neglected. Also, common to all lattice calculations, are the problems of discretization and finite-volume effects. Moreover, for technical reasons, the calculations are performed at unphysical quark masses (or pion masses) and some procedure to extrapolate the results to the physical mass must be defined. With the increase of computing power, however, these limitations are rapidly diminishing. An intrinsic problem of lattice methods is that it is difficult to unravel how bound-states are formed since, by the very nature of lattice calculations, they include all possible quantum correlations among quarks.

Hadron physics should, in principle, be describable directly from QCD as a continuum quantum field theory. A rigorous, and systematically improvable, method to extract hadron properties from the QCD Lagrangian (and widely used to extrapolate lattice calculations to the physical pion mass) is chiral perturbation theory (for a pedagogical introduction see, e.g., \cite{Scherer2010,Scherer2003}). However, this approach is limited to the light quark and low momentum region.

A complete description of a continuum quantum field theory, and in particular of QCD, is given when all the (infinitely many) Green's functions of the theory are known. Functional methods  (e.g., Functional Renormalization Group and Dyson-Schwinger equations) provide such a description. In particular, Dyson-Schwinger equations (DSEs) \cite{Dyson:1949ha,Schwinger:1951ex} constitute an infinite set of coupled, non-linear integral equations for the full Green's functions of the theory (in a quantum field theory one must distinguish between bare Green's functions, which are derived directly from the Lagrangian and are purely classical, and the full or dressed Green's functions, which include all quantum effects). These are the same Green's functions that one studies in lattice QCD, of which therefore DSEs offer a complementary approach. An interesting feature of these equations is that they can be solved exactly in the infrared-momentum region, thus providing insight into the non-perturbative regime of the theory. However, any feasible solution of DSEs in a general momentum range requires to truncate the system to a finite number of equations (for a review see, e.g,. \cite{Alkofer:2000wg,Fischer:2006ub}).

Hadrons are encoded in those Green's functions. In general, a bound state in quantum field theory corresponds to a pole in the full Green's function describing the evolution of the constituent particles (see e.g. \cite{0521670535}). This pole cannot be obtained from a perturbative expansion in Feynman diagrams but requires a non-perturbative treatment. In the DSE framework, bound-states are described by generalized covariant Bethe-Salpeter equations (BSEs). They can be defined from the corresponding Green's function and its defining DSE by a Laurent expansion around the bound-state pole. The bound-state is now represented by the so-called Bethe-Salpeter amplitudes, which are related to the residue of the Green's function at the pole and are the solutions of the BSEs. These equations require, as an input, some of the Green's functions obtained from the full DSE system. Naturally, if a truncation of the DSEs is performed, a consistent truncation in the BSE of interest is required. Nevertheless, once such a truncation scheme is fixed, BSEs provide information both about the particle spectrum and their internal composition.

Mesons have been thoroughly studied within the DSE/BSE framework, mostly using the simplest of the truncations, so-called Rainbow-Ladder truncation. In this truncation scheme, of all the possible interactions between the two quarks forming the meson, only a single dressed gluon-exchange is taken into account. Moreover, the quark-gluon interaction vertex and the gluon propagator are modelled (and restricted to depend only on the momentum of the exchanged gluon) and only the quark propagator is solved self-consistently from its DSE (see, e.g., \cite{Maris:2005tt,Maris:2006ea,Krassnigg:2009zh}). There are also some studies about the role of beyond Rainbow-Ladder effects \cite{Fischer2008,Fischer2009a,Chang2009a}. 

Baryons, being a three-body system, are much more complicated and therefore have been less studied so far but, nevertheless, significant progress using also the Rainbow-Ladder truncation has been achieved recently \cite{Eichmann2010,Eichmann2011a,SanchisAlepuz:2011jn,SanchisAlepuz:2011aa}. Since the Rainbow-Ladder truncation entails the choice of a model for some dressing functions, it is not entirely transparent what features of the calculation are due to the modeling and which ones to the truncation itself. The goal of this thesis is to make some model-independent statements about the study of baryon properties using covariant Bethe-Salpeter equations within the Rainbow-Ladder truncation scheme. To do this we perform the calculations using two models very different in nature.

Another interesting aspect of bound-states in QCD is to study the nature of exotic hadrons. Since QCD is a non-abelian Yang-Mills theory, the gauge bosons have self-interactions. For this reason, QCD predicts that there should exist bound states formed by gluons only, so-called glueballs \cite{Fritzsch:1975tx}.  Although there is an intense experimental effort to discover glueballs, for the moment there is no direct evidence of them. In Chapter \ref{ch:glueballs} we propose a BSE for a two-gluon system. To derive it, we start from the DSE for the four-gluon Green's function, which describes the propagation of two gluons in spacetime, and select the diagrams which would develop a pole if a bound-state is formed. A consistent solution of this equation, which is beyond the scope of this thesis, would require the knowledge of several full, i.e. dressed, Green's functions, namely the gluon propagator, and full three-gluon and four-gluon vertices. These Green's functions can, in principle, be obtained by solving the corresponding DSEs.

In the next section we give a brief description of how to derive the Dyson-Schwinger equations from a given Lagrangian

\section{Dyson-Schwinger and Bethe-Salpeter equations in a nutshell}\label{sec:DSEBSE}

As already mentioned above, a bound state in quantum field theory corresponds to a pole in the corresponding Green's function. Those poles cannot appear in a perturbative series, but are an essentially non-perturbative phenomenon. The residue of the Green's function at the the pole allows to define the Bethe-Salpeter amplitudes (see e.g. \cite{0521670535}).

The steps to write down a relativistic equation for bound states can be summarized as:
\begin{itemize}
 \item Derive a non-perturbative equation describing the relevant Green's function.
 \item Examine the structure of the equation to find those terms which will develop a pole when a bound-state is formed (pole ansatz).
 \item Performing a Laurent expansion of the Green's function around the pole one finds an homogeneous equation for the Bethe-Salpeter amplitudes, the Bethe-Salpeter equation, and a normalization condition.
\end{itemize}
In the rest of this section we shortly describe the first point. The other two will be developed, when necessary, along the thesis.

\subsection{Equation for the Green's function: Diagrammatic derivation}\label{sec:DSE_diag}

In this subsection we describe a diagrammatic derivation of an equation for a given Green's function (for details see e.g. \cite{Loring:2001kv}). Equations of this type are commonly referred to as Dyson equations. Although the formal derivation given in the next subsection is more powerful and does not rely on any kind of perturbative expansion, the steps described here can provide a more intuitive, or at least complementary, picture of how the different interaction terms appear.

A $2n$-points Green's function is given by the (physical-) vacuum expectation value of the time-ordered product of $2n$ Heisenberg field operators
\begin{align}
 G_{a_1\dots a_n;a'_1\dots a'_n}(x_1\dots x_n;x'_1\dots x'_n)\equiv
\langle 0_{out}| T\left[\psi_{a_1}^{1}(x_1)\dots\psi_{a_n}^{n}(x_n)\bar{\psi}_{a'_1}^{1}(x'_1)\dots\bar{\psi}_{a'_n}^{n}(x'_n)\right] |0_{in}\rangle~,
\end{align}
where the generic indices $a$ represent all possible indices carried by the fields. By using the iterative expansion of the time evolution operator (see, e.g \cite{Lurie}), the above expression can be rewritten in terms of interaction-picture operators:
\begin{multline}\label{eq:Green_definition}
 G_{a_1\dots a_n;a'_1\dots a'_n}(x_1\dots x_n;x'_1\dots x'_n)\equiv  \\
 \sum_{k=0}^{\infty}\frac{(-i)^k}{k!}\int d^4y_1\dots d^4y_k\langle 0| T\left[\psi_{I,a_1}^{1}(x_1)\dots\psi_{I,a_n}^{n}(x_n)\right.\\
\left.\times\bar{\psi}_{I,a'_1}^{1}(x'_1)\dots\bar{\psi}_{I,a'_n}^{n}(x'_n)\mathcal{H}_I(y_1)\dots\mathcal{H}_I(y_k)\right] |0\rangle~,
\end{multline}
where the $k=0$ case represent the free propagation of fields. This perturbative expression is suitable to study, for instance, scattering processes. However, if the system develops bound states the Green's function will have poles. A pole in the Green's function will never appear by summing any finite number of diagrams in this expansion, but instead the whole series, or an infinite subset of it must be considered.

\begin{figure}[ht!]
 \begin{center}
  \includegraphics[width=0.56\textwidth,clip]{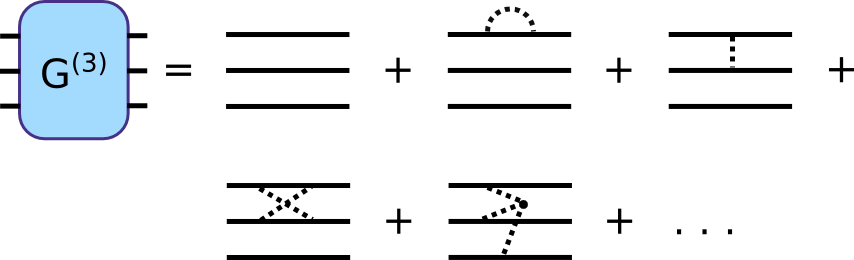}
 \end{center}
 \caption{Some examples of diagrams that would be generated by Wick contractions in Equation (\ref{eq:Green_definition}) for the case of a 6-points Green's function. The full series contains all possible one-particle- (e.g. second diagram), two-particle- (e.g. third and fourth diagrams) and three-particle-irreducible (e.g fifth diagram) terms.}\label{fig:diagrams}
\end{figure}

Applying Wick's theorem to the elements on the right-hand side of (\ref{eq:Green_definition}) one generates all possible diagrams (see Figure \ref{fig:diagrams}). The (infinite) sum of all $\ell$-particle-irreducible connected terms (with $1< \ell\le n$) is called irreducible $\ell$-particle interaction kernel $-iK^{(\ell)}$ and the sum of all one-particle terms gives the full propagators $S$. All reducible diagrams can be generated by iteration of the irreducible ones, using the full propagators as internal lines. Defining the modified kernels as
\begin{equation}\label{eq:ModKernel}
 \widetilde{K}^{(\ell)}=\sum_{
\begin{array}{c}
 \textnormal{{\tiny cyclic~perm.}}\\
 \textnormal{{\tiny $\ell$ elements}}
\end{array}}K^{(\ell)}\underbrace{S^{-1}\dots S^{-1}}_{n-\ell}~,
\end{equation}
we can introduce a single interaction kernel
\begin{equation}
 K\equiv\sum_{\ell=2\dots n}\widetilde{K}^{(\ell)}
\end{equation}
and write an inhomogenous integral equation for the Green's function
\begin{flalign}\label{eq:Green_integral}
& G_{a_1\dots a_n;a'_1\dots a'_n}(x_1\dots x_n;x'_1\dots x'_n)= S_{a_1a'_1}(x_1,x'_1)\dots S_{a_na'_n}(x_n,x'_n) \nonumber \\
& -i\int d^4y_1\dots d^4y_n S_{a_1b_1}(x_1,y_1)\dots S_{a_nb_n}(x_n,y_n)\nonumber \\
& \times\int d^4y'_1\dots d^4y'_n K_{b_1\dots b_n;b'_1\dots b'_n}(y_1\dots y_n;y'_1\dots y'_n)G_{b'_1\dots b'_n;a'_1\dots a'_n}(y'_1\dots y'_n;x'_1\dots x'_n)
\end{flalign}
or, symbolically
\begin{equation}\label{eq:compactGreen}
 G=G_0-iG_0KG
\end{equation}
where $G_0$ represents the product of full propagators. This is the Dyson equation for the Green's function $G$.

It is convenient to work in terms of the scattering matrix $T$, defined by amputating all incoming and outgoing legs
\begin{equation}\label{eq:defT}
 G\equiv G_0+G_0TG_0
\end{equation}
and substituting in (\ref{eq:compactGreen}), one gets
\begin{equation}\label{eq:GreenScattering}
 T=-iK-iKG_0T~.
\end{equation}
The advantage of having an integral equation for the Green's function is that, even making approximations for the interaction kernel $K$, one is considering an infinite set of diagrams and therefore is useful to study bound states.

\subsection{Dyson-Schwinger equations: Formal derivation}\label{sec:DSE_formal}

In this section we describe the derivation of Dyson-Schwinger equations from the action of a given quantum field theory. These equations relate a particular Green's function with higher order Green's functions which, in turn, fulfill their own Dyson-Schwinger equations. They constitute an infinite and coupled system of integral equations for the Green's functions of the theory which do not rely on any perturbative expansion.

A quantum field theory is defined (in Euclidean spacetime) by the generating functional
\begin{equation}\label{eq:generatingfunctional}
 Z[J]=\int \mathcal{D}\Phi e^{-S[\Phi]+\int d^4x J_a(x)\Phi_a(x)}\equiv G_{i_1\dots i_n}J_{i_1}\dots J_{i_n}~,
\end{equation}
where $\Phi_a$ are generic fields, $J_a$ are the sources associated to these fields and we defined here the full Green's functions $G_{i_1\dots i_n}$ as the moments in an expansion in terms of the sources, with the indices $i_j$ denoting any possible discrete index of the field as well as spacetime variables.

Assuming that the functional integration $\int \mathcal{D}\Phi$ is defined such that the integral of a total derivative vanishes, we obtain
\begin{equation}\label{eq:DSEgenerator}
 0=\int \mathcal{D}\Phi\frac{\delta}{\delta\Phi_i} ~e^{-S[\Phi]+J_i\Phi_i}=\int\mathcal{D}\Phi ~e^{-S[\Phi]+J_i\Phi_i}\left(\frac{\delta S[\Phi]}{\delta\Phi_i}-J_i\right)=\left\langle\frac{\delta S[\Phi]}{\delta\Phi_i}-J_i\right\rangle_{[J]}~,
\end{equation}
where the brackets in the last term denote the vaccuum expectation value with non-zero sources.
We can take the term in parentheses out of the integral by noting that, acting on $Z[J]$, we can make the identification $\Phi_i\rightarrow \delta/\delta J_i$. We have then the following identity
\begin{equation}\label{eq:DSEgenerator2}
 \left(-\left.\frac{\delta S}{\delta\Phi_i}\right|_{\Phi_i\rightarrow \frac{\delta}{\delta J_i}}+J_i\right)Z[J]=0~.
\end{equation}
This equation provides the \textit{seed} to generate all Dyson-Schwinger equations for the full Green's functions, upon taking appropriate functional derivatives with respect to the sources and afterwards setting them to zero.

As an example, we will derive the Dyson-Schwinger equation for the quark propagator
\begin{equation}
 S(x-y)=\langle 0| q(x)\bar{q}(y)|0\rangle~,
\end{equation}
 since it is the essential element in any covariant bound-state calculation for mesons and baryons. We start with a particular case of (\ref{eq:DSEgenerator})
\begin{equation}
 \left\langle\frac{\delta S_{QCD}}{\delta\bar{q}(x)}-j(x)\right\rangle_{[J]}=0~,
\end{equation}
with $\bar{q}(x)$ the anti-quark field and $j(x)$ its source. Taking a further derivative with respect to $j(y)$ we obtain
\begin{equation}
 \left\langle\frac{\delta S_{QCD}}{\delta\bar{q}(x)}\bar{q}(y)\right\rangle_{[J]}=\delta^{(4)}(x-y)~.
\end{equation}
Using the explicit expression of the Lagrangian (\ref{eq:QCD_lagrangian}) to perform the derivative $\delta S_{QCD}/\delta\bar{q}$ and afterwards setting the sources to zero, we obtain the quark-propagator DSE
\begin{multline}
 \delta^{(4)}(x-y)=Z_2(-\Slash{\partial}+Z_mm)S(x-y)\\
-igZ_{1f}\int d^4z d^4z'\delta^{(4)}(x-z)\delta^{(4)}(x-z')\gamma^\mu\frac{\lambda^a}{2}\langle q(z)\bar{q}(y)A^a_{\mu}(z')\rangle~,
\end{multline}
where the full quark-gluon Green's function $\langle q(z)\bar{q}(y)A^a_{\mu}(z')\rangle$ can be written as a proper (one-particle irreducible) quark-gluon vertex $\Gamma^a_{\mu}(x,y,z)$ with two quark propagators and one gluon propagator attached. In momentum space, the quark DSE reads
\begin{equation}\label{eq:quarkDSE0}
	S^{-1}(p) = Z_2\left(i\Slash{p}+Z_m m_q\right)+ig Z_{1f}\int\frac{d^4k}{\left( 2\pi
	\right)^4}\frac{t^a}{2}\gamma_\mu S(k)\Gamma^a_\nu(k,p)D_{\mu\nu}(q)~.
\end{equation}

\chapter{Baryon Masses}\label{ch:baryons}

In this chapter we calculate the masses of spin-$\nicefrac{1}{2}$ and spin-$\nicefrac{3}{2}$ ground-state baryons using a covariant three-body Bethe-Salpeter equation, also called covariant Faddeev equation. This equation assumes that three-body-irreducible interactions are negligible. Moreover, of all possible two-body-irreducible interactions only a single dressed-gluon exchange is kept. This truncation induces the necessity of a model for the quark-gluon interaction. Using two models, very different in nature, for this interaction we intend to make model-independent statements. We will find that this simple setup gives a good description of baryon masses, up to a $10\%$ accuracy.

\section{Covariant three-body Bethe-Salpeter equation}

\begin{figure}[ht!]
 \begin{center}
  \includegraphics[width=0.5\textwidth,clip]{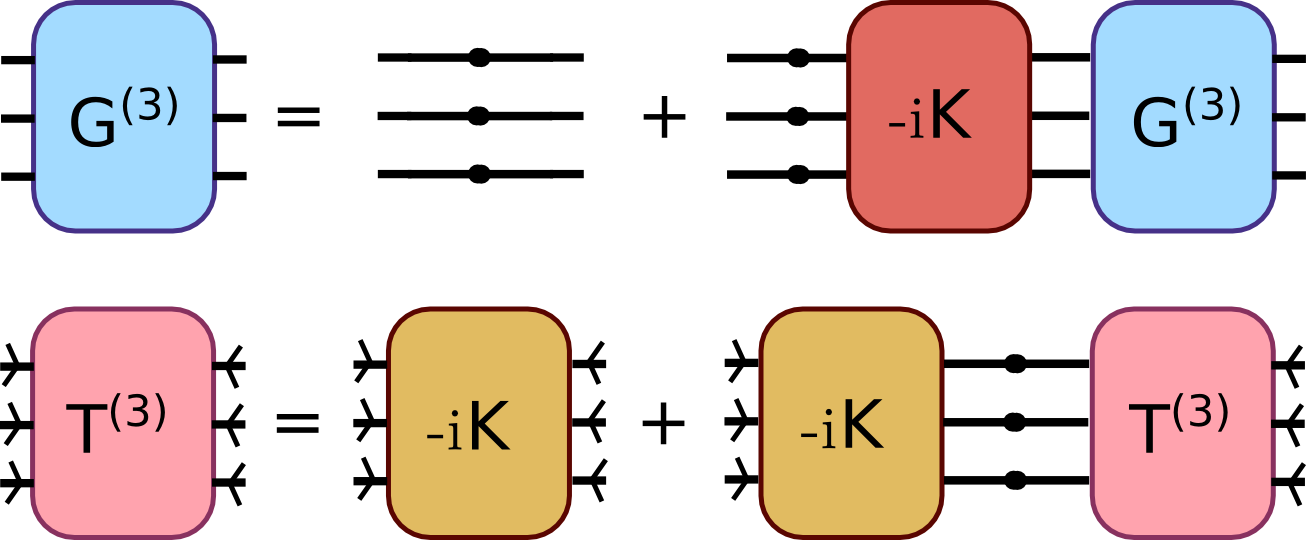}
 \end{center}
 \caption{Dyson equation (\ref{eq:Dysoneqs}) for a three-particle system. The system is equivalently described by the Green's function $G^{(3)}$ or the scattering matrix $T^{(3)}$. Lines with blobs represent fully dressed propagators and $K$ is the interaction kernel.}\label{fig:3bDSE}
\end{figure}

The evolution of a three-quark system is encoded in the six-point Green's function $G^{(3)}$ or equivalently in the six-point scattering matrix $T^{(3)}$. As explained in previous chapter, both are described by the corresponding Dyson equations (see Figure \ref{fig:3bDSE}) 
\begin{equation}\label{eq:Dysoneqs}
 \begin{array}{rcl}
 G^{(3)}&=&G^{(3)}_0-iG^{(3)}_0KG^{(3)} \\
 T^{(3)}&=&-iK-iKG^{(3)}_0T^{(3)}
\end{array}
\end{equation}
where, using the definitions in (\ref{eq:ModKernel}), the interaction kernel $K$ can be decomposed into four terms $K^{(3)}$ and $K^{(2)}_{(a)}$ (with $a=1,2,3$ denoting the spectator quark), containing only three- and two-particle irreducible graphs, respectively.

We already discussed in previous chapter that when the three quarks form a baryon, the scattering matrix (and the Green's function) develops a pole at $P^2=-M^2$, where $P$ is the total momentum of the bound-state
\begin{equation}
 P=p_1+p_2+p_3~,
\end{equation}
with $p_i$ the quark momenta, and $M$ the baryon mass. On the baryon mass-shell we write the scattering matrix as
\begin{equation}\label{eq:Psi_definition}
 T\sim C_s~\frac{\Psi\bar{\Psi}}{P^2+M^2}
\end{equation}
which defines the Bethe-Salpeter amplitude $\Psi$ and its conjugate $\bar{\Psi}$. The factor $C_s$, which depends on the baryon spin $s$, is
\begin{flalign}
 s=\frac{1}{2}~\textnormal{:}&~~~~~C_{s}=2M\Lambda_+(\hat{\textnormal{P}})~,\\
 s=\frac{3}{2}~\textnormal{:}&~~~~~C_{s}=2M\mathbb{P}_+^{\mu\nu}(\hat{\textnormal{P}})~,
\end{flalign}
where we introduced the positive-energy Dirac projector $\Lambda_+$ and the Rarita-Schwinger positive-energy projector $\mathbb{P}^{\mu\nu}$
\begin{flalign}\label{eq:def_projectors}
\Lambda^+(\hat{P})=&\frac{1}{2}\left(\mathbb{1}+\Slash{\hat{P}}\right)~,\\
 \mathbb{P}_+^{\mu\nu}(\hat{\textnormal{P}})=&~\Lambda_+(\hat{\textnormal{P}})
 \left(T_P^{\mu\nu}-\frac{1}{3}\gamma^\mu_T\gamma^\nu_T\right)~,
\end{flalign}
with $\gamma^\mu_T=T_P^{\mu\nu}\gamma^\nu$, $T_P^{\mu\nu}$ the transverse projector
\begin{equation}
 T_{\mu\nu}(q)=\delta_{\mu\nu}-\frac{q_\mu q_\nu}{q^2}
\end{equation} 
and the hat denotes a unit vector
\begin{equation}
 \hat{v}^\mu=\frac{v^\mu}{\sqrt{v^2}}~.
\end{equation}

\begin{figure}[ht!]
 \begin{center}
  \includegraphics[width=\textwidth,clip]{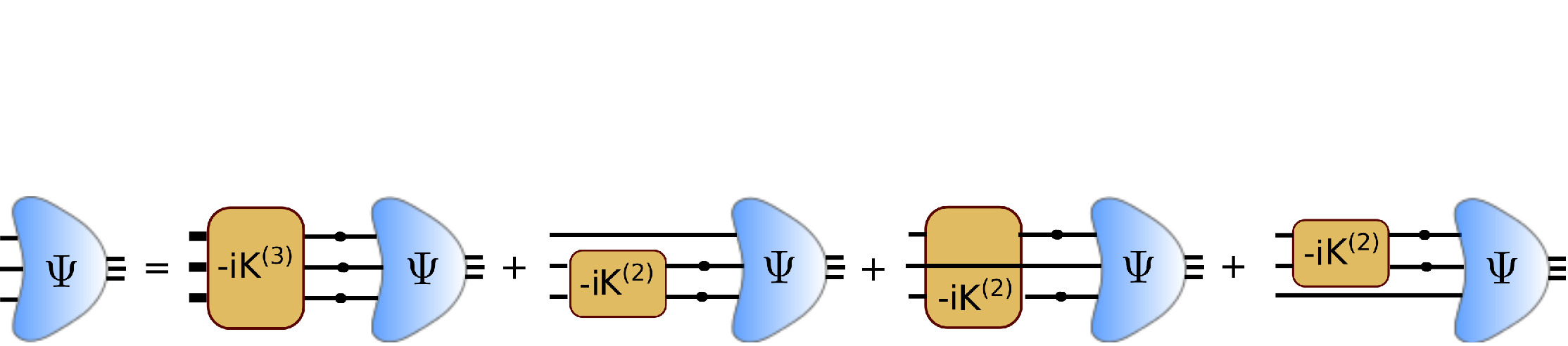}
 \end{center}
 \caption{Pictorial representation of the three-body Bethe-Salpeter equation (\ref{eq:3bBSEcompact}). Lines with blobs represent fully dressed propagators. The kernels $K^{(3)}$ and $K^{(2)}$ contain only three- and two-particle irreducible graphs, respectively.}\label{fig:3bBSE}
\end{figure}

Inserting this expression in (\ref{eq:Dysoneqs}) and neglecting regular terms, we can write an equation for $\Psi$, the three-body Bethe-Salpeter equation (see Figure \ref{fig:3bBSE}) 
\begin{equation}\label{eq:3bBSEcompact}
\Psi = -i\widetilde{K}^{(3)}~G_0^{(3)}~\Psi + \sum_{a=1}^3 -i\widetilde{K}_{(a)}^{(2)}~G_0^{(3)}~\Psi\,,
\end{equation}
where we used (\ref{eq:ModKernel}) to introduce de modified kernels $\widetilde{K}$. 

The nature of the BS amplitudes $\Psi$ depends on the baryon of interest (see Appendix \ref{sec:basis}): for spin-$\nicefrac{1}{2}$ baryons it is a rank-4 Dirac tensor and for spin-$\nicefrac{3}{2}$ baryons it is a mixed tensor with four Dirac and one Lorentz indices (and on top of this they have flavor and color structure). Its general structure can be determined imposing only Poincar\'e covariance and parity invariance. To solve this equation, one needs to specify the interaction kernels and the full quark propagator.

\section{Quark propagator and Rainbow truncation}

The full quark propagator is an essential element for any covariant bound-state calculation in QCD. The general structure of this propagator is
\begin{equation}
(S^{-1})^{AB}_{ab}(p) = A(p^2) \left(  i\Slash{p} + M(p^2) \right)_{ab}\delta^{AB}\label{eqn:inverse_quark_propagator}
\end{equation}
where $1/A(p^2)$ is the quark wave-function renormalization and $M(p^2)$ is the quark mass function. The indices $a,b$ and $A,B$ are Dirac and color indices, respectively. 

In this work, the quark propagator is obtained by solving the quark Dyson-Schwinger equation (\ref{eq:quarkDSE0}),
\begin{multline}\label{eq:quarkDSE}
	(S^{-1})^{AB}_{ab}(p) = Z_2\left(i\Slash{p}+Z_m m_q\right)_{ab} \\- Z_{1f}\int \frac{d^4k}{\left( 2\pi
	\right)^4} \left(-ig t_r^{AC}\gamma^\mu_{ac}\right)~S^{CD}_{cd}(k) ~\left(-ig t_s^{DB}\Gamma^\nu_{db}(k,p)\right)~D^{rs}_{\mu\nu}(q)~,
\end{multline}
where $q=k-p$ is the momentum of the exchanged gluon and here $r,s$ represent color indices in the adjoint representation. The color traces can be worked out (see Appendix \ref{sec:appendix_color}) to give $(N_c^2-1)/(2N_c)$, with $N_c$ the number of colors. The renormalization constants $Z_{1f}(\mu^2,\Lambda_{reg.}^2)$, $Z_{2}(\mu^2,\Lambda_{reg.}^2)$ and $Z_{m}(\mu^2,\Lambda_{reg.}^2)$ depend on the renormalization scale $\mu$ and on an ultraviolet scale $\Lambda_{reg.}$ required to regularize the integral. The renormalized mass $m_q$ is related to the bare mass $m_0$ by $m_0=Z_mm_q$. Equation (\ref{eq:quarkDSE}) is solved imposing the renormalization conditions
\begin{equation}
 \begin{array}{rcl}
   A(p^2=\mu^2)&=&1~, \\
   M(p^2=\mu^2)&=&m_0~,
 \end{array}
\end{equation}
at a sufficiently large $\mu$ and for a fixed value for $m_0$.

\begin{figure}[ht!]
 \begin{center}
  \includegraphics[width=0.6\textwidth,clip]{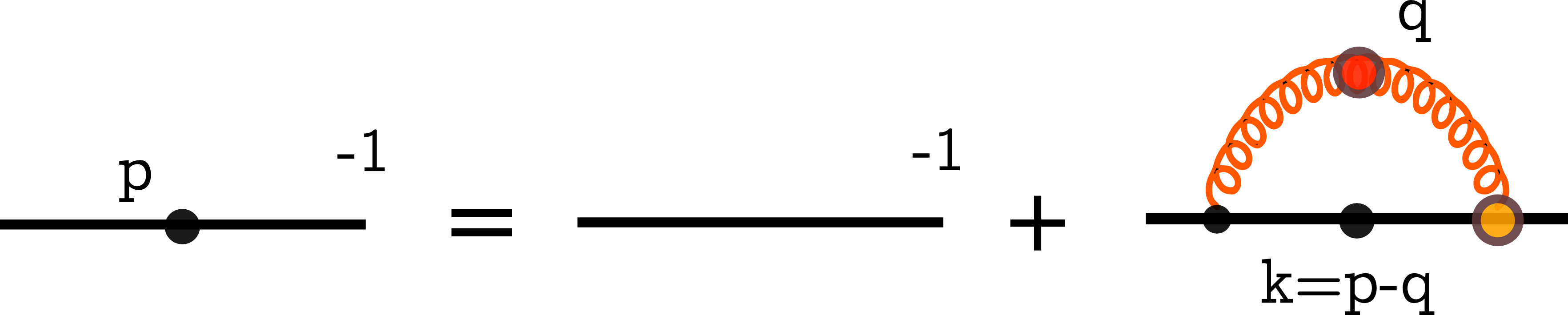}
 \end{center}
 \caption{Pictorial representation of the quark Dyson-Schwinger equation (\ref{eq:quarkDSE}). Blobs represent fully dressed propagators or vertices.}\label{fig:quarkDSE}
\end{figure}

A diagrammatic representation of this equation is given in Figure \ref{fig:quarkDSE}. In Figure \ref{fig:quarkmasses} we illustrate the dynamical generation of quark mass when the quark DSE is solved using appropriate models for the quark-gluon vertex, as explained in the next sections. We also show how this feature disappears if the interaction is not strong enough. 

%
%\subsection{Rainbow truncation}

The Green's functions required to solve the quark propagator DSE (\ref{eq:quarkDSE}) are the full gluon propagator $D_{\mu\nu}(q)$ and the full quark-gluon vertex $\Gamma^\mu(p,q)$ (we omit here all other indices). Slavnov-Taylor identities restrict the full gluon propagator, in Landau gauge, to be
\begin{equation}\label{eq:GpropLandauGauge}
 D^{rs}_{\mu\nu}(q)=\delta^{rs}\left(\delta_{\mu\nu}-\frac{q_\mu q_\nu}{q^2}\right)\frac{Z(q^2)}{q^2}
\end{equation}
with only one unknown dressing function $Z(q^2)$. On the other hand, using symmetry arguments, one can decompose the full quark-gluon vertex into 12 linearly independent Lorentz covariants $T_i^\mu$
\begin{equation}\label{eq:qgvertexdecomposition}
 \Gamma^\mu(p,q)=\sum_{i=1}^{12}f_i(p^2,q^2,p\cdot q)T_i^\mu(p,q)
\end{equation}
where $f_i$ are Lorentz-invariant dressing functions. Eight of these twelve components are purely transverse.

A completely consistent solution (that is, using a gluon propagator and quark-gluon vertex obtained as solutions of their own DSEs) is not possible in a general momentum range\footnote{It is possible, nevertheless, to solve the full set of DSEs in the deep infrared momentum-regime, but this regime is not expected to be of relevance to hadron physics.} and some truncation scheme must be chosen. The simplest possibility in the context of the quark DSE is to use for the full quark-gluon vertex the tree-level vertex $\gamma^\mu$ times a dressing function. More specifically, if we single out the tree-level part in the full quark-gluon vertex as 
\begin{equation}
 \Gamma^\mu(p,q) = Z_{1f}\gamma^\mu + \Lambda^\mu~,
\end{equation}
with all other structures gathered in $\Lambda^\mu$, one can truncate the full vertex by projecting it onto the tree-level part and limiting it to depend only on the gluon momentum $q$,
\begin{equation}\label{eq:truncvertex}
\Gamma^\mu(p,q)\rightarrow\left( Z_{1f} + \Lambda(q^2) \right)
\gamma^\mu~.
\end{equation}
This is known as the \textbf{Rainbow truncation} of the quark DSE. The flavor and color parts of the vertex are the tree-level ones. Within this truncation, only the dressing functions $Z(q^2)$ and $\Lambda(q^2)$ have to be   fixed. It has been customary in previous phenomenological studies of hadron properties in the DSE/BSE approach to combine all scalar dressings into a renormalization-group invariant\footnote{This can be seen taking into account that, under a change of the renormalization scale, $g$ scales as $1/Z_g^2$, the quark-gluon vertex as $1/Z_{1f}$ and the gluon propagator as $1/Z_3$, and using the Slavnov-Taylor identity $Z_{1f}=Z_gZ_2Z_3^{\nicefrac{1}{2}}$} effective running
coupling, $\alpha_{eff}(q^2)$
\begin{equation}\label{eq:Rainbow}
Z_{1f}~\frac{g^2}{4\pi} ~D_{\mu\nu}(q) ~\Gamma_\nu(k,p)
  \rightarrow Z_2^2 ~ T_{\mu\nu}(q) ~\frac{\alpha_{eff}(q^2)}{q^2}~\gamma_\nu~.
\end{equation}
If, instead, one wishes to draw a distinction between the gluon and the quark-gluon vertex dressings, it can be useful to model them separately. This 
distinction can be important because the gluon propagator is by now
well-known from both lattice studies and other functional approaches.

A truncation of the full quark-gluon vertex means that, in a diagrammatic expansion of the vertex (or skeleton expansion), only one (or several) of those terms is taken into account. Therefore, once such a truncation for the quark DSE is chosen, a consistent truncation of the interaction kernel in the BSE must be defined, as described in the next section.

As will be explained later in this chapter, the momenta of the internal quark propagators in a BSE are complex. The numerical techniques used to solve the quark DSE in the complex plane have been described, for instance, in \cite{Eichmann2009a,Fischer2009}. In this work we use the method of \cite{Fischer2009} and we refer the reader to this article for details. It is worth noting that the resulting quark propagator shows complex conjugate poles in the complex plane. Whether this is an artifact of the truncation or not is not clear since this feature is found in more complex truncation schemes \cite{Alkofer2009,Fischer2009}. In any case, the appearance of these poles results in a limitation of the maximum bound-state mass one can study (see \cite{Eichmann2009a}).

\section{Rainbow-Ladder truncation}

The QCD Lagrangian features an approximate $SU_R(N_f)\otimes SU_L(N_f)$ symmetry, known as chiral symmetry (which becomes exact when quarks are considered massless). This symmetry would imply, for example, that the $\pi$ and $\sigma$ mesons or the $\rho$ and $a_1$ mesons are degenerate in mass, which is not the case. In nature, this symmetry is spontaneously broken due to the dynamical generation of quark mass and the associated Nambu-Goldstone bosons are the pions. 

If mesons are studied using a covariant quark-antiquark BSE, and for the quark propagators one takes the solutions of the quark DSE, it has been shown \cite{Delbourgo:1979me,Maris:1997hd} that the crucial relation that ensures a correct implementation of chiral symmetry and its dynamical breaking is the axial-vector Ward-Takahashi identity (AxVWTI)
\begin{equation}\label{eq:AxVWTI}
 -i~P_\mu\Gamma^j_{5\mu}(p;P)=S^{-1}(p+P/2)\gamma_5\frac{\tau^j}{2}+\gamma_5\frac{\tau^j}{2}S^{-1}(p+P/2)~,
\end{equation}
which relates the axial-vector vertex $\Gamma_{5\mu}(p;P)$ (and, indirectly from its BSE, the quark-antiquark kernel) to the quark self-energy. In this equation $\tau^j$ are the Gell-Mann flavor matrices and $P$ and $p$ represent the total and relative quark momentum, 
\begin{equation}
 \begin{aligned}
  P&=p_1+p_2~, \\
  p&=\frac{p_1-p_2}{2}~,
 \end{aligned}
\end{equation}
with $p_i$ the quark momenta. It can be proven that, in the chiral limit, if (\ref{eq:AxVWTI}) is fulfilled, the dynamical generation of quark mass is accompanied by the appearance of massless pseudoscalar bound-states (the pions).

From (\ref{eq:AxVWTI}), and given a truncation of the quark DSE, one can find the corresponding quark-antiquark interaction kernel. In the case of the Rainbow truncation it corresponds to a single dressed-gluon exchange between the quark and the antiquark, with the gluon interacting with quarks via a vector coupling. In terms of
$\alpha_{eff}$, it reads
\begin{equation}\label{eq:ladder}
	-iK^{\textrm{q}-\bar{\textrm{q}}}= 4\pi ~Z_2^2 ~\frac{\alpha_{eff}(q^2)}{q^2}~
	T_{\mu\nu}(q)~i\gamma^\mu \otimes i\gamma^\nu~
\end{equation}
where, as before, the flavor and color parts of the vertex are the tree-level ones. This, together with (\ref{eq:Rainbow}), constitute what is known as the \textbf{Rainbow-Ladder (RL) truncation} of the quark-DSE/meson-BSE system.

The Rainbow-Ladder truncation provides a flavor-blind and quark-mass independent interactions. Corrections beyond Rainbow-Ladder, which may include corrections to the effective coupling and the inclusion of additional structures beyond a vector-vector interaction, are expected to change this. An open question is how important are these effects. We will see later that a quark-mass dependent interaction is essential to achieve a precise description of baryon spectra for all current-quark masses and to observe non-analiticities in the current-quark mass evolution of baryon masses due to the opening of decay channels.

As a side remark, a systematic way to relate truncations in the quark DSE with truncations of the BSE kernel such that they preserve (\ref{eq:AxVWTI}) has been described in \cite{Munczek:1994zz}. In this paper both the quark DSE and the meson BSE are derived from a chirally-symmetric 2PI effective action so that they automatically fulfill (\ref{eq:AxVWTI}). Thus, if a chiral-symmetry preserving approximation is performed for the effective action, the truncated DSE and BSE derived from this action will still fulfill the AxVWTI. As a rule of thumb, if the quark-gluon vertex in the quark DSE is truncated in a certain way (that is, keeping only some diagrams in its skeleton expansion), a symmetry-preserving BSE kernel can be obtained by cutting one internal quark line of those diagrams in all possible ways.

The goal of this work, however, is the study of baryons within the DSE/BSE framework. To this end one needs to fix the three-quark and two-quark irreducible interaction kernels. In particular, the two-quark (or quark-quark) kernel is not restricted by (\ref{eq:AxVWTI}) and one is, in principle, free to choose any other truncation scheme. There is, nevertheless, extensive literature on the calculation of meson properties using the covariant DSE/BSE approach within Rainbow-Ladder (see e.g. \cite{Maris:2005tt,Maris:2006ea,Krassnigg:2009zh} and references therein). With the idea of having a common approach to hadron properties, the Rainbow-Ladder truncation has been also used in baryon studies \cite{Hellstern1997,Oettel1998,Nicmorus2009,Eichmann2009,Nicmorus2010,Eichmann2010,Eichmann2010a,Nicmorus2011,Mader2011,Eichmann2011,Eichmann2011a,Eichmann2011b,Eichmann2012a}. In this thesis we continue this trend and use (\ref{eq:ladder}) as the quark-quark interaction kernel.

On the other hand, the study of baryons with covariant bound-state equations was performed, until recently, by reducing the three-body problem to a two-body problem using the so-called diquark ansatz \cite{Buck:1992wz,Hellstern1997,Oettel1998}. It assumes that two-body correlations are dominant in baryons and that two of the three valence quarks are bound into an object called diquark. This diquark is in turn bound to the remaining quark to form the baryon. This approach has been very successful describing baryon phenomenology. 

Moreover, recent calculations of heavy-baryon masses using perturbative non-relativistic QCD (pNRQCD) \cite{Llanes-Estrada2011} suggest that irreducible three-body interactions contribute only $\sim25$~MeV to the total mass. These results, therefore, give support to the approximation of neglecting three-particle irreducible correlations $K^{(3)}$ (\textbf{Faddeev approximation}). The corresponding three-body Bethe-Salpeter equation is known as covariant Faddeev equation (see Figure \ref{fig:FaddeevRLeq}), for historical reasons, and in what follows we will refer to the amplitudes $\Psi$ as Faddeev amplitudes.

\section{Effective interactions}\label{sec:effective_interactions}

As explained in previous section, to completely specify the covariant Faddeev equation one needs to model the gluon propagator dressing function and the quark-gluon interaction. In fact, this is the only model input of the approach. Therefore, to assess the model-independent features of the Rainbow-Ladder truncation, in this work we use two different models for the effective interactions. 

The common feature between QCD and both effective interactions is that they reproduce the one-loop behavior of the QCD running coupling at high momentum
\begin{equation}\label{eq:UVrunning}
 \alpha(q^2)\rightarrow\frac{\pi\gamma_m}{\textnormal{ln}~q^2/\Lambda^2_{QCD}}
\end{equation}
with $\gamma_m=12/(11N_C-2N_f)$ the anomalous dimension of the quark propagator. The models differ in their infrared behavior.

The first model we use is known as Maris-Tandy model \cite{Maris1997,Maris1999} and it has dominated hadron studies within Rainbow-Ladder. This dominance is well-earned
since this ansatz performs very well when it comes to the purely phenomenological calculation of ground-state meson and baryon properties. However, this model has no clear connection to QCD in the infrared and is, therefore, not entirely satisfactory to gain understanding of the formation of hadronic bound-states in QCD. On
the other hand, with the rapid improvement in our knowledge of QCD Green's
functions from both lattice and functional approaches, it is possible to
define different effective interactions which, presumably, capture more
faithfully some of QCD's features. Based on this, an effective interaction has been proposed in \cite{Alkofer2008}.

\begin{figure}[ht!]
 \begin{center}
  \includegraphics[width=.95\textwidth,clip]{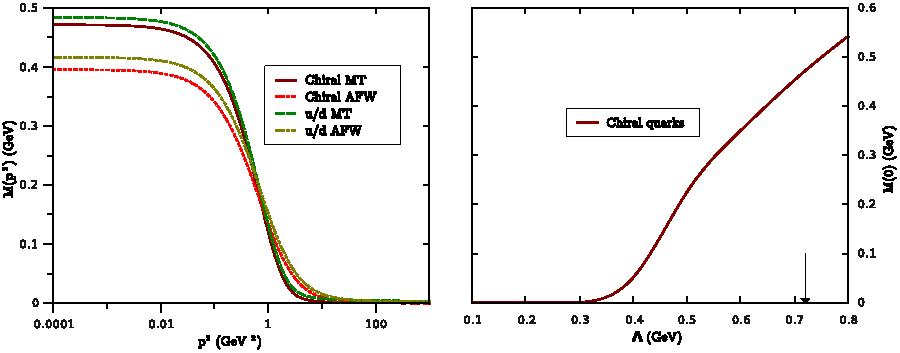}
 \end{center}
 \caption{\textit{Left panel}: Mass function $M(p^2)$ for chiral and u/d-quark bare masses using the MT and AFW models (the precise value of the bare quark-mass depends on the model). \textit{Right panel}: To illustrate the necessity of a strong enough effective interaction to account for dynamical chiral symmetry breaking, we plot $M(p^2=0)$ as a function of the interaction strength $\Lambda$ in the MT model. The arrow indicates the value of $\Lambda$ in our bound-state calculations.}\label{fig:quarkmasses}
\end{figure}

Note that the fact that an effective interaction captures more features of QCD does not necessarily mean that it will perform better phenomenologically. This is because the interaction is used within a given truncation scheme and, therefore, if one wants to reproduce hadron properties the model has to be tuned to account for the effect of the missing contributions. In particular, it has been shown \cite{Alkofer2009} that dynamical quark-mass generation is accompanied by the appearance of scalar components in the quark-gluon vertex. These components, of course, are missing in the Rainbow-Ladder truncation and, therefore, the effective interaction must somehow mimic its effects. This can already be seen in Figure \ref{fig:quarkmasses}, where dynamical chiral symmetry breaking is missing if the effective interaction is too weak.

\begin{figure}[ht!]
 \begin{center}
  \includegraphics[width=0.9\textwidth,clip]{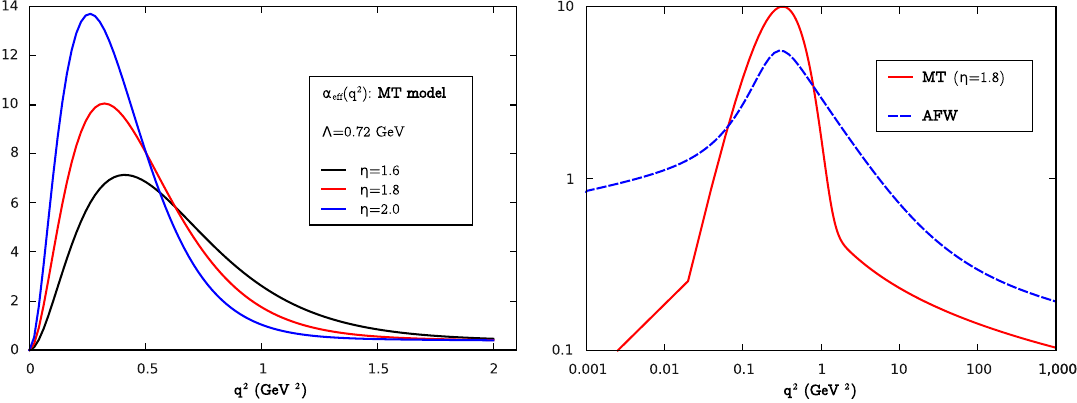}
 \end{center}
 \caption{\textit{Left panel}: Maris-Tandy effective interaction for different values of $\eta$. \textit{Right panel}: Comparison of the AFW and the MT (for $\eta=1.8$ models using a log-log scale to stress the qualitatively different behavior in the deep infrared.}\label{fig:interaction_panel}
\end{figure}

In this respect, both models described below are designed to reproduce correctly dynamical chiral-symmetry breaking as well as pion properties at the physical $u/d$ mass. This means that they capture beyond Rainbow-Ladder effects at this quark mass. As a consequence, both interactions have similar strength at the intermediate momentum region $\sim0.5~-~1~$GeV (see Figure \ref{fig:interaction_panel}). To analyze whether the effects beyond Rainbow-Ladder are analogous for mesons and for light baryons or, more precisely, whether baryon spectra is also well reproduced at this, or at higher, quark mass is one of the goals of this chapter.

\subsection{Maris-Tandy model}

In the Maris-Tandy (MT) model \cite{Maris1997,Maris1999} the effective running
coupling is given by
\begin{equation}\label{eq:MTmodel}
\alpha_{eff}(q^2) =
 \pi\eta^7\left(\frac{q^2}{\Lambda^2}\right)^2 e^{-\eta^2\frac{q^2}{\Lambda^2}}+\frac{2\pi\gamma_m \big(1-e^{-q^2/\Lambda_{t}^2}\big)}{\textnormal{ln}[e^2-1+(1+q^2/\Lambda_{QCD}^2)^2]}\,, 
\end{equation}
which, behaves as (\ref{eq:UVrunning}) in the UV and features a Gaussian distribution in the infrared (see Figure \ref{fig:interaction_panel}) that provides dynamical chiral symmetry breaking. The scale $\Lambda_t=1$~GeV is introduced for technical reasons and has no impact on the results. Therefore, the interaction strength is characterized by an energy scale $\Lambda$, fixed to $\Lambda=0.74$ GeV to reproduce correctly the pion decay constant from the RL-truncated meson-BSE. The dimensionless parameter $\eta$ controls the width of the interaction (see Figure \ref{fig:interaction_panel}). 
Many ground-state hadron observables have been found to be almost insensitive to the value of $\eta$ around $\eta=1.8$ \cite{Krassnigg:2009zh,Nicmorus2011,Eichmann2011a}. This has been used as an argument in favor of the model independence of Rainbow-Ladder results. Instead of pursuing this line of research, we prefer to introduce a new, not-related model to evaluate the validity of those assertions.
For the anomalous dimension we use $\gamma_m=12/(11N_C-2N_f)=12/25$, corresponding to $N_f=4$ flavors and $N_c=3$ colors. For the QCD scale $\Lambda_{QCD}=0.234$ GeV.

Note that in the numerical resolution of the quark DSE we employ the Pauli-Villars regularization method of the integrals, with a mass scale of $200$~GeV.
Moreover, for this model, we fit the quark masses, at the renormalization scale $\mu=19$~GeV, to be $3.7$, $85.2$, $869$ and $3750$ MeV for the
$u/d$, $s$, $c$, and $b$ quarks, respectively.

\subsection{Alkofer-Fischer-Williams model}

The Alkofer-Fischer-Williams (AFW) model \cite{Alkofer2008} is motivated by the desire to account
for the $U_A(1)$-anomaly by the Kogut-Susskind
mechanism \cite{Kogut1974b}. The effective coupling is
constructed as the product of the gluon dressing \cite{Alkofer:2003jk,Alkofer2004}
 and a model for
the non-perturbative behavior of the quark-gluon
vertex \cite{Alkofer2009},
\begin{equation}
  \alpha_{eff}(q^2) =  \mathcal{C} \left(\frac{x}{1+x}\right)^{2\kappa}
  \left(\frac{y}{1+y}\right)^{-\kappa-1/2}
  \left( \frac{\alpha_0+a_{UV}\,x}{1+x} \right)^{-\gamma_0}
  \left( \lambda +\frac{a_{UV}\,x}{1+x} \right)^{-2\delta_0}\,\,.
\end{equation}
The four terms in parentheses are: the IR scaling of the gluon
propagator; IR scaling of the quark-gluon vertex; logarithmic running of
the gluon propagator; and the logarithmic running of the quark-gluon
vertex. Additionally, the last two are constructed to interpolate between the IR and UV
behavior. The remaining terms are defined as follows:
\begin{equation}
    \lambda = \frac{\lambda_S}{1+y} + \frac{\lambda_B \,y}{1+(y-1)^2}\,, \quad
    a_{UV}= \pi  \gamma_m \left( \frac{1}{\ln{z}}-\frac{1}{z-1} \right), \quad
    \begin{array}{rl}
       x &= q^2/\Lambda_{YM}^2\,, \\
       y &= q^2/\Lambda_{IR}^2\,, \\
       z &= q^2/\Lambda_{MOM}^2\,,
    \end{array}
\end{equation}
and $\alpha_0=8.915/N_C$. Here, $\Lambda_{YM}= 0.71$ GeV is the
dynamically generated Yang-Mills scale, while $\Lambda_{MOM}= 0.5$ GeV corresponds
to the one-loop perturbative running. The IR scaling exponent is
$\kappa=0.595353$, and the one-loop anomalous dimensions are
related via $1+\gamma_0 = -2\delta_0 = \frac{3}{8}\,N_C \,\gamma_m$, with $\gamma_m=12/(11N_C-2N_f)$.
We choose $N_f=5$ active quark flavors at the
renormalization point $\mu=19$ GeV. The constant
$\mathcal{C}=0.968$ is chosen such that $\alpha_{eff}$ runs appropriately in
the UV. Finally, $\Lambda_{IR}=0.42$ GeV, $\lambda_S=6.25$, and $\lambda_{B}=21.83$
determine the IR properties of the quark-gluon vertex and are fitted such that the properties of $\pi$, $K$ and $\rho$
mesons are all reasonably well reproduced.  The quark
masses at $\mu=19$~GeV are $2.76$, $55.3$, $688$ and $3410$ MeV for the
$u/d$, $s$, $c$, and $b$ quarks, respectively.

\section{Baryon masses from the covariant Faddeev equation}

\begin{figure}[ht!]
 \begin{center}
  \includegraphics[width=0.9\textwidth,clip]{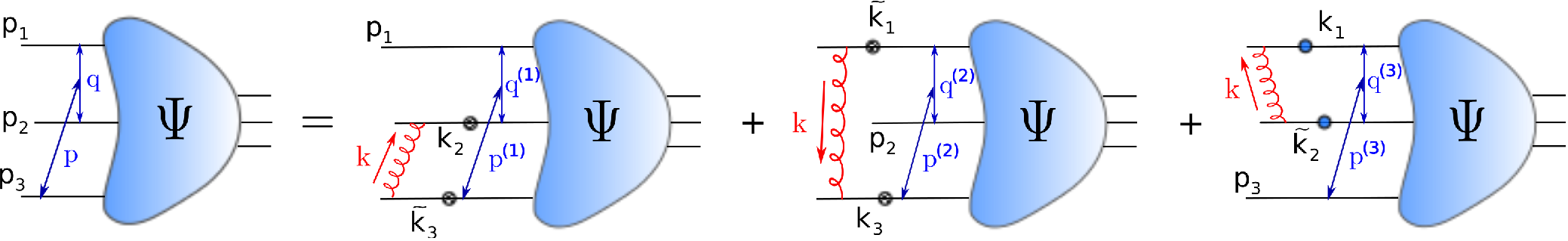}
 \end{center}
 \caption{Diagrammatic representation of the covariant Faddeev equation in the Rainbow-Ladder truncation (\ref{eq:faddeev_eq}).}\label{fig:FaddeevRLeq}
\end{figure}

The three-body Bethe-Salpeter equation (\ref{eq:3bBSEcompact}) in the Rainbow-Ladder truncation (see Figure \ref{fig:FaddeevRLeq}) reads 
\begin{flalign}\label{eq:faddeev_eq}
\Psi_{\alpha\beta\gamma\mathcal{I}}(p,q,P) ={}&
\int_k  \left[ K_{\beta\beta'\gamma\gamma'}(k)~S_{\beta'\beta''}(k_2)
S_{\gamma'\gamma''}(\tilde{k}_3)~
\Psi_{\alpha\beta''\gamma''\mathcal{I}}(p^{(1)},q^{(1)},P)\right.\nonumber\\
&\quad \left. +K_{\alpha\alpha'\gamma\gamma'}(-k)~S_{\gamma'\gamma''}(k_3)
S_{\alpha'\alpha''} (\tilde{k}_1)~
\Psi_{\alpha''\beta\gamma''\mathcal{I}}(p^{(2)},q^{(2)},P)
\right. \nonumber\\
&\quad  \left. + K_{\alpha\alpha'\beta\beta'}(k)~S_{\alpha'\alpha''}(k_1)
S_{\beta'\beta''}(\tilde{k}_2)~
\Psi_{\alpha''\beta''\gamma\mathcal{I}}(p^{(3)},q^{(3)},P)\right] ~,\nonumber\\
\end{flalign}
where we have absorbed the $-i$ factor into the definition $K$, so that it is defined as in (\ref{eq:ladder}),
\begin{equation}
	K_{\alpha\alpha'\beta\beta'}(k)= -4\pi C~Z_2^2 ~\frac{\alpha_{eff}(k^2)}{k^2}~
	T_{\mu\nu}(k)~\gamma^\mu_{\alpha\alpha'} \otimes \gamma^\nu_{\beta\beta'}~,
\end{equation}
where $C=-2/3$ stems from the traces of the color matrices (see Appendix \ref{sec:appendix_color}). The Faddeev amplitudes depend on the quark momenta $p_1$, $p_2$ and $p_3$, but this dependence can be reexpressed in terms of the total momentum $P$ and two relative momenta $p$ and $q$:
\begin{equation}\label{eq:defpq}
\begin{array}{rl@{\quad}rl}
        p &= (1-\zeta)\,p_3 - \zeta (p_1+p_2)\,, &  p_1 &=  -q -\dfrac{p}{2} + \dfrac{1-\zeta}{2} P\,, \\[0.25cm]
        q &= \dfrac{p_2-p_1}{2}\,,         &  p_2 &=   q -\dfrac{p}{2} + \dfrac{1-\zeta}{2} P\,, \\[0.25cm]
        P &= p_1+p_2+p_3\,,                &  p_3 &=   p + \zeta  P~,
\end{array}
\end{equation}
with $\zeta$ is a free momentum partitioning parameter. The total momentum is constrained by $P^2=-M^2$, with $M$ the baryon mass.

Our calculations are performed in Euclidean spacetime. Using the conventions of Appendix \ref{sec:conventions} it is easy to realize that the arguments of the quark dressing functions, $p_i^2$, are complex. For instance,
\begin{equation}
 p_3^2=(p + \zeta  P)\cdot(p + \zeta  P)=p^2-\zeta^2 M^2+2\zeta~i~M\sqrt{p^2}~z_1~.
\end{equation}
where $z_1=\hat{p}\cdot\hat{P}$ (see Appendix \ref{sec:conventions}). Throughout this thesis we use the value $\zeta=\nicefrac{1}{3}$. This choice allows for a tremendous simplification of the bound-state equation's solution method (see Appendix \ref{sec:numdetails}). Moreover, as mentioned before, the non-trivial analytic structure of the quark propagator in the complex plane limits the maximum bound-state available in the approach. The choice $\zeta=\nicefrac{1}{3}$ for the momentum partitioning parameter maximizes the accessible bound-state
mass range before hitting the poles of the quark-propagator \cite{Eichmann2009a}.

The internal quark propagators $S$
depend on the internal quark momenta $k_i=p_i-k$ and $\tilde{k}_i=p_i+k$, with $k$ the gluon momentum.
The internal relative momenta, for each of the three terms in the Faddeev equation, are
\begin{equation}\label{internal-relative-momenta}
\begin{array}{l@{\quad}l@{\quad}l}
p^{(1)} = p+k,& p^{(2)} = p-k,& p^{(3)} = p,\\
q^{(1)} = q-k/2,& q^{(2)} = q-k/2, & q^{(3)} = q+k.
\end{array}
\end{equation}

The tensor structure of the Faddeev amplitudes $\Psi$ is described in Appendix \ref{sec:basis}. The numerical resolution of the Faddeev equation (for details on this, see Appendix \ref{sec:numdetails}) is simplified if one expands the spin part of the amplitudes in an orthonormal basis $\{\tau^{(i)}\}$
\begin{equation}\label{eq:basisexpansion}
 \Psi_{\alpha\beta\gamma\mathcal{I}}(p,q,P)=f^{(i)}(p^2,q^2,z_0,z_1,z_2)~\tau^{(i)}_{\alpha\beta\gamma\mathcal{I}}(p,q,P)\otimes FLAVOR \otimes COLOR~,
\end{equation}
with $z_0=\widehat{p_T}\cdot\widehat{q_T}$,  $z_1=\widehat{p}\cdot\widehat{P}$ and  $z_2=\widehat{q}\cdot\widehat{P}$ and where the subscript $T$ denotes
a transverse projection with respect to the total momentum $P$ and the hat, as before, denotes unit four-vectors. The color and spin parts of the basis are those of the quark model and, since the kernel is flavor and color independent, they factor out.
We then solve for the Lorentz-invariant coefficients $f^{(i)}(p^2,q^2,z_0,z_1,z_2)$. 

From the solution of the Faddeev equation we obtain the bound-state mass and amplitudes. The latter provide information about the internal structure of the baryon that can be used to calculate baryon form factors, as shown in the next chapter.

\subsection{Discussion of the results}

In Table \ref{tab:results} we summarize our results for the vector-meson, spin-$\nicefrac{1}{2}$ and spin-$\nicefrac{3}{2}$ baryons at the $u/d$, $s$, $c$ and $b$ quark masses, and compare with lattice or experimental results, if available. Note that these results differ slightly from those presented in \cite{SanchisAlepuz:2011aa} since, in connection with the calculation of form factors (see next chapter), we needed to repeat the calculations at much higher precision (see Appendix \ref{sec:numdetails}).
\begin{table*}[htp]
\begin{center}
\begin{tabular}{|c@{\!\;\;}l|ccc|} \hline
\multicolumn{2}{|c|}{$J^{PC}=0^{-+}$}  & MT         & AFW         & exp. \\ \hline
$n\overline{n}$ &$(\pi)$               & 0.140\dag  & 0.139\dag   & 0.138 \\
$n\overline{s}$ &$(K)$                 & 0.496\dag  & 0.497\dag   & 0.496 \\
$s\overline{s}$ &                      & 0.697      & 0.686       & -- \\
$c\overline{c}$ &$(\eta_c)$            & 2.979\dag  & 2.980\dag   & 2.980 \\
$b\overline{b}$ &$(\eta_b)$            & 9.388\dag  & 9.390\dag   & 9.391 \\ \hline
\multicolumn{2}{|c|}{$J^{PC}=1^{--}$}  & MT         & AFW         & exp. \\ \hline
$n\overline{n}$ &$(\rho)$              & 0.743      & 0.710       & 0.775 \\
$n\overline{s}$ &$(K^\star)$           & 0.942      & 0.961       & 0.892 \\
$s\overline{s}$ &$(\phi)$              & 1.075      & 1.114       & 1.020 \\
$c\overline{c}$ &$(J/\psi)$            & 3.163      & 3.302       & 3.097 \\
$b\overline{b}$ &$(\Upsilon)$          & 9.466      & 9.621       & 9.460 \\ \hline
\end{tabular} \hspace{5mm}
\begin{tabular}{|c|cccc|} \hline
                    & MT    & AFW   & exp. &   \\ \hline
$N$                 & 0.94  & 0.97  & 0.94 &      \\
$\Delta$            & 1.22  & 1.22  & 1.23 &      \\
$\Omega$            & 1.65  & 1.80  & 1.67 &       \\ \hline
                    & MT    & AFW   & lattice  & LPW  \\ \hline
$\Omega_{ccc}$      & 4.4   & 4.9   & 4.7      & 4.9(0.25)        \\
$\Omega_{bbb}$      & 13.7  & 13.8  & 14.4     & 14.5(0.25)        \\
\hline
\end{tabular}
\caption{Computed meson and baryon masses (in GeV) for both MT and AFW
interactions, compared to experiment. Quantities fitted to their experimental values are
indicated by a $\dag$. Since the heavy-Omega baryons have not been observed yet, we compare
to lattice calculations \cite{Chiu2005a,Lewis2011}
and a recent study from pNRQCD \cite{Llanes-Estrada2011}.
}\label{tab:results}
\end{center}
\end{table*}

Let us analyze first the results at the $u/d$-quark mass. As explained before, both effective interactions are conceived to reproduce pseudoscalar ground-state meson properties at this quark mass and to this end they must include corrections to the quark-antiquark kernel beyond Rainbow-Ladder. The calculated mass for nucleon and Delta agrees very well with the physical mass; the difference between the calculated and the physical mass, and between the result for the two models, is smaller than $1\%$. Assuming that beyond Rainbow-Ladder corrections are the same for both the quark-antiquark and the quark-quark kernels, this seems to indicate that irreducible three-body contributions are negligible for light quarks (in agreement with the success of the quark-diquark approach to baryon properties). The mass of the $\rho$-meson is also well reproduced, although the interaction appears slightly too attractive and the calculated masses are about $4-8\%$ smaller than the physical mass, depending on the model.

\begin{figure}[ht!]
 \begin{center}
  \includegraphics[width=0.6\textwidth,clip]{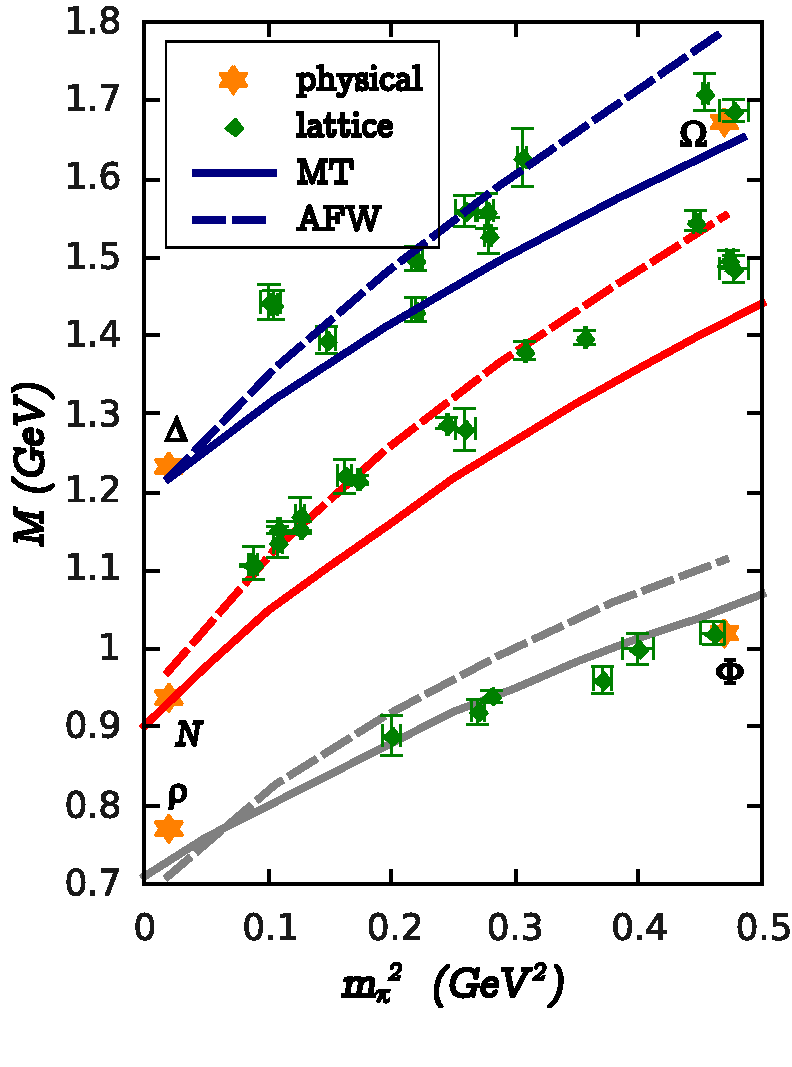}
 \end{center}
 \caption{Evolution of the vector-meson and spin-$\nicefrac{1}{2}$ and spin-$\nicefrac{3}{2}$ baryon masses as a function of the pion mass, for the MT and the AFW models. Results are compared to lattice data \cite{Lin2010,Alexandrou2006b,Bratt2010,Syritsyn2010,Yamazaki2009,Engel2010,Alexandrou2009a,Gattringer:2008vj,Zanotti:2003fx}. Experimental values are denoted by stars.}\label{fig:Massevolution}
\end{figure}

Figure \ref{fig:Massevolution} shows the evolution of the bound-state masses with respect to the squared pion mass. This, by virtue of the Gell-Mann-Oakes-Renner relation \cite{Gell-Mann1968} is equivalent to plotting them with respect the current quark mass; the plot shows results from $u/d$ to $s$ quark masses. Our results are compared with lattice data \cite{Lin2010,Alexandrou2006b,Bratt2010,Syritsyn2010,Yamazaki2009,Engel2010,Alexandrou2009a,Gattringer:2008vj,Zanotti:2003fx} at intermediate quark masses and with experimental values for the $\Phi$-meson and $\Omega$-baryon at the $s$-quark mass. Here one sees that the results for different interactions start to differ as the quark mass increases. AFW provides always a weaker binding and, for baryons, MT results appears to be systematically below the lattice data or the physical Omega mass. Nevertheless, one can say that both interactions compare reasonably well with lattice/experimental data within a $10\%$ accuracy. Thus, from these results we can establish the accuracy of RL calculations of hadron masses to be $\sim 10\%$. Another way of expressing this is that hadron masses seems to be determined at $\sim90\%$ by a single dressed-gluon exchange.

However, there is a missing feature in the quark-mass evolution of the $\Delta$ and $\rho$ masses. In our approach, all particles are considered stable. However, $\Delta$ and $\rho$ can decay into a nucleon or into pions, respectively, via strong interactions and therefore have a significant width. Such a decay would
manifest itself in a non-analytical behavior of the  $\Delta$/$\rho$-masses as
a function of the pion mass when the decay channel would open \cite{Armour2006,Guo2011a,Mader2011}. Our truncation scheme does not provide such a mechanism. The possibility of a decay can be classified as a beyond Rainbow-Ladder effect and, in particular, would lead to a quark-mass dependent interaction.

To get a glimpse of the behavior of the bound-state spectra at even higher quark masses, we calculate the $J/\psi$ and $\Upsilon$ vector-mesons, and triple-charm and triple-beauty Omega-like baryon masses. The results are shown in Table \ref{tab:results}. For the Omega-like baryons the are no experimental values and we compare with lattice calculations \cite{Chiu2005a,Lewis2011} and a perturbative non-relativistic QCD calculation \cite{Llanes-Estrada2011}. At the charm quark-mass the trend of both interactions is maintained; namely, vector-meson masses are overestimated whereas for the Omega-like state the MT and AFW models underestimate and overestimate the lattice result, respectively. Also, we can say that both models agree between each other and with lattice/physical values within $\sim10\%$. The result at the beauty quark-mass is surprising, since both models give essentially the same result. The reason for this is unclear at the moment and deserves further study.

\chapter{Spin-$\nicefrac{3}{2}$ electromagnetic form factors}\label{ch:formfactors}

In the previous chapter we calculated baryon and vector-meson masses using two different effective interactions for the Rainbow-Ladder-truncated three-body Bethe-Salpeter equation. We quantified the model-independent accuracy of Rainbow-Ladder results to be $\sim10$\%. We extend here this study to the electromagnetic properties of spin-$\nicefrac{3}{2}$ baryons with spacelike photon momentum. On one hand we try to confirm the validity of our estimate for the accuracy of Rainbow-Ladder calculations. On the other hand, due to the absence of experimental data and the limitations of theoretical calculations on Delta and Omega electromagnetic form factors, our results constitute qualitative predictions of the shape of these baryons.

\section{Coupling photons to baryons}

A procedure to couple a gauge field to a Green's function described by integral equations, in a way such that gauge symmetry is preserved and there is no overcounting of diagrams has been described in \cite{Kvinikhidze1999b,Kvinikhidze1999a} and named by the authors as \textit{gauging} of equations. Here we reproduce the main steps to derive an expression for a baryon coupled to an external electromagnetic field.

The idea is that, starting with a $2n$-points Green's function
\begin{align}
 G^{(2n)}(x_1\dots x_n;x'_1\dots x'_n)\equiv
\langle 0_{out}| T\left[\psi^{1}(x_1)\dots\psi^{n}(x_n)\bar{\psi}^{1}(x'_1)\dots\bar{\psi}^{n}(x'_n)\right] |0_{in}\rangle~,
\end{align}
to couple the system to an external gauge field $A_\mu$ we need an expression for
\begin{align}
 G^{(2n),\mu}(x_1\dots x_n;x'_1\dots x'_n;y)\equiv
\langle 0_{out}| T\left[\psi^{1}(x_1)\dots\psi^{n}(x_n)\bar{\psi}^{1}(x'_1)\dots\bar{\psi}^{n}(x'_n)\mathcal{J}^\mu(y)\right] |0_{in}\rangle~,
\end{align}
with $\mathcal{J}^\mu$ being the current that couples to the external field. The simplest example of this is the quark propagator
\begin{equation}
 S_i(x_1-x_2)=\langle 0_{out}|T\left[\psi_i(x_1)\bar{\psi}_i(x_2)\right]|0_{in}\rangle~,
\end{equation}
for a quark of species $i$, which is \textit{gauged} to
\begin{equation}
 S_i^\mu(x_1-x_2;y)=\langle 0_{out}|T\left[\psi_i(x_1)\bar{\psi}_i(x_2)\mathcal{J}^\mu(y)\right]|0_{in}\rangle~.
\end{equation}
This also allows to introduce the proper (or amputated) vertex $\Gamma_i^\mu$
\begin{equation}\label{eq:quark_vertex}
 S_i(p_f-p_i)^\mu\equiv S_i(p_f) \Gamma_i^\mu(p_f-p_i) S_i(p_i)~,
\end{equation}
where we used the definition of Green's functions in momentum spacelike
\begin{multline}
 G(p_1\dots p_n;p'_1\dots p'_n)\equiv\\
\int d^4x_1\dots d^4x_nd^4x'_1\dots d^4x'_n e^{-ip_1x_1}\dots e^{-ip_nx_n}e^{ip'_1x'_1}\dots e^{ip'_nx'_n}
G(x_1\dots x_n;x'_1\dots x'_n)~,
\end{multline}
and translation invariance. 

This gauged Green's functions can be obtained by adding a term $-\int \mathcal{J}^\mu(y)A_\mu (y)$ and taking a functional derivative of the corresponding Green's function
\begin{equation}
 G^\mu=-\left.\frac{\delta}{\delta A_\mu}G\right\lvert_{A=0}~.
\end{equation}
Therefore, gauging a Green's function is equivalent to taking a derivative and, thus, follows the same rules when acting, for instance, on products of functions.

The coupling of a baryon (through its constituents) to an external field $A_\mu$ can therefore be obtained by gauging the $6$-points Green's function $G^{(6)}$. Using equation (\ref{eq:compactGreen}) and the \textit{product rule} for gauging a Green's function (and, for the moment, dropping all momentum dependence) we obtain
\begin{equation}
 G^\mu=G_0^\mu-iG_0^\mu KG-iG_0 K^\mu G-iG_0KG^\mu~.
\end{equation}
We can rewrite this as
\begin{flalign}\label{eq:gaugedG}
 G^\mu&=\left(\mathbb{1}+iG_0K\right)^{-1}\left(G_0^\mu-iG_0^\mu KG-iG_0 K^\mu G\right) \nonumber\\
      &=G\left(G_0^{-1}G_0^\mu G_0^{-1}-iK^\mu\right)G
\end{flalign}
where we have used (\ref{eq:compactGreen}) to go from the first to the second line. The proper vertex, $J^\mu$, for the coupling of the three-quark system to the field $A_\mu$ is defined as
\begin{equation}
 G^\mu=G_0J^\mu G_0~.
\end{equation}

If the three quarks form a bound state, both before and after interacting with $A_\mu$, the Green's functions $G$ appearing in (\ref{eq:gaugedG}) will develop a pole. In this work we are interested in virtual\footnote{This is equivalent to say that the momentum \textit{injected} by the external field is spacelike.} interactions with the external field, in which the initial and final baryons are the same. Therefore, taking only the pole contribution, and writing it in terms of the scattering matrix $T$ by means of (\ref{eq:defT}), Equation (\ref{eq:gaugedG}) becomes
\begin{equation}
 J^\mu\sim T\left(G_0^\mu-iG_0 K^\mu G_0\right)T
\end{equation}
or, in terms of the bound-state amplitudes $\Psi$ (see Equation (\ref{eq:Psi_definition}))
\begin{equation}\label{eq:FFeq_compact}
 J^\mu=\bar{\Psi}_f \left(G_0^\mu-iG_0 K^\mu G_0\right)\Psi_i~.
\end{equation}

\begin{figure}[ht!]
 \begin{center}
  \includegraphics[width=0.8\textwidth,clip]{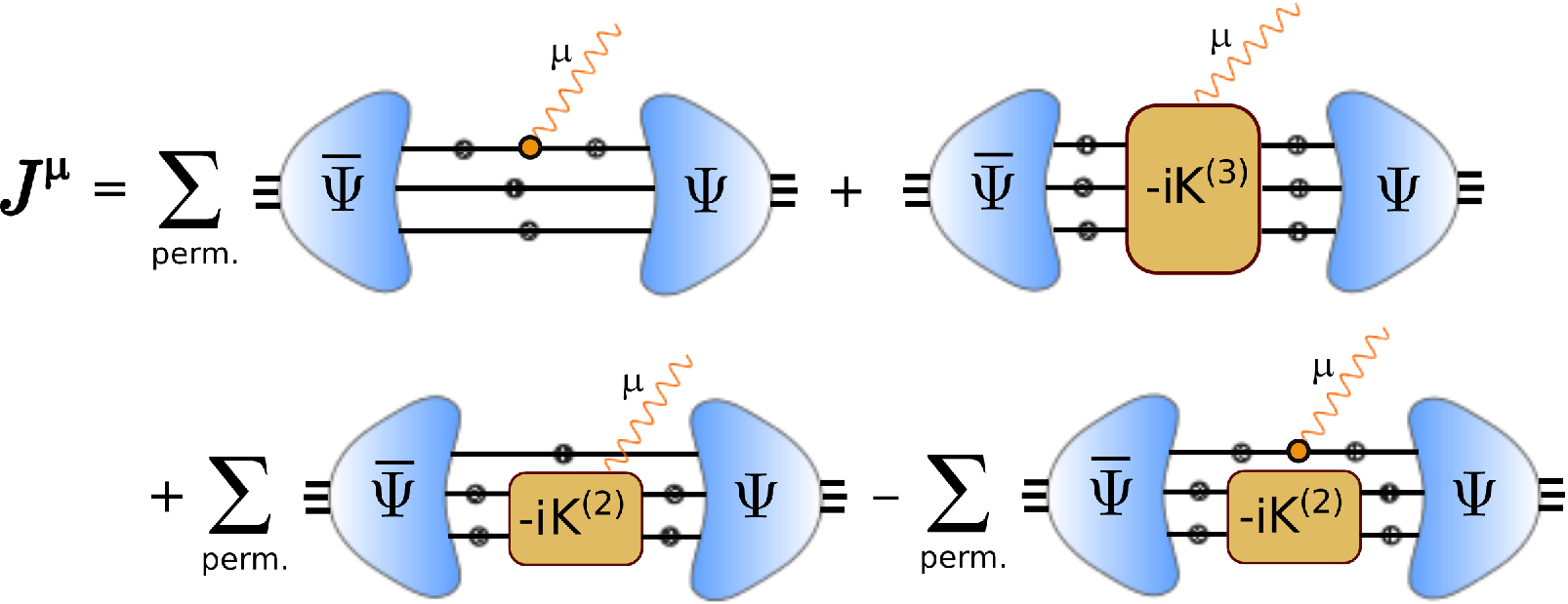}
 \end{center}
 \caption{Diagrammatic representation of Equation (\ref{eq:FFeq_compactlong}).}\label{fig:FFeq_full}
\end{figure}

We need now explicit expressions for $G_0^\mu$ and $K^\mu$. Using (\ref{eq:quark_vertex}) and again the product rule for gauging the equations, one gets
\begin{flalign}
 G_0^\mu=(S_1S_2S_3)^\mu=S_1^\mu S_2S_3+S_1S_2^\mu S_3+S_1S_2S_3^\mu=\nonumber \\
S_1^f\Gamma_1^\mu S_1^i S_2S_3+S_1S_2^f\Gamma_2^\mu S_2^i S_3+ S_1 S_2S_3^f\Gamma_3^\mu S_3^i~.
\end{flalign}
To find an expression for $K^\mu$, one must recall that $K$ is defined as the sum
\begin{equation}
 K\equiv\sum_{\ell=2,3}\widetilde{K}^{(\ell)}
\end{equation}
with
\begin{align}
 \widetilde{K}^{(\ell)}=\sum_{
\begin{array}{c}
 \textnormal{{\tiny cyclic~perm.}}\\
 \textnormal{{\tiny $\ell$ elements}}
\end{array}}K^{(\ell)}\underbrace{S^{-1}\dots S^{-1}}_{3-\ell}~.
\end{align}
Therefore, to gauge $\widetilde{K}^{(2)}$ one must also gauge the inverse quark propagator $S^{-1}$. To do this we define $\mathbb{1}^\mu=0$, and thus
\begin{equation}
 \mathbb{1}^\mu=\left(SS^{-1}\right)^\mu=S^\mu S^{-1}+S(S^{-1})^\mu~,
\end{equation}
and from here
\begin{equation}
 (S^{-1})^\mu=-S^{-1}S^\mu S^{-1}~.
\end{equation}
We can now write (\ref{eq:FFeq_compact}) as
\begin{equation}\label{eq:FFeq_compactlong}
 J^\mu=\bar{\Psi}_f \left(G_0^\mu-iG_0 K^{(3),\mu} G_0-i\sum_{\textnormal{{\tiny perm.}}}SS^fS^fK^{(2),\mu}S^iS^i
+i\sum_{\textnormal{{\tiny perm.}}}S^fS^fS^fK^{(2)}\Gamma^{\mu}S^iS^iS^i\right)\Psi_i~.
\end{equation}
A diagrammatic representation of this equation is shown in Figure \ref{fig:FFeq_full}.

\begin{figure}[hbtp]
 \begin{center}
  \includegraphics[width=0.8\textwidth,clip]{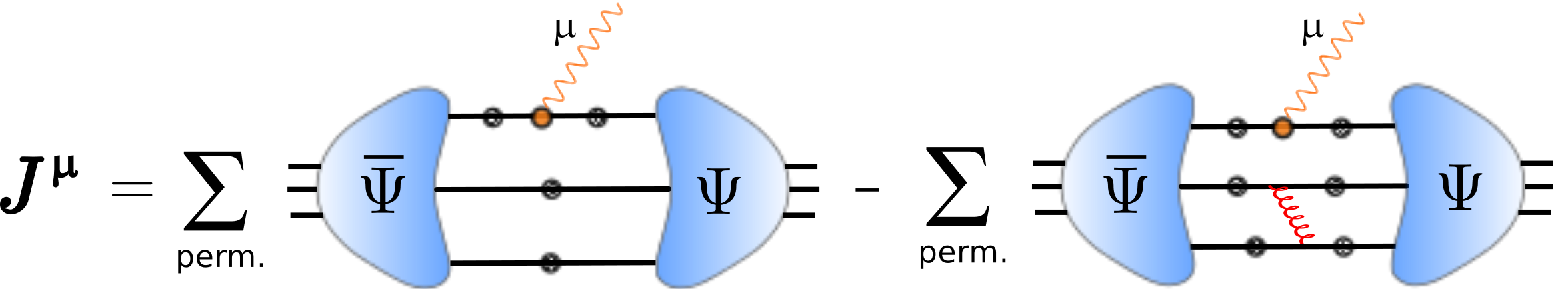}
 \end{center}
 \caption{Diagrammatic representation of Equation (\ref{eq:FFeq_compactlong}) in the Rainbow-Ladder truncation.}\label{fig:FFeq_RL}
\end{figure}

The expression for $J^\mu$ in the Rainbow-Ladder truncation scheme is considerably simpler. First of all, the three-body irreducible kernel $K^{(3)}$ is absent and so is $K^{(3),\mu}$. Furthermore, since the two-body kernel is reduced to a single gluon exchange with all dressings embraced in the effective coupling $\alpha_{eff}$, it does not couple to the external field $A_\mu$, which means that the term $K^{(2),\mu}$ is also absent. This simplified expression is illustrated in Figure \ref{fig:FFeq_RL}. This scheme has been used in \cite{Eichmann2011a,Eichmann2012a} to calculate nucleon electromagnetic, axial and pseudoscalar form factors using the covariant Faddeev equation. 

For spin-$\nicefrac{3}{2}$ baryons, and restoring all indices and momentum dependence, it reads
\begin{flalign}\label{eq:FFeqRL}
 J_{\mathcal{I}'\mathcal{I}}^\mu=\int_p\int_q\bar{\Psi}_{\beta'\alpha'\mathcal{I}'\gamma'}(p_f^{\{1\}},q^{\{1\}}_f,P_f)\left[\left(S(p_1^f)\Gamma^\mu(p_1,Q)S(p_1^i)\right)_{\alpha'\alpha}S_{\beta'\beta}(p_2)S_{\gamma'\gamma}(p_3)\right]\times\nonumber\\
~~~~~~~~~~~~~~~~~~~~~~~~~~~~~~\left(\Psi_{\alpha\beta\gamma\mathcal{I}}(p^{\{1\}}_i,q^{\{1\}}_i,P_i)-\Psi^{\{1\}}_{\alpha\beta\gamma\mathcal{I}}(p^{\{1\}}_i,q^{\{1\}}_i,P_i)\right)\nonumber\\
+\int_p\int_q\bar{\Psi}_{\beta'\alpha'\mathcal{I}'\gamma'}(p^{\{2\}}_f,q^{\{2\}}_f,P_f)\left[S_{\alpha'\alpha}(p_1)\left(S(p_2^f)\Gamma^\mu(p_2,Q)S(p_2^i)\right)_{\beta'\beta}S_{\gamma'\gamma}(p_3)\right]\times\nonumber\\
~~~~~~~~~~~~~~~~~~~~~~~~~~~~~~\left(\Psi_{\alpha\beta\gamma\mathcal{I}}(p^{\{2\}}_i,q^{\{2\}}_i,P_i)-\Psi^{\{2\}}_{\alpha\beta\gamma\mathcal{I}}(p^{\{2\}}_i,q^{\{2\}}_i,P_i)\right)\nonumber\\
+\int_p\int_q\bar{\Psi}_{\beta'\alpha'\mathcal{I}'\gamma'}(p^{\{3\}}_f,q^{\{3\}}_f,P_f)\left[S_{\alpha'\alpha}(p_1)S_{\beta'\beta}(p_2)\left(S(p_3^f)\Gamma^\mu(p_3,Q)S(p_3^i)\right)_{\gamma'\gamma}\right]\times\nonumber\\
~~~~~~~~~~~~~~~~~~~~~~~~~~~~~~\left(\Psi_{\alpha\beta\gamma\mathcal{I}}(p^{\{3\}}_i,q^{\{3\}}_i,P_i)-\Psi^{\{3\}}_{\alpha\beta\gamma\mathcal{I}}(p^{\{3\}}_i,q^{\{3\}}_i,P_i)\right)~,
\end{flalign}
where we define
\begin{flalign}
 \Psi^{\{1\}}_{\alpha\beta\gamma\mathcal{I}}=
\int_k  K_{\beta\beta'\gamma\gamma'}(k)~S_{\beta'\beta''}(p_2-k)
S_{\gamma'\gamma''}(p_3+k)~
\Psi_{\alpha\beta''\gamma''\mathcal{I}}(p+k,q-k/2,P)~,
\end{flalign}
as the result of the first term in the Faddeev equation (\ref{eq:faddeev_eq}) and in a similar fashion we define $\Psi^{\{2\}}$ and $\Psi^{\{3\}}$. We have introduced the \textit{injected} momentum $Q$ via the final and initial momenta of the interacting quark $\kappa$
\begin{equation}
 p_\kappa^{\nicefrac{f}{i}}=p_\kappa\pm\frac{Q}{2}
\end{equation}
The relative momenta in the respective terms of (\ref{eq:FFeqRL}) are, using the definitions in (\ref{eq:defpq}),
\begin{align}\label{eq:relative_momenta_withQ}
%\begin{array}{lcl}
 p_{\nicefrac{f}{i}}^{\{1\}}&=p\mp\zeta\frac{Q}{2}~,&q_{\nicefrac{f}{i}}^{\{1\}}&=q\mp\frac{Q}{4} \nonumber\\
 p_{\nicefrac{f}{i}}^{\{2\}}&=p\mp\zeta\frac{Q}{2}~,&q_{\nicefrac{f}{i}}^{\{2\}}&=q\pm\frac{Q}{4} \\
 p_{\nicefrac{f}{i}}^{\{3\}}&=p\pm(1-\zeta)\frac{Q}{2}~,&q_{\nicefrac{f}{i}}^{\{3\}}&=q \nonumber
%\end{array}
\end{align}
and since the initial and final states are on-shell, the total momenta are constrained to $P_i^2=P_f^2=-M^2$, with $M$ the mass of the bound state. As is the case for the Faddeev equation, we show in Appendix \ref{sec:numdetails} that the three terms in (\ref{eq:FFeqRL}) are formally the same when the momentum partitioning parameter is chosen $\zeta=1/3$.

\subsection{Normalization of the Bethe-Salpeter amplitudes}

The Bethe-Salpeter equation (\ref{eq:3bBSEcompact}), and its Rainbow-Ladder truncated version (\ref{eq:faddeev_eq}), are homogeneous and linear integral equations for $\Psi$. Therefore, its solutions $\Psi$ are defined up to an irrelevant constant. This is not the case, however, for the calculation of the current (\ref{eq:FFeq_compact}). Nevertheless, the pole ansatz for the scattering matrix $T$ (see Equation (\ref{eq:Psi_definition})) also fixes a condition for the normalization of the Bethe-Salpeter amplitudes. This condition can be expressed in two different, but equivalent, ways \cite{Cutkosky1964,Nakanishi1965}.

We start with the equation for the scattering matrix
\begin{equation}
 T=-iK-iKG_0T~.
\end{equation}
As explained in Appendix \ref{sec:numdetails}, the Bethe-Salpeter equation is solved as an eigenvalue equation, the real solution corresponding to the eigenvalue $\lambda=1$. Thus, we introduce a fictitious parameter $\lambda$, which would correspond the the eigenvalue of the bound-state equation when $T$ develops a pole, in the equation above (which we can set to $1$ at the end)
\begin{equation}
 T=-iK-i\lambda KG_0T~,
\end{equation}
or
\begin{equation}
 T=\left(\mathbb{1}+i\lambda KG_0\right)^{-1}\left(-iK\right)~.
\end{equation}
Taking a derivative with respect to $\lambda$
\begin{equation}\label{eq:step1normalization}
 \frac{dT}{d\lambda}=-\left(\mathbb{1}+i\lambda KG_0\right)^{-2}\left(iKG_0\right)\left(-iK\right)=-\frac{KG_0K}{\left(\mathbb{1}+i\lambda KG_0\right)^{2}}=TG_0T~.
\end{equation}
Now we want to write this in terms of the Bethe-Salpeter amplitudes. On the bound-state pole we write
\begin{equation}\label{eq:Psi_withlambda}
 T\sim \frac{\Psi\bar{\Psi}}{P^2+M^2(\lambda)}~,
\end{equation}
where the dependency on the eigenvalue appears through the bound-state mass $M(\lambda)$. Equation (\ref{eq:step1normalization}) becomes
\begin{equation}
 -\frac{\Psi\bar{\Psi}}{(P^2+M^2(\lambda))^2}\frac{dM^2}{d\lambda}=\frac{\Psi\bar{\Psi}G_0\Psi\bar{\Psi}}{(P^2+M^2(\lambda))^2}~,
\end{equation}
and from here
\begin{equation}\label{eq:nakanishi_norm}
 -\frac{dM^2}{d\lambda}=\bar{\Psi}G_0\Psi~.
\end{equation}
This is the Nakanishi condition \cite{Nakanishi1965} for the Bethe-Salpeter amplitudes $\Psi$.

Equation (\ref{eq:nakanishi_norm}) can be written in a way such that the eigenvalue $\lambda$ does not appear explicitly. Using the Bethe-Salpeter equation
\begin{equation}
 \Psi=-i\lambda KG_0\Psi
\end{equation}
we obtain,
\begin{equation}
 i\bar{\Psi}K^{-1}\Psi=\lambda\bar{\Psi}G_0\Psi~.
\end{equation}
Now, since $P^2=-M^2$, the eigenvalue $\lambda(M^2)$ can be equivalently considered as a function of $P^2$. Taking a derivative with respect to $P^2$,  we get
\begin{equation}
\frac{d\lambda}{dP^2}\bar{\Psi}G_0\Psi=\bar{\Psi}\left(i\frac{dK^{-1}}{dP^2}-\lambda\frac{dG_0}{dP^2}\right)\Psi~.
\end{equation}
Finally, using the normalization condition (\ref{eq:nakanishi_norm}), and afterwards setting $\lambda=1$ we find another condition for the normalization of the Bethe-Salpeter amplitudes
\begin{equation}
\bar{\Psi}\left(\frac{dG_0}{dP^2}-i\frac{dK^{-1}}{dP^2}\right)\Psi=1.
\end{equation}
This is the Leon-Cutkosky normalization condition \cite{Cutkosky1964}.

\subsection{Quark-Photon vertex}

To solve Equation (\ref{eq:FFeqRL}), the quark-photon vertex $\Gamma^\mu(k_1,k_2,Q)$ has to be specified. To put the calculation of form factors into the same frame as meson and baryon calculations using covariant bound-state equations, we determine the quark-photon vertex as a solution of an inhomogeneous Bethe-Salpeter equation \cite{Roberts1994,Maris:1997hd,Maris2000b}
\begin{equation}\label{eq:vertex}
 \Gamma^\mu_{ab}(k,Q)=Z_2\gamma^\mu_{ab}+\int_r\Gamma^\mu_{a''b''}(r,Q)S_{a''a'}(k_1-r)S_{b''b'}(k_2+r)K_{a''a,b''b}(k_1,k_2;r)
\end{equation}
with $k=(k_1+k_2)/2$, $Q=k_2-k_1$ and $Z_2$ the quark wave-function renormalization constant. We defined
\begin{equation}
 \Gamma^\mu(k_1,k_2,Q)=-i\mathcal{Q}_i(2\pi)^4\delta{(4)}(k_1-k_2+Q) \Gamma^\mu(k,Q)
\end{equation}
with $\mathcal{Q}_i$ the electric charge of the quark $i$.

\begin{figure}[hbtp]
 \begin{center}
  \includegraphics[width=0.53\textwidth,clip]{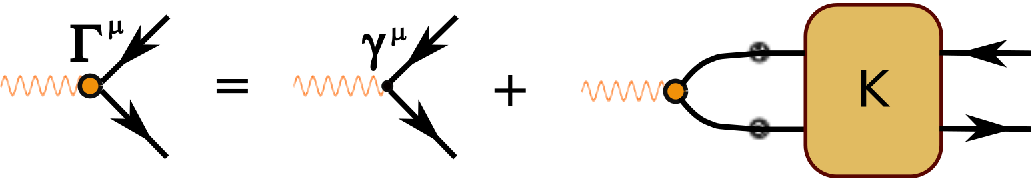}
 \end{center}
 \caption{Diagrammatic representation of the inhomogeneous Bethe-Salpeter equation for the quark-photon vertex.}
\end{figure}

The kernel $K$ is the quark-antiquark scattering kernel. Conservation of the electromagnetic current (\ref{eq:FFeqRL}) requires that the bound-state amplitudes are normalized according to (\ref{eq:nakanishi_norm}) and the the interaction kernel in the quark-photon vertex is truncated in the same way \cite{Maris2000b} as in (\ref{eq:FFeqRL}). Therefore, to solve the quark-photon vertex we also use a Rainbow-Ladder quark-antiquark interaction kernel. 

\section{Electromagnetic current}

In the last section we derived an expression for the electromagnetic current in terms of the photon interaction with the quarks forming a baryon. On the other hand, the form of the current is constrained by Lorentz covariance and current conservation to be a linear combination of a finite numbers of Lorentz covariants with scalar coefficients (see, e.g., \cite{0521670535}). These coefficients are the form factors.

The electromagnetic current for a spin-$\nicefrac{3}{2}$ particle is characterized by four form factors $F_i(Q^2)$ \cite{Nozawa1990}. An expression useful for our purposes has been derived in \cite{Nicmorus2010}. It reads
\begin{flalign}
 J^{\mu,\alpha\beta}(P,Q)={}&\mathbb{P}^{\alpha\alpha'}(P_f)\left[\left((F_1+F_2)i\gamma^\mu-F_2\frac{P^\mu}{M}\right)\delta^{\alpha'\beta'}\right.\nonumber \\
                          &+\left.\left((F_3+F_4)i\gamma^\mu-F_4\frac{P^\mu}{M}\right)\frac{Q^{\alpha'}Q^{\beta'}}{4M^2}\right]\mathbb{P}^{\beta'\beta}(P_i)
\end{flalign}
where $\mathbb{P}$ is the Rarita-Schwinger projector (\ref{eq:def_projectors}), $P_i$ and $P_f$ are the initial and final baryon total momenta, respectively,  $Q=P_f-P_i$ is the photon momentum (see Appendix \ref{sec:conventions}), $M$ is the baryon mass and $P=(P_f+P_i)/2$. The form factors that are measured experimentally are the electric monopole ($G_{E_0}(Q^2)$), magnetic dipole ($G_{M_1}(Q^2)$), electric quadrupole ($G_{E_2}(Q^2)$) and magnetic octupole ($G_{M_3}(Q^2)$) form factors. They are related to the $F_i's$ via
\begin{equation}
\begin{aligned}
G_{E_0} &= \left(1+\frac{2\tau}{3}\right) ( F_1 - \tau F_2)  - \frac{\tau}{3} (1+\tau) \,( F_3 - \tau F_4) \,, \\
G_{M_1} &= \left(1+\frac{4\tau}{5}\right) (F_1+F_2) - \frac{2\tau}{5} (1+\tau)\, (F_3 + F_4)\,,\\
G_{E_2} &= (F_1 - \tau F_2) - \frac{1}{2}\,(1+\tau) \,(F_3 - \tau F_4)\,,\\
G_{M_3} &= (F_1 + F_2) - \frac{1}{2}\,(1+\tau) \,(F_3 + F_4)\,,
\end{aligned} 
\end{equation}
with $\tau=Q^2/4M^2$. It is shown in \cite{Nozawa1990} that if the baryon is spherically symmetric $G_{E_2}$ and $G_{M_3}$ must vanish; therefore they measure the deformation of the object. At $Q^2=0$ the form factors define the electric charge ($e_{\nicefrac{3}{2}}$), magnetic dipole moment ($\mu_{\nicefrac{3}{2}}$), electric quadrupole moment ($\mathcal{Q}_{\nicefrac{3}{2}}$) and magnetic octupole moment ($\mathcal{O}_{\nicefrac{3}{2}}$) of a spin-$\nicefrac{3}{2}$ particle,
\begin{equation}
\begin{aligned}
e_{\nicefrac{3}{2}}&=G_{E_0}(0)\,, \\
\mu_{\nicefrac{3}{2}}&=\frac{e}{2M}G_{M_1}(0)\,,\\
\mathcal{Q}_{\nicefrac{3}{2}}&=\frac{e}{M^2}G_{E_2}(0)\,,\\
\mathcal{O}_{\nicefrac{3}{2}}&=\frac{e}{2M^3}G_{M_3}(0)\,.\\
\end{aligned} 
\end{equation}

Once the electromagnetic current is calculated from (\ref{eq:FFeqRL}), the form factors can be extracted using the expressions \cite{Nicmorus2010}
\begin{equation}
\begin{aligned}\label{eq:expressionGs}
G_{E_0} &= \frac{s_2-2s_1}{4i\sqrt{1+\tau}}\,, \\
G_{M_1} &= \frac{9i}{40\,\tau}\left(s_4-2s_3\right)\,, \\
G_{E_2} &= \frac{3}{8i\,\tau^2\sqrt{1+\tau}} \left[ 2s_1 \left(\tau+\frac{3}{2}\right) - \tau s_2 \right], \\
G_{M_3} &= \frac{3i}{16\,\tau^3} \left[ 2s_3 \left(\tau+\frac{5}{4}\right) - \tau s_4 \right]~,
\end{aligned}
\end{equation}
where
\begin{equation} \label{sses}
\begin{aligned} 
s_1 &= \textnormal{Tr}\left\{ J^{\mu,\alpha\beta}  \hat{P}^\mu  \hat{P}^\alpha  \hat{P}^\beta \right\}~, \\
s_2 &= \textnormal{Tr}\left\{ J^{\mu,\alpha\alpha}  \hat{P}^\mu \right\}~, \\
s_3 &= \textnormal{Tr}\left\{ J^{\mu,\alpha\beta} \,\gamma^\mu_T  \hat{P}^\alpha  \hat{P}^\beta \right\}~, \\
s_4 &= \textnormal{Tr}\left\{ J^{\mu,\alpha\alpha} \,\gamma^\mu_T \right\} \,,
\end{aligned}
\end{equation}
with $\gamma^\mu_T=T_P^{\mu\nu}\gamma^\nu$.

\section{Results}

Having solved the covariant Faddeev equation (\ref{eq:faddeev_eq}), we have available not only the baryon mass but also all the information about its internal structure through the Faddeev amplitudes $\Psi$. In this section we use this information to calculate the electromagnetic form factors of the Delta baryon. As in previous chapter, we repeat the calculation for the two different models for the effective interaction described in Section \ref{sec:effective_interactions}.

\begin{figure}[hbtp]
 \begin{center}
  \includegraphics[width=\textwidth,clip]{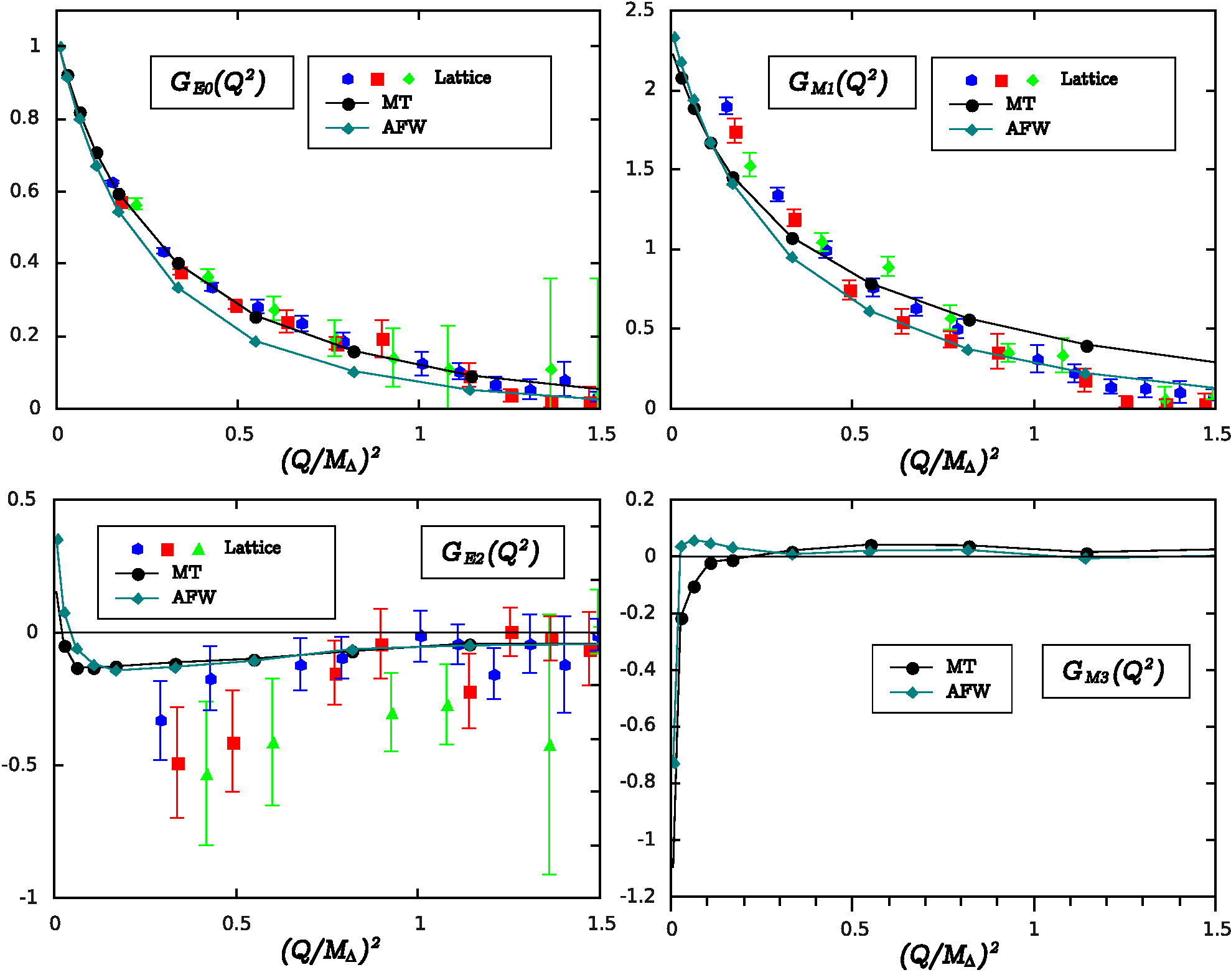}
 \end{center}
 \caption{Electromagnetic form factors for the $\Delta^+$ for the MT and AFW models. The results are compared to lattice data \cite{Alexandrou2009a} for dynamical Wilson fermions at $m_\pi=384$~MeV (green), $m_\pi=509$~MeV (red) and $m_\pi=691$~MeV (blue).}\label{fig:FFdelta}
\end{figure}

In Figure \ref{fig:FFdelta} we show our results for the $\Delta^+$ form factors. It is worth noting that, due to isospin symmetry, the form factors for $\Delta^{++}$, $\Delta^{0}$ and $\Delta^{-}$ differ from those of the $\Delta^+$ only by a factor corresponding to their charge, as shown in Appendix \ref{sec:numdetails}. This means, in particular, that all form factors for $\Delta^{0}$ vanish identically in our approach\footnote{This is not the case for spin-\nicefrac{1}{2} baryons. In this case, due to the mixed-symmetry properties of the flavor part of the Faddeev amplitude, the neutron has non-vanishing form factors \cite{Eichmann2011a}.}.

Due to the short lifetime of the Delta resonance, it is very difficult to study its properties experimentally. In fact, the experimental information on the Delta electromagnetic properties is restricted to the $\Delta^{++}$ and $\Delta^+$ magnetic dipoles and, even in these cases, the experimental uncertainties make these values unreliable. For this reason, we compare our results to a lattice calculation with dynamical Wilson fermions at different pion masses \cite{Alexandrou2009a}. Lattice calculations, however, also suffer from large errors, especially for the electric quadrupole form factor. For the magnetic octupole there are no lattice results whatsoever. Moreover, the low photon-momentum regime is inaccessible to lattice calculations. An interesting calculation using a constituent spectator quark model has been presented in \cite{Ramalho2009,Ramalho2010}. In this work the authors restrict the Delta to be an admixture of s- and d-waves, and fit their parametrization of the wave function to lattice data. Using these fits they are able to calculate the form factors, thereby providing a bridge from lattice to continuum calculations. When lattice data is not available, we will use these fits to compare qualitatively with our results.

The evolution of the electric monopole form factor $G_{E0}$ with the photon momentum $Q^2$ is shown in the upper-left panel of Figure \ref{fig:FFdelta}. Note that it is ambiguous to compare the result of different calculations because each of them gives a different result for the baryon mass. Therefore, to eliminate this scale dependence we plot the results with respect to $Q^2/M^2$. For $G_{E0}$, there is a good agreement of our results with lattice data as well as a qualitative model independence. As we illustrate in the next section, this result is a manifestation of the fact that $G_{E0}$ is relatively insensitive to the detailed internal structure of the baryon.

The calculated charge radius
\begin{equation}
 <r_{E0}^2>=-\frac{6}{G_{E0}(0)}\frac{dG_{E0}(Q^2)}{dQ^2}
\end{equation}
is shown in Table \ref{tab:delta_moments} for the MT and AFW models. Compared to the lattice results, our values appear considerably overestimated. A possible explanation is the pion-mass dependence of the charge radius, which grows as the pion mass approaches the physical value from above. Moreover, chiral perturbation theory shows than when the $\Delta\to N\pi$ decay channel opens the charge radius changes abruptly to a lower value \cite{Ledwig2012}. Since in our approach we do not have a mechanism for the Delta to decay, it is therefore reasonable that we obtain a higher result for $<r_{E0}^2>$. A combination of these effects could lead to a better agreement between lattice results and ours. Incidentally, a number of effective models such as Goldstone-boson exchange models and many others (see \cite{Ramalho2010} for a collection of results), give also a large value for $<r_{E0}^2>$.

\begin{table*}[htp]
\begin{center}
\small
\renewcommand{\arraystretch}{1.2}
\begin{tabular}{c|cc|ccc|c|} 
~ & F-MT & F-AFW & DW1 & DW2 & DW3 & Exp.\\
\hline\hline
$<r_{E0}^2> (\textnormal{fm}^2)$ & 0.67 & 0.60 & 0.373~(21) & 0.353~(12) & 0.279~(6) &
\\
\hline
$G_{M1}(0)$ & 2.22 & 2.33 & 2.35~(16) & 2.68~(13) & 2.589~(78) & 3.54$^{+4.59}_{-4.72}$ \\ \hline
\end{tabular}
\caption{Comparison of results for the charge radius $<r_{E0}^2>$ and for $G_{M1}(0)~~(\propto\mu)$. We compare our results for the MT model (F-MT) and for the AFW model (F-AFW) to a lattice calculation with dynamical Wilson fermions at $m_\pi=384$~MeV (DW1), $m_\pi=509$~MeV (DW2) and $m_\pi=691$~MeV (DW3) \cite{Alexandrou2009c,Alexandrou2009a}. For $G_{M1}(0)$ we also compare to the experimental value \cite{Kotulla2002a,Nakamura2010o}.}\label{tab:delta_moments}
\end{center}
\end{table*}

In the upper-right panel of Figure \ref{fig:FFdelta} we plot the magnetic dipole form factor. The dimensionless magnetic dipole moment $G_{M1}(0)$ is $~2.22~$ and $~2.33~$ for MT and AFW, respectively. This is to be compared with the experimental value $~3.54^{+4.59}_{-4.72}$ \cite{Kotulla2002a,Nakamura2010o} or the lattice extrapolated result $~2.35\pm0.16$ \cite{Alexandrou2009a} (see Table \ref{tab:delta_moments}). For the $Q^2$-evolution we observe again a qualitative model independence, but the behavior in both cases is significantly different to that of lattice data.

The deformation of the Delta is signaled by the electric quadrupole and magnetic octupole form factors. In the non-relativistic limit, a negative or positive electric quadrupole moment corresponds to an oblate or a prolate charge distribution, respectively. In our calculation, and for both effective interactions, $G_{E2}$ starts with a positive value at low-$Q^2$, changes sign and reaches a minimum at $Q^2/M_{\Delta}^2\sim 0.1$ and then approaches zero from the negative sign, in agreement with lattice data. The behavior at low-$Q^2$ is in contradiction with the constituent-quark calculation in \cite{Ramalho2009,Ramalho2010}, with an extrapolation of lattice data \cite{Alexandrou2009a} and with a covariant quark-diquark calculation \cite{Nicmorus2010}. In this respect it is important to point out the $1/Q^4$ factor in the expression for $G_{E2}$ in (\ref{eq:expressionGs}). This entails that very precise cancellations must take place in the denominator in (\ref{eq:expressionGs}) to cancel the $Q^4$ factor and get finite results close to $Q=0$. This is difficult to achieve numerically and, in fact, we needed to solve the Faddeev equation at the highest precision available to obtain reliable results at a reasonable small $Q$. Using as a measure of the numerical accuracy the value of the imaginary parts of the form factors (which, ideally, should be zero), we plot those values for which the ratio between the imaginary and the real parts of all form factors is smaller than $10^{-9}$; the minimum $Q^2$-value we are able to achieve in this way is $\sim0.01$. In any case, this low-$Q^2$ behavior must be considered with caution and further studies are required. If this feature of $G_{E2}$ is confirmed, it will be a true prediction relying on the high orbital angular momentum components of the Faddeev amplitude, as shown in next section.

A similar situation appears for the magnetic octupole form factor $G_{M3}$. In this case we obtain a small, but non-zero, positive form factor at high-$Q^2$, with a similar behavior for both models. At low-$Q^2$ the models differ, but both seem to feature a pump and a zero crossing (i.e. a change of sign of the form factor). At $Q\sim0$, a factor $1/Q^6$ must be canceled by the numerator in (\ref{eq:expressionGs}) and, therefore, the result here must be taken with due caution. It is remarkable, nevertheless, that this behavior of $G_{M3}$ is qualitatively similar to that of \cite{Ramalho2009,Ramalho2010}. These features at low-$Q^2$ are also absent if one ignores the subleading components in the Faddeev amplitude (see next section).

In summary, we find a reasonable model independence of our results, which gives us confidence that we can make statements about the Rainbow-Ladder truncation of the baryon BSE. The models slightly disagree at low-$Q^2$ for the electric quadrupole and the magnetic octupole form factors but, as we discussed above, it is not clear at the moment to what extent this is a numerical artifact.

To conclude, it is noteworthy that in a covariant approach there is no freedom to choose the relative importance of the different quark-spin ar quark orbital angular momentum components. Once the parameters of the interaction are chosen, the contribution of each component is determined by the Faddeev equation. This is in contrast with, for example, the constituent-quark calculation in \cite{Ramalho2009,Ramalho2010} where the different angular-momentum contributions have to be fixed \textit{a priori} and fitted afterwards to lattice data.

\subsection{Role of the subleading components}

\begin{figure}[hbtp]
 \begin{center}
  \includegraphics[width=\textwidth,clip]{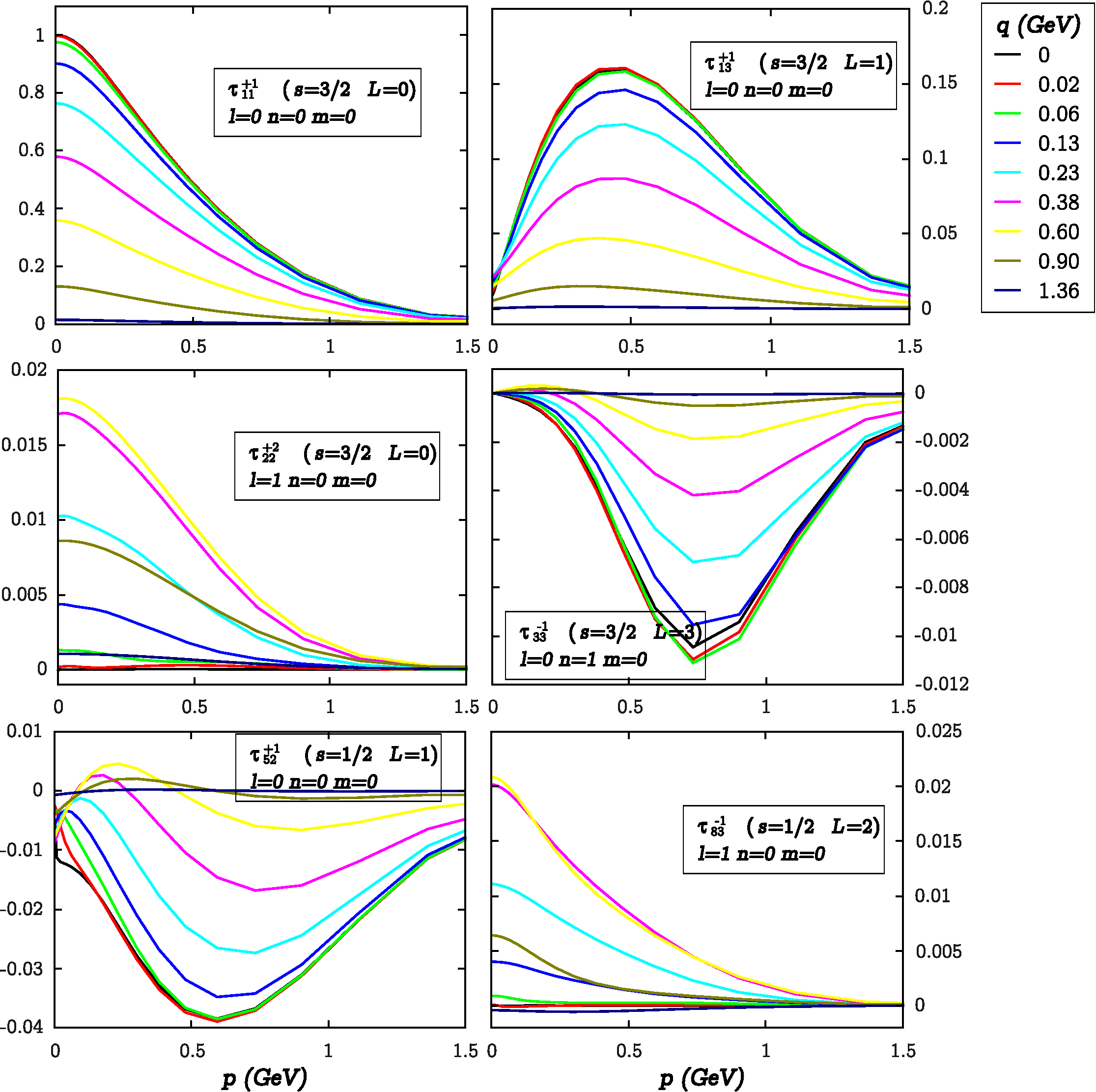}
 \end{center}
 \caption{Dominant Chebyshev moments in each of the $(s,L)$ sectors (we change here the notation for the orbital angular momentum from $\ell$ to $L$ for clarity), as a function of $p$ and for different values of $q$ in the Maris-Tandy model. The amplitudes are normalized such that $f^{(1)}_{000}(q^2=0,p^2=0)=1$.}\label{fig:amplitudesMT}
\end{figure}

The Poincar\'e-covariant description of spin-$\nicefrac{3}{2}$ baryons forces us to use a basis of dimension 128 for the Faddeev amplitudes\footnote{For spin-$\nicefrac{1}{2}$ baryons the dimension of the basis is 64.}, as shown in Appendix \ref{sec:basis}. This basis can, in turn, be classified in terms of the quark spin ($s$) and orbital angular momentum ($\ell$). At first sight, this seems an exceedingly complicated way of describing a baryon. In fact, for spin-$\nicefrac{3}{2}$ baryons the s-wave sector ($s=0$ and $\ell=0$) is dominant. To illustrate this we have plotted in Figure \ref{fig:amplitudesMT} the dominant amplitudes in each of the $(s,\ell)$ sectors (specifically, we plot the dominant Chebyshev moment, as defined in (\ref{eq:cheby_expansion}), of this amplitude)  for the Delta in the MT model, although a the same reasoning applies for the AFW model.

To evaluate the relevance of the subleading components we performed the following calculation. After solving the Faddeev equation for the MT model and using all 128 basis elements we isolate those components $f^{(i)}_{lnm}$ such that\footnote{Note that the normalization of the Faddeev amplitudes for this exercise is arbitrarily fixed such that $f^{(1)}_{000}(p^2=0,q^2=0)=1$. This normalization is not related to the physical normalization given by (\ref{eq:nakanishi_norm}), which is used for the calculation of form factors.}
\begin{equation}\label{eq:criteria_zero}
 \frac{1}{N_{points}}\sum_{ij}\sqrt{\left(f^{(i)}_{lnm}(p^2_i,q^2_j)\right)^2}< 10^{-6}
\end{equation}
with $\{p_i,q_j\}$ the quadrature points used in the numerical integration and $N_{points}$ the number of quadrature points used. Note that this condition especially implies that each of these components is a factor of, at least, $10^{-7}$ smaller than the dominant one.

We then remove this components from the basis and solve again the Faddeev equation using the remaining basis subset. The resulting Delta mass is $1.30 GeV$, to be compared with the result of the full calculation $M_\Delta=1.22 GeV$. That is, the difference between both results is about $6\%$, and the mass with the reduced basis is only $2.3\%$ above the physical Delta mass. This might suggest that, indeed, only the dominant components (i.e. s-waves) are necessary in the calculation (especially if, as we concluded in the last chapter, the results in Rainbow-Ladder are expected to be precise up to $\sim 10\%$).

\begin{figure}[hbtp]
 \begin{center}
  \includegraphics[width=\textwidth,clip]{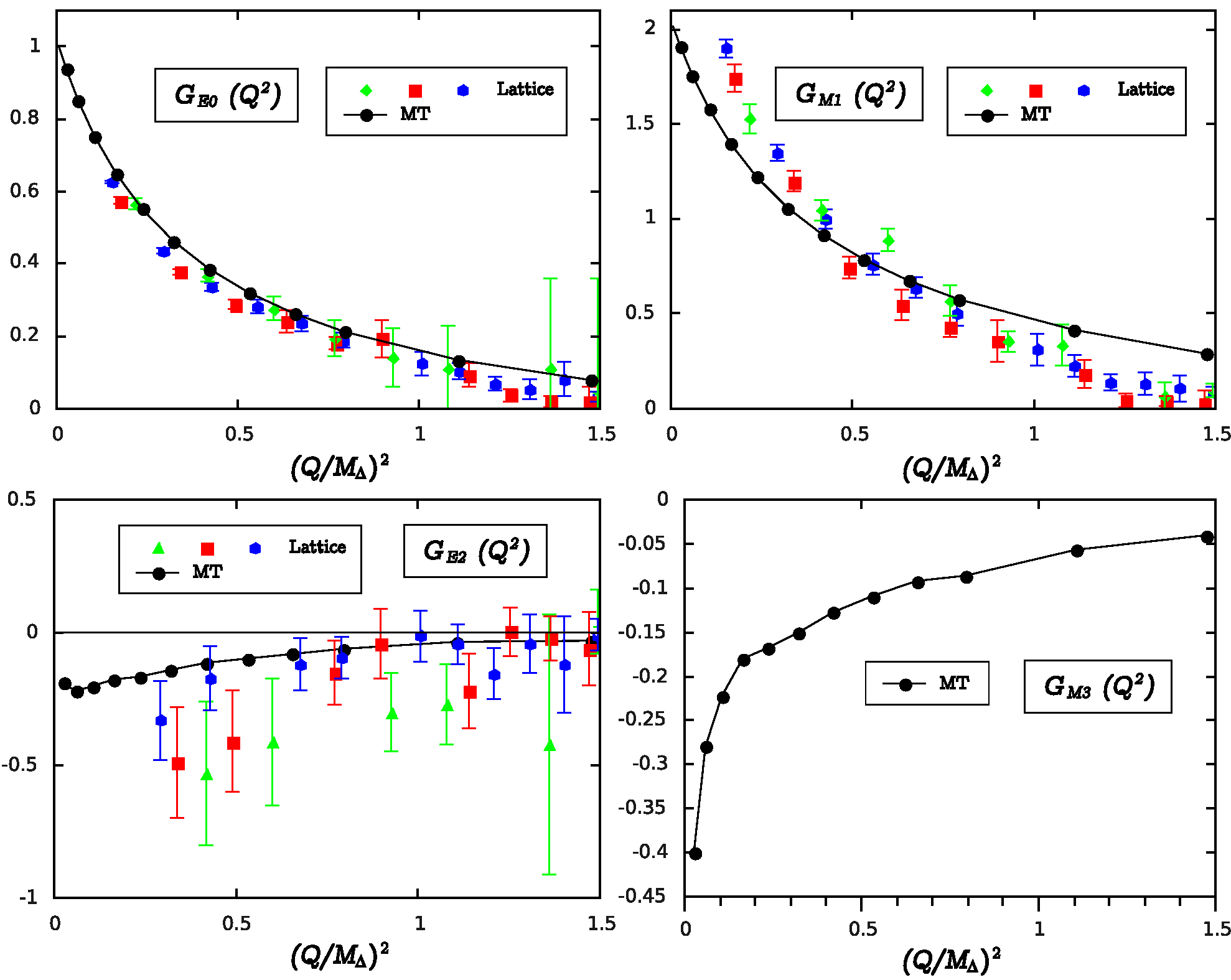}
 \end{center}
 \caption{Result for the Delta form factors when the subleading components are removed.}\label{fig:FFsTrick}
\end{figure}

The result of the calculation of the Delta electromagnetic form factors using the reduced basis is shown in Figure \ref{fig:FFsTrick}. The result for the electric monopole $G_{E0}$ and magnetic dipole $G_{M1}$ form factors are similar to the ones in the full calculation. The electric quadrupole form factor $G_{E2}$ is also qualitatively similar for the full and the reduced calculation, although a bit featureless in the latter case. Finally, the behavior of the magnetic octupole form factor $G_{M3}$, which is indicative of the baryon deformation, is very different to the corresponding behavior in the full solution. Therefore, we conclude that although a rough description of baryon properties is achievable using a simplified setup, the study of fine details such as deformation requires to keep all components, as dictated by symmetry.

Yet another manifestation of the importance of subleading components comes from the comparison of the results presented here with those obtained using the quark-diquark approximation \cite{Nicmorus2010}. The description of baryons using diquark effective degrees of freedom naturally misses some of the elements of a full three-body description. As a consequence, the qualitative behavior of $G_{E2}$ and $G_{M3}$ in the quark-diquark and in the Faddeev approach is different.

\chapter{A note on glueballs}\label{ch:glueballs}

The quark model allows two families of hadrons: mesons, formed by a quark and an antiquark and baryons, formed by three quarks. States not describable by the quark model are called exotic states. In particular, a longstanding prediction of QCD is that there should exist bound states formed by gluons only, so-called glueballs \cite{Fritzsch:1975tx}. 

An enormous experimental effort has been put on studying meson spectroscopy (see, e.g., \cite{Crede:2008vw} for a review) resulting on the discovery of exotic hadrons at Belle \cite{Garmash:2004wa,Garmash:2006fh} and Babar \cite{Aubert:2005wb} collaborations, but for the moment there is no direct evidence of glueballs. The reason for this is that the nature of these exotic states is not well understood; one possibility is that they are normal quark-antiquark states mixed with glueballs since they have the same quantum numbers. Glueballs could also be \textit{hidden} in the background of other mesons which are more abundantly produced. To improve this situation, there are several planned experimental facilities with the glueball search as one of the primary goals. GlueX at Jefferson Laboratory (USA) is expected to run in 2014 and will focus on mapping the spectrum of exotic states and on the search of light-quark hybrids which may serve to study glueballs via their decays. PANDA at FAIR/GSI (Germany) will study non-exotic hadrons up to charmonium states and from their decays will search gluonic states. However, a deeper theoretical understanding of glueball and other exotic-state formation must proceed in parallel with these experimental endeavors. For example, a calculation of glueball masses from first principles and the description of possible mixtures of glueballs and standard quark-antiquark bound states would be desirable.

Glueballs are also interesting from a purely theoretical point of view. The reason why QCD predicts the existence of glueballs is because it is a non-abelian gauge (or Yang-Mills) theory, which in particular implies that, as opposed to abelian gauge theories like Quantum Electrodynamics, the gluons (or gauge bosons) can interact among themselves and therefore can form pure gluonic (i.e., without quarks) bound states. Thus, glueballs are an excellent probe into the non-perturbative, or low-energy, regime of non-abelian gauge theories. However, most of the glueball spectrum calculations have been done by modeling QCD with effective degrees of freedom (see, e.g., \cite{Mathieu2009} for a review). The first calculations of glueball masses were performed using bag models with constituent massless gluons \cite{Robson1980,Jaffe:1975fd,Carlson1983,Carlson1984,Hansson1982,Chanowitz1983a}, where gluons are confined into a \textit{potential bag}. Models with massive constituent gluons have also been studied \cite{Hou2001,Barnes1981a,Cornwall1983,Szczepaniak1996}, however they introduce spurious states due to the unphysical longitudinal degree of freedom of a spin-1 constituent gluon. Other approaches used to calculate the glueball spectrum are QCD sum rules \cite{Shifman1981b,Novikov1980,Shifman1979b}, flux-tubes \cite{Isgur1983} or AdS/QCD \cite{Boschi-Filho2006}. There is no general agreement about the glueball masses among different approaches, although some qualitative features can be extracted \cite{Close1988,Jaffe1986}. Moreover, it is not clear whether these models capture all features, or any, of QCD dynamics.

A connection of the constituent gluon models with QCD was attempted in \cite{Kaidalov2000} by deriving an effective Hamiltonian from the QCD Lagrangian. Later, a calculation of the low-lying glueball masses based on continuum QCD was presented in \cite{Szczepaniak2003}. Specifically, the authors construct a Fock space of constituent quarks and gluons and an instantaneous potential between them using Coulomb gauge QCD.
So far, only lattice QCD calculations provide values of the glueball spectrum using the fundamental degrees of freedom of QCD \cite{Gregory2006a,Chen2006,Morningstar1999a,Hart:2001fp,Hart2006}. These calculations have nevertheless some drawbacks. First of all, most of them are performed in the so-called quenched approximation, in which the quantum fluctuations involving quarks are neglected, and there is no consensus about the effect of including those quark loops \cite{Hart:2001fp,Hart2006,Gregory2006a}. Also, common to all lattice calculations, are the problems of discretization and finite-volume effects which indeed yield to different glueball masses in different calculations. Finally, it is difficult from lattice calculations to unravel how gluonic bound-states are formed.

To improve the understanding of how glueballs emerge from QCD it is convenient to have an approach complementary to lattice-QCD. This could help to identify the leading mechanisms responsible for the formation of bound-states out of the elementary degrees of freedom of the theory and therefore simplify the theoretical treatment by keeping only those mechanisms in the calculations. This is especially important if one eventually attempts to study the mixing of gluonic and ordinary bound-states. This complementary approach is provided by functional methods like Dyson-Schwinger or Functional Renormalization Group equations.

However, so far the calculation of glueball masses using a covariant Bethe-Salpeter approach has been very little explored. In this chapter we propose a BSE for the simplest glueball, namely with no admixture of quarkionic states and formed only by two valence gluons.

\section{Glueball Bethe-Salpeter equation}

Throughout this work we have stressed that the quark-antiquark or the three-quark Green's functions develop a pole when the system forms a bound state. We have used this fact to go from the Dyson equation for the Green's function (\ref{eq:compactGreen}), or for the scattering matrix (\ref{eq:GreenScattering}), to the Bethe-Salpeter equation for the bound-state amplitude. These equations rely upon an, in general, unknown interaction kernel $K$ which, nevertheless, can be expanded diagrammatically in a (more or less) straightforward way.

Green's functions are equivalently defined by Dyson-Schwinger equations. Thereby, the equation is completely fixed, without the presence of undetermined interaction kernels. Instead, the equation depends on the knowledge of higher order Green's functions and, therefore, one has to solve an infinite and coupled system of Dyson-Schwinger equations. In any case, a truncation (or ansatz) must be chosen either for the interaction kernel $K$ or for the high-order Green's functions.

In this section we use the pole assumption to derive a Bethe-Salpeter equation for a system of two gluons forming a glueball, using the Dyson-Schwinger equation for the two-gluon four-points Green's function. Before that, however, we illustrate the procedure using the meson Dyson-Schwinger and Bethe-Salpeter equations as an example.

\subsection{Warm-up: meson Bethe-Salpeter equation}

\begin{figure}[ht!]
 \begin{center}
  \includegraphics[width=\textwidth,clip]{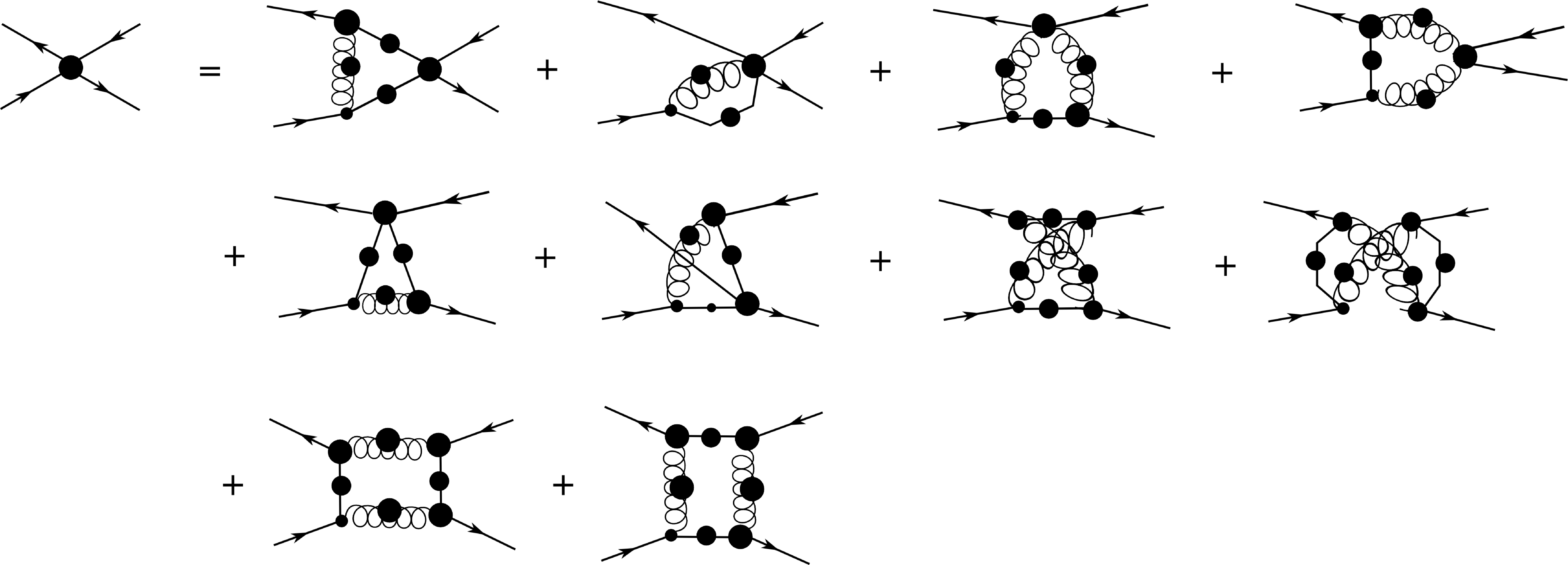}
 \end{center}
 \caption{Dyson-Schwinger equation for the quark-antiquark four-points amputated Green's function. Big blobs denote full Green's functions.}\label{fig:DSEmeson}
\end{figure}

The evolution of a quark-antiquark system is described by the four-points Green's function
\begin{equation}\label{eq:Greenmeson}
 G_{q\bar{q}}(x'_1,x'_2,x_1,x_2)=\langle 0|\bar{q}(x'_1)q(x'_2)q(x_1)\bar{q}(x_2)|0\rangle~,
\end{equation}
or, more conveniently for our purposes, by its momentum-space counterpart 
\begin{equation}
G_{q\bar{q}}(p'_1,p'_2,p_1,p_2)~. 
\end{equation}
The derivation of the Dyson-Schwinger equation for this Green's functions is very involved and for these reason we use DoFun, a Mathematica{\tiny \texttrademark}~ package described in \cite{Alkofer2009a,Huber2012}. The result (more precisely, the equation for the amputated Green's function, or scattering matrix $T$) is shown in Figure \ref{fig:DSEmeson}.

To obtain a Bethe-Salpeter equation we proceed as follows. Whenever the incoming quark and antiquark momenta $p_1$ and $p_2$ are such that $(p_1+p_2)^2=P^2=-M^2$, where $M$ is some meson (i.e. bound state) mass, the momentum-space Green's function $G_{q\bar{q}}(p'_1,p'_2,p_1,p_2)$ will have a pole. Of the diagrams in Figure \ref{fig:DSEmeson}, only those in which the incoming quark and antiquark enter a quark-antiquark four-points Green's function\footnote{In principle, a skeleton expansion (that is, in terms of primitive Green's functions) of higher-order Green's function may contain sub-structures which consist of quark-antiquark four-points Green's functions and should, therefore, be included as well. We ignore here this possibility.} in the s-channel will potentially develop a pole when the bound-state is formed. This procedure has been applied in \cite{Alkofer2011,Alkofer2010,Alkofer2011a} to study an explicit implementation of the BRST quartet mechanism \cite{Nakanishi1990}.

\begin{figure}[ht!]
 \begin{center}
  \includegraphics[width=0.4\textwidth,clip]{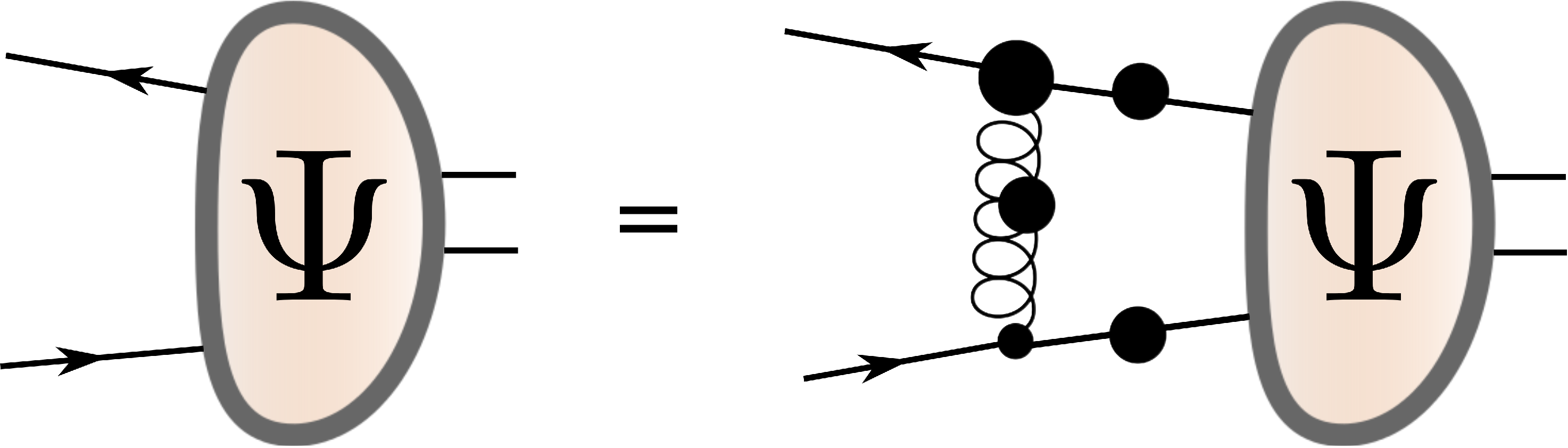}
 \end{center}
 \caption{Bethe-Salpeter equation for the meson. Big blobs denote full propagators or vertices.}\label{fig:BSEmeson}
\end{figure}

It is clear that only the first diagram in Figure \ref{fig:DSEmeson} fulfills the conditions stated above. Keeping only this diagram and expanding, as usual, the amputated Green's function around the bound-state pole to introduce the Bethe-Salpeter amplitudes $\Psi$
\begin{equation}
 T\sim\frac{\Psi\bar{\Psi}}{P^2+M^2}~,
\end{equation}
we obtain the Bethe-Salpeter equation for the meson, as shown in Figure \ref{fig:BSEmeson}.

The conventional Bethe-Salpeter equation for a meson would be
\begin{equation}
 \Psi=K_{q\bar{q}}G_0^{(2)}\Psi
\end{equation}
with an undetermined kernel $K$. The relation between this equation and the one derived here (Figure \ref{fig:BSEmeson}) should be obtained, in principle, by using a skeleton expansion of the full quark-gluon vertex. It is not clear, however, how this expansion would reproduce all diagrams in $K$\footnote{Possibly including some of the terms in a skeleton expansion of higher-order Green's functions in the equation.}. It is obvious, nevertheless, that within the Rainbow-Ladder truncation scheme both equations are identical.

\subsection{Glueball Bethe-Salpeter equation}

As illustrated in previous section, the first step to study a given state in a quantum field theory is to determine the operator which, acting on the Hilbert-space vacuum, creates that state. In the case of QCD, observable particles correspond to gauge invariant local operators. Based on this idea, an analysis of the possible glueball quantum numbers was performed in Refs. \cite{Fritzsch:1975tx,Jaffe1986}. It is also possible to mimic the simplistic picture of the quark model and classify the lightest glueballs as composed of two or three gluons. This widely used classification makes implicitly a difference between the so-called valence gluons and the gluons which bind those into a bound-state. In this project we assume this picture and will concentrate on two-gluon glueballs. 

The simplest operator describing a glueball is:
\begin{equation}
\langle 0|F^{\mu\nu}_a(y)F_{\mu\nu}^a(x)|0\rangle~.
 \label{eq:glueball}
\end{equation}
This operator provides information about the gluon propagation as well as its self-interactions. In particular it contains a four-gluon operator which will give rise to the glueball.

\begin{figure}[ht!]
 \begin{center}
  \includegraphics[width=0.9\textwidth,clip]{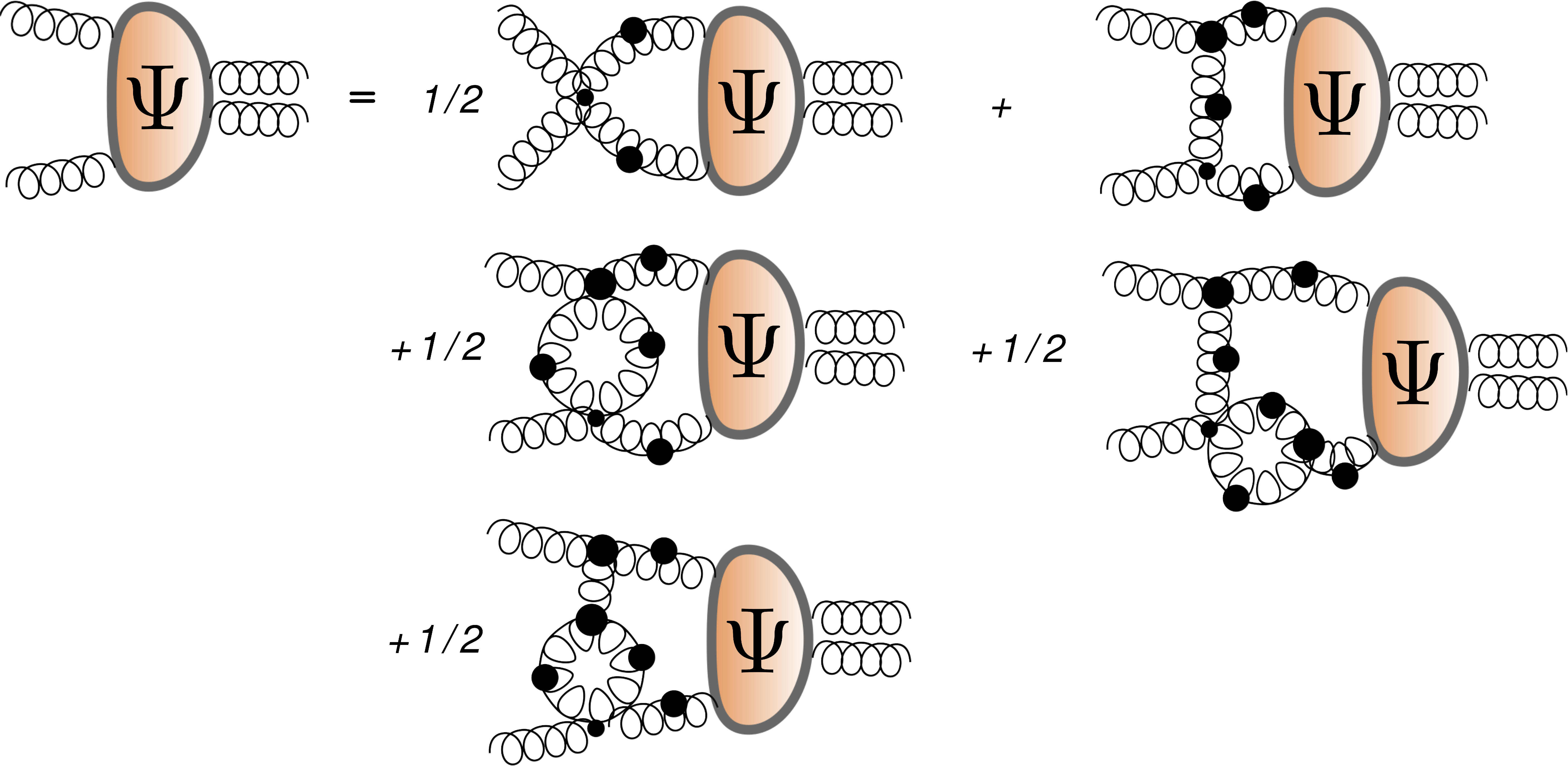}
 \end{center}
 \caption{Proposed Bethe-Salpeter equation for the glueball. Big blobs denote full propagators and vertices.}\label{fig:BSE-full}
\end{figure}

We use DoFun to calculate the Dyson-Schwinger equation for the two-gluon four-points Green's function. The output contains 79 diagrams which we will not show here. Applying the procedure discussed in previous section we end up with the Bethe-Salpeter equation shown in Figure \ref{fig:BSE-full}, which consists of five diagrams.

The glueball Bethe-Salpeter amplitude can be written, in general, as 
\begin{equation}\label{eq:Glueballamplitude}
 \Psi^{ab}_{\mu\nu,\mu_1\dots\mu_J}(p,P)~,
\end{equation}
where the $\mu,\nu$ Lorentz indices and the $a,b$ color indices refer to the valence gluons and the rest of Lorentz indices describe the angular momentum of the glueball. The relative and total momenta are
\begin{equation}
\begin{aligned}
P&=p_1+p_2=p'_1+p'_2~,\\
p&=\frac{p_1-p_2}{2}~,
\end{aligned}
\end{equation}
with $p_1$ and $p_2$ (and $p'_1$ and $p'_2$) are the valence-gluon momenta.

\begin{figure}[ht]
\begin{center}
  \includegraphics[width=0.8\textwidth]{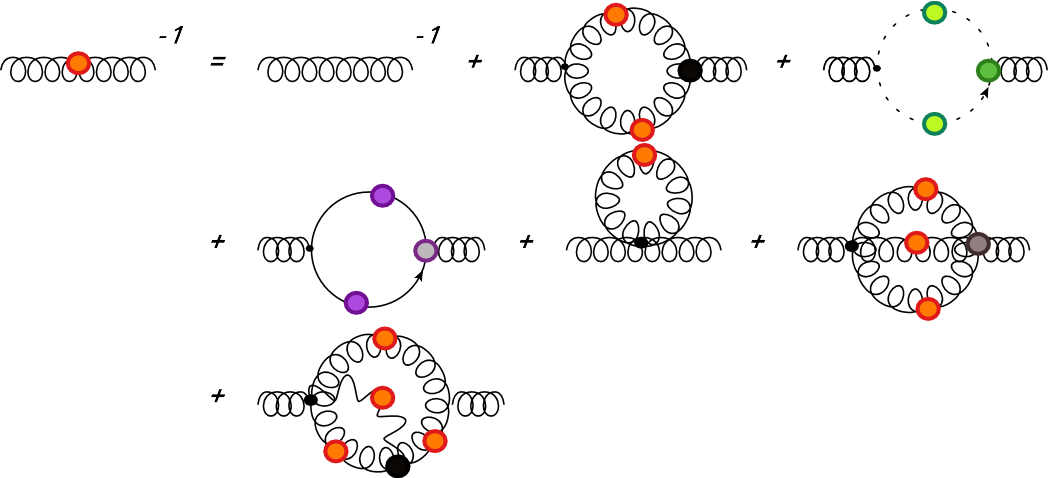}
\end{center}
\caption{Gluon DSE. Wiggly, solid and dotted lines represent gluon, quark and ghost propagators, respectively. Dressed Green's functions (propagators or vertices) are denoted by blobs. To solve the equation one needs to know the dressed quark and ghost propagators and three-gluon, four-gluon, ghost-gluon and quark-gluon vertices which are given by the corresponding DSEs that in turn depend on other Green's functions.}\label{fig:gluonDSE}
\end{figure}

A remarkable, and somewhat unexpected, feature of this equation is that ghosts do not appear explicitly. There are essentially two ways ghost lines could be explicit. The first one is through a skeleton expansion of higher order Green's function in the four-gluon DSE but, as explained before, our truncation scheme enforces to neglect these diagrams. The second possibility is to assume \textit{a priori} that ghosts must appear explicitly and study the four-gluon and two-gluon -- two-ghost coupled system. It is important to notice, however, that in the equation we propose ghosts can play an important role in the dynamics of the system. For example ghost propagators and ghost-gluon vertices are essential elements in the gluon DSE (see Figure \ref{fig:gluonDSE}). The same can be said about the DSE for the three-gluon vertex.

The Bethe-Salpeter equation obtained in this way is extremely complicated. First of all, the diagrams with two gluon-loops will be UV-divergent and some regularization procedure must be used. On the other hand, the third diagram in Figure \ref{fig:BSE-full} contains the off-shell full four-gluon vertex. Since the BSE is derived from the full four-gluon vertex DSE assuming this vertex shows a pole, to include this diagram we need an ansatz which is consistent with the pole assumption. This seems a highly intractable problem.

\begin{figure}[ht!]
 \begin{center}
  \includegraphics[width=0.8\textwidth,clip]{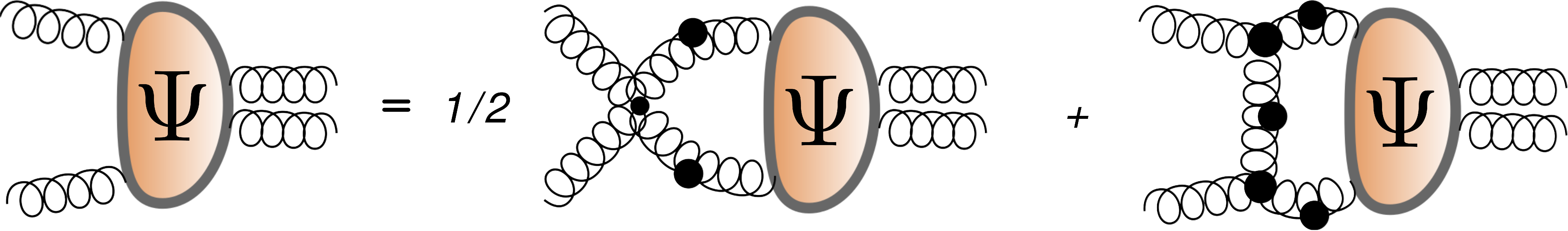}
 \end{center}
 \caption{Simplified Bethe-Salpeter equation for the glueball. Big blobs denote full propagators and vertices.}\label{fig:BSE-simple}
\end{figure}

In the absence of a way to deal with those problems, we simply ignore the abovementioned diagrams and define our glueball BSE with only the first two diagrams in Figure \ref{fig:BSE-full}. However, we still face another problem. To see this we expand the amplitude (\ref{eq:Glueballamplitude}) in a basis
\begin{equation}\label{eq:glueball_expansion_basis}
 \Psi^{ab}_{\mu\nu,\mu_1\dots\mu_J}(p,P)=f^{(i)}(p^2,p\cdot P;P^2) \tau^{(i)}_{\mu\nu,\mu_1\dots\mu_J}(p,P)\otimes\frac{\delta^{ab}}{\sqrt{N_c^2-1}}
\end{equation}
with $\sum_if^{(i)}\tau^{(i)}$ the spin part of the amplitude and the color part $\frac{\delta^{ab}}{\sqrt{N_c^2-1}}$, with $\{a,b\}$ indices in the adjoint representation, is fixed by requiring the glueball to be a color singlet. Since gluons are bosons, the total amplitude (\ref{eq:Glueballamplitude}) must be symmetric upon the interchange of the valence-gluon indices and momenta. Therefore, the color part $\delta^{ab}$ forces the spin part to be symmetric. Now, consider one has a symmetric amplitude as the input for the second diagram in Figure \ref{fig:BSE-full}. Since only one of the three-gluon vertices is full, the output of this diagram will not be symmetric and therefore this term cannot appear in the glueball BSE if we want to keep Bose-symmetry. We propose to fix this problem by using full vertices for both three-gluon vertices in this diagram. The resulting glueball BSE is shown in Figure \ref{fig:BSE-simple}.

A realistic description of glueballs must deal with the mixing of them with quark-antiquark states.  This is an issue of exceptional importance in the experimental search of glueballs and so far there is no clear theoretical account for it. In the DSE/BSE formalism the mixing of quarkionic and gluonic states would be studied by solving a coupled system of BSEs. The equation derived here assumed that only the four-gluon Green's functions develops a pole. However, in the case of a mixed quarkionic-gluonic state, the poles corresponding to a bound-state would appear simultaneously in the four-gluon, four-quark, and the quark-quark-gluon-gluon Green's functions. Therefore the BSE should be modified. 

It is also important to remark here that the full Green's functions contain information about bound-states with all possible quantum numbers. A specific choice of quantum numbers for the bound-state of interest is done via the symmetries and structure of the BS amplitudes.

\section{Steps to solve the glueball Bethe-Salpeter equation}

To write an explicit expression for the glueball Bethe-Salpeter equation, we need to introduce the definitions of the different Green's functions involved. The tree-level gluon propagator in Landau gauge is
\begin{equation}\label{eq:GpropLandauGauge2}
 D^{ab}_{\mu\nu}(p)=\frac{1}{Z_3}\left(g_{\mu\nu}-\frac{p_\mu p_\nu}{p^2}\right)\frac{Z(p^2)}{p^2}\delta^{ab}
\end{equation}
with $Z_3$ the gluon wave-function renormalization constant and $Z(p^2)$ the gluon propagator dressing function. The renormalized four-gluon vertex is
\begin{flalign}
 \Gamma_{\mu\nu\rho\sigma}^{abcd}(p_1,p_2,p_3,p_4)={}&-g^2Z_4(2\pi)^4\delta^{(4)}(p_1+p_2+p_3+p_4)\left[f^{abe}f^{cde}\left(g_{\mu\rho}g_{\nu\sigma}-g_{\mu\sigma}g_{\nu\rho}\right)\right. \nonumber \\
&~~~~~~~~~~~~~~~~~~~~~~~~~~~~~~~~~~~~~~~~~~~~+ f^{ace}f^{bde}\left(g_{\mu\nu}g_{\rho\sigma}-g_{\mu\sigma}g_{\nu\rho}\right) \nonumber \\
&~~~~~~~~~~~~~~~~~~~~~~~~~~~~~~~~~~~~~~~~~~~~+\left. f^{ade}f^{bce}\left(g_{\mu\nu}g_{\rho\sigma}-g_{\mu\rho}g_{\nu\sigma}\right)\right]
\end{flalign}
where $f^{abc}$ are the $SU(N_c)$ structure constants, $Z_4$ the four-gluon renormalization constant and all momenta are considered as outgoing. Finally, for the renormalized full three-gluon, we assume that the color structure is the same as the tree-level one and we write it as
\begin{equation}
 \Gamma^{abc}_{\mu\nu\rho}(k,p,q)=-igZ_1(2\pi)^4\delta^{(4)}(p+q+k)\Gamma_{\mu\nu\rho}(k,p,q)
\end{equation}
so that for the bare vertex one has $\Gamma_{\mu\nu\rho}(k,p,q)=\Gamma^{(0)}_{\mu\nu\rho}(k,p,q)=(k-p)_\rho g_{\mu\nu}+(p-q)_\mu g_{\nu\rho}+(q-k)_\nu g_{\mu\rho}$. $Z_1$ is the three-gluon renormalization constant and, again, all momenta are taken as outgoing.

It is also necessary to define a covariant basis $\tau^{(i)}$ for the expansion (\ref{eq:glueball_expansion_basis}). A systematic derivation of such a basis for glueballs of any spin has been described in \cite{KellermannPhD}. We shortly describe here the main ideas to obtain this basis. The first step is to couple the two valence gluons (that is, their Lorentz indices) into a Lorentz invariant object. The only possibilities are
\begin{equation}
 g^{\mu\nu}\quad,\quad \hat{p}^{\rho}\hat{P}^{\sigma}\epsilon^{\rho\sigma\mu\nu}~.
\end{equation}
The scalar glueball $0^{++}$ is represented by $g^{\mu\nu}/2$ whereas the appropriate basis element for the pseudoscalar glueball $0^{-+}$ is
\begin{equation}
\mathcal{N}^{-1}(\hat{p}\cdot\hat{P})~\hat{p}^{\rho}\hat{P}^{\sigma}\epsilon^{\rho\sigma\mu\nu}~,\qquad \mathcal{N}=\sqrt{(\hat{p}\cdot\hat{P})^2\left(1-(\hat{p}\cdot\hat{P})^2\right)}~,
\end{equation}
where the factor $(\hat{p}\cdot\hat{P})$ is included to obtain a state of positive charge conjugation, since this operation amounts to a sign flip in the relative momentum for a system of two gluons. Basis elements for higher angular-momentum states are constructed with tensor products of the aforementioned Lorentz scalars with Lorentz tensors with the appropriate number of degrees of freedom. To represent a state of angular momentum $J$, the tensor must have $2J+1$ independent components. As described in \cite{KellermannPhD}, this is achieved with tensors that are symmetric, traceless in all pair of indices and orthogonal to the total momentum in all indices. The construction of these states grows very fast in complexity.

With these definitions, the first diagram reads
\begin{flalign}\label{eq:4gluon}
 f^{(i)}(p^2,p\cdot P;P^2)={}&-\frac{1}{2}g(\mu)^2\int\frac{d^4p'}{(2\pi)^4}\bar{\tau}^{(i)}_{\mu\nu,\mu_1\dots\mu_J}(p,P)\frac{\delta^{ab}}{\sqrt{N_c^2-1}}\nonumber\\
&\times\left\{f^{abe}f^{cde}\left(g_{\mu\lambda}g_{\nu\sigma}-g_{\mu\sigma}g_{\nu\lambda}\right)\right. \nonumber \\
&~~+ f^{ace}f^{bde}\left(g_{\mu\nu}g_{\lambda\sigma}-g_{\mu\sigma}g_{\nu\lambda}\right) \nonumber \\
&~~+\left. f^{ade}f^{bce}\left(g_{\mu\nu}g_{\lambda\sigma}-g_{\mu\lambda}g_{\nu\sigma}\right)\right\}\nonumber \\
&\times D_{\lambda\lambda'}^{c c'}(p'_1)D_{\sigma\sigma'}^{d d'}(p'_2)\nonumber \\
&\times \tau^{(j)}_{\lambda'\sigma',\mu_1\dots\mu_J}(p',P)\frac{\delta^{c'd'}}{\sqrt{N_c^2-1}}f^{(j)}(p'^2,p'\cdot P;P^2)~,\nonumber\\
\end{flalign}
where we used the Slavnov-Taylor identity $Z_4=Z_3^2Z_g^2$ and introduced the renormalized coupling constant $g(\mu)=Z_gg$. The color traces for the three terms in the integral are
\begin{equation}
 \begin{aligned}
  \delta^{ab}f^{abe}f^{cde}\delta^{cc'}\delta^{dd'}\delta^{c'd'}&=\delta^{ab}f^{abe}f^{cce}=0~,\\
  \delta^{ab}f^{ace}f^{bde}\delta^{cc'}\delta^{dd'}\delta^{c'd'}&=\delta^{ab}f^{ace}f^{bce}=\delta^{ab}\delta^{ab}N_c=(N_c^2-1)N_c~,\\
  \delta^{ab}f^{ade}f^{bce}\delta^{cc'}\delta^{dd'}\delta^{c'd'}&=\delta^{ab}f^{ace}f^{bce}=\delta^{ab}\delta^{ab}N_c=(N_c^2-1)N_c~,
 \end{aligned}
\end{equation}
where we used the identity $f^{acd}f^{bcd}=N_c\delta^{ab}$.

The gluon-exchange diagram reads
\begin{flalign}\label{eq:gexchange}
f^{(i)}(p^2,p\cdot P;P^2)={}&-g(\mu)^2\int\frac{d^4p'}{(2\pi)^4}\bar{\tau}^{(i)}_{\mu\nu,\mu_1\dots\mu_J}(p,P)\frac{\delta^{ab}}{\sqrt{N_c^2-1}}\nonumber\\
&\times f^{brs}\Gamma_{\nu\alpha\beta}(p_1,-p'_1,k)f^{atm}\Gamma_{\mu\delta\rho}(p_2,-k,-p'_2)D_{\beta\delta}^{st}(k)D_{\alpha\gamma}^{ru}(p'_1)D_{\rho\sigma}^{ml}(p'_2)\nonumber \\
&\times \tau^{(j)}_{\gamma\sigma,\mu_1\dots\mu_J}(p',P)\frac{\delta^{ul}}{\sqrt{N_c^2-1}}f^{(j)}(p'^2,p'\cdot P;P^2)~,\nonumber\\
\end{flalign}
where we used the Slavnov-Taylor identity $Z_1=Z_gZ_3^{3/2}$ and $g(\mu)=Z_gg$. The color trace gives
\begin{equation}
\delta^{ab}f^{atm}f^{brs}\delta^{st}\delta^{ml}\delta^{ru}\delta^{lu}=\delta^{ab}f^{atm}f^{bmt}=-\delta^{ab}f^{atm}f^{bmt}=-\delta^{ab}\delta^{ab}N_c=-(N_c^2-1)N_c~.
\end{equation}
It is important to notice that, if one is given with expressions for the propagator and vertex dressing functions, Equations (\ref{eq:4gluon}) and (\ref{eq:gexchange}) have only $g(\mu)$ as a free parameter.

For a self-consistent glueball calculation in the DSE/BSE formalism it is essential to solve the gluon propagator from the gluon DSE (see Fig. \ref{fig:gluonDSE}). The gluon propagator has been studied numerically in \cite{Alkofer:2003jk,Alkofer2004,Strauss2012239} using several approximations. In both works it was concluded that the gluon propagator in the complex plane has a non-trivial analytic structure, although in contrast to the quark propagator used in previous chapters, it has only branch-cuts with absence of poles. Non-analiticities of the gluon propagator will presumably be of relevance in the resolution of the glueball BSE.

To solve the gluon DSE one needs to know the dressed quark and ghost propagators and three-gluon, four-gluon, ghost-gluon and quark-gluon vertices. Exact expressions for these quarks and vertices are generally unknown and models, at least for the vertices, must be used. Restrictions on those models can be based on infrared analysis of Yang-Mills vertices \cite{Alkofer2010a,Alkofer2009c} and on lattice studies \cite{Cucchieri2008,Cucchieri2006}, as well as from the underlying symmetries of QCD (formalized by the Slavnov-Taylor identities). A self-consistent solution of the glueball BSE is, nevertheless, beyond the scope of this work.

\chapter{Summary}\label{ch:summary}

The calculation of hadron properties from QCD remains an open issue due to the fundamental non-perturbative nature of the problem. Covariant Bethe-Salpeter equations, in combination with Dyson-Schwinger equations for QCD Green's functions, constitute an excellent tool for hadron studies in continuum quantum field theory. They allow, in particular, a systematic analysis of the mechanisms in QCD that lead to the formation of bound states. Moreover its applicability is not restricted, in principle, to any momentum or quark-mass range, which is a limitation in other approaches such as lattice QCD.

The limitations of the approach come from the fact that a complete solution of the Bethe-Salpeter equation would require to solve an infinite set of coupled Dyson-Schwinger equations for the QCD Green's functions. A realistic calculation thus requires a truncation of the system, which means that of all the possible correlations among particles provided by QCD, one considers only a subset. Any truncation, in turn, induces the necessity of some modeling. It is therefore necessary to disentangle from the results the model-dependent and the truncation-dependent features. The latter carry the information about the relevance of the interaction terms considered in the calculation.

In this thesis we focused on the study of baryon properties using a three-body covariant Bethe-Salpeter calculation. We used the Rainbow-Ladder truncation for the quark-quark kernel. This truncation consists of a vector-vector dressed-gluon exchange between quarks. Since we do not solve the corresponding Dyson-Schwinger equations for the gluon propagator and the quark-gluon vertex, a choice of a model for their dressing functions is necessary. Moreover, we neglect  three-body irreducible interactions.

We performed our calculations using two different models. One is the Maris-Tandy model, which is a purely phenomenological one, designed to provide dynamical chiral symmetry breaking and a good description of ground-state meson properties. The second model, which we called the Alkofer-Fischer-Williams model, attempts to capture some of the QCD dynamics in the infrared and, therefore, might be a closer approximation to a calculation from first principles in QCD.

Using those two models, in Chapter 2 we calculate the masses of spin-$\nicefrac{1}{2}$ and spin-$\nicefrac{3}{2}$ baryons for a range of quark masses from $u/d$ up to $b$ quarks. The conclusion of this chapter is no strong model dependence is manifested in the baryon spectrum. The difference between models, and between the models and the physical or lattice values is always smaller than $10$\%. Our results also indicate that irreducible three-body interactions are subdominant in baryons. However, a more precise determination of its relevance would require to study also quark-quark interaction terms beyond the Rainbow-Ladder truncation. 

In Chapter 3 we study the electromagnetic form factors of the Delta(1232), again using the two models for the effective interaction. In this context it is important to remark that, in a covariant approach, all partial waves allowed by symmetry are present in the calculation and that their relative importance is dictated by the dynamics. In the case of spin-$\nicefrac{3}{2}$ baryons, this implies the presence of s-, p-, d- and f-waves. The most important result in this chapter is that the Delta has non-zero electric quadrupole and magnetic octupole form factors, which is a signal of deformation from sphericity. Although the Delta(1232) is dominated by s-waves, we also show the importance of the subleading components in the electromagnetic properties. In the calculated form factors we also observe a qualitative model independence.

Most of the experimental information about the electromagnetic properties of the Delta and the details of its deformed shape is obtained, however, from the electromagnetic decay $N\to\Delta\gamma$. In a covariant approach, the nucleon and Delta Bethe-Salpeter amplitudes and the quark-photon vertex used for the form factor calculations are also the elements needed to study this process, which will be the object of future work.

Covariant Bethe-Salpeter equations are not limited to the calculation of hadron properties in QCD. They can be extended to study bound-states in any quantum gauge-field theory and also to the study of exotic states in QCD. In Chapter 4 we take a first step in the calculation of glueball properties in the BSE/DSE framework. We propose a Bethe-Salpeter equation for glueballs based on the four-gluon Dyson-Schwinger equation in QCD. However, an explicit solution of this equation is very difficult and is part of current and future investigation.

\begin{appendix}
\chapter{}\label{ch:appendix}

%%%%%%%%%%%%%%%%%%%%%%%%%%%%%%%%%%%%%%%%%%%%%%%%%%%%%%%%%%
\section{Conventions and reference frames}\label{sec:conventions}
%%%%%%%%%%%%%%%%%%%%%%%%%%%%%%%%%%%%%%%%%%%%%%%%%%%%%%%%%%

In this thesis we work in Euclidean spacetime. Four vectors can be expressed using hyperspherical coordinates:
\begin{equation}
 p^\mu=\sqrt{p^2}\left(
 \begin{array}{c}
  \sin\alpha~\sin\beta~\sin\varphi \\
  \sin\alpha~\sin\beta~\cos\varphi \\
  \sin\alpha~\cos\beta \\
  \cos\alpha
 \end{array}
\right)\equiv\sqrt{p^2}\left(
 \begin{array}{c}
  \sqrt{1-z^2}~\sqrt{1-y^2}~\sin\varphi \\
  \sqrt{1-z^2}~\sqrt{1-y^2}~\cos\varphi \\
  \sqrt{1-z^2}~y \\
  z
 \end{array}
\right)
\end{equation}
in which the four-momentum integration is written as
\begin{equation}
 \int \frac{d^4p}{(2\pi)^4}\equiv\int \widetilde{d^4p}\equiv\int_p\dots=\frac{1}{2(2\pi)^4}\int_o^\infty dp^2~p^2\int_{-1}^1dz\sqrt{1-z^2}\int_{-1}^1dy\int_0^{2\pi}d\varphi~.
\end{equation}

In the resolution of the Faddeev equation one has two external (i.e. they are not integration variables) momenta $p$, $q$. In this a case one can choose a reference frame such that they can be written as
\begin{equation}\label{eq:def_external_momenta}
 p^\mu=\sqrt{p^2}\left(
 \begin{array}{c}
  0 \\
  0 \\
  \sqrt{1-z_1^2} \\
  z_1
 \end{array}
\right)~,\qquad 
 q^\mu=\sqrt{q^2}\left(
 \begin{array}{c}
  0 \\
  \sqrt{1-z_2^2}~\sqrt{1-z_0^2} \\
  \sqrt{1-z_2^2}~z_0 \\
  z_2
 \end{array}
\right)~.
\end{equation}
With this choice and on the baryon's rest frame $P^\mu=\left(0,0,0,i~M\right)$, the normalized momenta $\widehat{p_T}^\mu$ and $\widehat{q_t}^\mu$ needed for the construction of the basis for the Faddeev amplitudes (see Appendix\ref{sec:basis}) simplify to
\begin{equation}
 \widehat{p_T}^\mu=\left(
 \begin{array}{c}
  0 \\
  0 \\
  1 \\
  0
 \end{array}
\right)~,\qquad 
 \widehat{q_t}^\mu=\left(
 \begin{array}{c}
  0 \\
  1 \\
  0 \\
  0
 \end{array}
\right)~,\qquad 
 \widehat{P}^\mu=\left(
 \begin{array}{c}
  0 \\
  0 \\
  0 \\
  1
 \end{array}
\right)~.
\end{equation}

For the calculation of form factors, however, the baryon's rest frame is not convenient. We use instead the \textbf{Breit frame} (or z-Breit frame). In this reference frame the baryon has initially a four-momentum $P_i^\mu$ and interacts elastically with a photon with four-momentum $Q$ to give a baryon with final four-momentum $P_f^\mu$, such that
\begin{equation}
  P_i^\mu=-P_f^\mu~,\qquad Q^\mu=P_f^\mu-P_i^\mu~,
\end{equation}
\begin{equation}
  P_{\nicefrac{i}{f}}^\mu=i~M\left(
 \begin{array}{c}
  0 \\
  0 \\
  \pm i~Q/2~M \\
  \sqrt{1+Q^2/4~M^2}
 \end{array}
\right)~,\qquad
Q^\mu=\left(
 \begin{array}{c}
  0 \\
  0 \\
  |Q| \\
  0
 \end{array}
\right)~.
\end{equation}
Note that $P_i^2=P_f^2=-M^2$ as it should be, since the baryon is on-shell in both cases.

For the Euclidean Dirac matrices we use the convention
\begin{equation}
 \gamma^0=\left(
 \begin{array}{cc}
  \mathbb{1} & 0 \\
      0      & \mathbb{1}
 \end{array}
\right)~,\quad
 \gamma^i=\left(
 \begin{array}{cc}
   0 & -i~\sigma^i\\
   i~\sigma^i & 0
 \end{array}
\right)~,\quad
 \gamma^4=\left(
 \begin{array}{cc}
  \mathbb{1} & 0 \\
      0      & -\mathbb{1}
 \end{array}
\right)~,\quad
 \gamma^5=\left(
 \begin{array}{cc}
  0 &\mathbb{1} \\
  \mathbb{1} & 0
 \end{array}
\right)
\end{equation}
with $\sigma^i$ the Pauli matrices. This choice of matrices is hermitian $\gamma^\mu=\left(\gamma^\mu\right)^\dagger$.

%%%%%%%%%%%%%%%%%%%%%%%%%%%%%%%%%%%%%%%%%%%%%%%%%%%%%%%%%%
\section{Covariant decomposition of spin-\nicefrac{3}{2} baryon amplitudes}\label{sec:basis}
%%%%%%%%%%%%%%%%%%%%%%%%%%%%%%%%%%%%%%%%%%%%%%%%%%%%%%%%%%

The Faddeev amplitudes are tensors that describe the baryon in terms of the valence quarks
\begin{equation}
 \Psi\sim\langle 0|q_\alpha q_\beta q_\gamma B^\dagger_{\mathcal{I}}|0\rangle~.
\end{equation}
where $B^\dagger$ is a baryon creation operator. Schematically they can be decomposed as
\begin{displaymath}
 SPIN \otimes FLAVOR \otimes COLOR~.
\end{displaymath}
The color part of the amplitude is fixed by the fact that the quarks must combine into a color singlet, and therefore for a system of three quarks it is given by the antisymmetric tensor
\begin{equation}
 COLOR = \frac{\epsilon_{ABC}}{\sqrt{6}}
\end{equation}
being $A$, $B$ and $C$ the quark color indices (in the fundamental representation) and the $\sqrt{6}$ is included to normalize it. As for the flavor part, it is given by the usual quark-model flavor states. 

The structure of the spin part depends on the baryon of interest. In general it can be written as
\begin{equation}
 \Psi_{\alpha\beta\gamma\mathcal{I}}(p,q;P)
\end{equation}
where $p$, $q$ and $P$ are the relative and total momenta, as defined in (\ref{eq:defpq}),
\begin{equation}
\begin{array}{rl@{\quad}rl}
        p &= (1-\zeta)\,p_3 - \zeta (p_1+p_2)\,, &  p_1 &=  -q -\dfrac{p}{2} + \dfrac{1-\zeta}{2} P\,, \\[0.25cm]
        q &= \dfrac{p_2-p_1}{2}\,,         &  p_2 &=   q -\dfrac{p}{2} + \dfrac{1-\zeta}{2} P\,, \\[0.25cm]
        P &= p_1+p_2+p_3\,,                &  p_3 &=   p + \zeta  P~,
\end{array}
\end{equation}
with $p_1$, $p_2$ and $p_3$ the quark momenta and $\zeta$ the momentum partitioning parameter. The indices $\alpha$, $\beta$ and $\gamma$ represent the valence quark and the generic index $\mathcal{I}$ indicates the baryon of interest. For spin-$\nicefrac{1}{2}$ baryons this index refers to a  Dirac field and, therefore, the Faddeev amplitude (since the color and flavor parts are trivial, when we talk about Faddeev amplitudes we refer to its spin part, unless otherwise stated) is a rank-4 Dirac tensor. 
For spin-$\nicefrac{3}{2}$ particles the index $\mathcal{I}$ refers to a Rarita-Schwinger field and therefore the Faddeev amplitude is
a mixed tensor with four Dirac indices and one Lorentz index. 

These tensors can be conveniently expressed in terms of a basis (see Equation \
(\ref{eq:basisexpansion})).  It can be shown that the positive
parity and positive energy subspace of the spin-$\nicefrac{1}{2}$ basis contains 64 linearly independent elements \cite{Eichmann2010a}. The positive
parity and positive energy subspace of the spin-$\nicefrac{3}{2}$ basis contains 128 elements \cite{SanchisAlepuz:2011jn}. In what follows, we show the steps for the construction of a covariant basis for the Faddeev amplitudes. 

The construction of the corresponding basis is in principle straightforward, but very cumbersome due to the large number of indices and basis elements involved. For this reason is most convenient to perform all the calculations using some symbolic programming language like Wolfram's Mathematica{\tiny \texttrademark}~. The most significant aspects of this construction are:
\begin{itemize}
 \item It is independent of any approximation in the three-body Bethe-Salpeter equation.
 \item Only Poincar\'e covariance as well as parity invariance are needed to
 construct the basis.
 \item The basis includes all possible internal quark-spin and orbital angular momentum
 values that can lead to the final spin of the baryon of interest.
\end{itemize}
The last point is very important. In a complete covariant calculation there is no freedom to choose the spin and orbital angular momentum composition of the baryon. All possibilities must be included, this is dictated by symmetry, and the relative importance of each of them will be determined by the interaction.

A basis for a rank-four Dirac tensor can be constructed out of the following
linearly dependent elements,
\begin{eqnarray}\label{eq:basis_dirac}
\left(\begin{array}{c}
\textnormal{S}_{ij}^\sigma\\
\textnormal{P}_{ij}^\sigma\\
\textnormal{V}_{ij}^\sigma\\
\textnormal{A}_{ij}^\sigma
\end{array}\right)=
\left(
\begin{array}{c}
\mathbb{1}\otimes\mathbb{1}\\
\gamma^5\otimes\gamma^5\\
\gamma^\mu_T\otimes\gamma^\mu_T\\
\gamma^\mu_T\gamma^5\otimes\gamma^\mu_T\gamma^5
\end{array}\right)\left(\Gamma_i\otimes \Gamma_j\right)
\Omega^{\sigma}(\hat{P})\,,
\end{eqnarray}
where $i,j=1\dots 4$ and
\begin{equation}
\Gamma_i=\left\{\mathbb{1},\frac{1}{2}\left[\widehat{\Slash{p}_T},\widehat{\Slash{q}_t}\right],\widehat{\Slash{p}_T},\widehat{\Slash{q}_t}\right\}~,
\end{equation}
where the hat denotes a normalized vector
\begin{equation}
 \hat{v}^\mu=\frac{v^\mu}{\sqrt{v^2}}
\end{equation}
and, for convenience, the 
unit transverse four-vectors $\widehat{p_T}$ and $\widehat{q_t}$ are used to contract the Poincar\'e indices 
\begin{flalign}
 p^\mu_T&=T_P^{\mu\nu}p^\nu~,\\  
q_t^\mu&=T_{p_T}^{\mu\lambda}\left(T_P^{\lambda\nu}q^\nu\right)=T_{p_T}^{\mu\lambda}q^{\lambda}_T~,
\end{flalign}
with
\begin{equation}
 T_v^{\mu\nu}=\delta^{\mu\nu}-\hat{v}^\mu\hat{v}^\nu
\end{equation}
the transverse projector. We also defined
\begin{equation}
 \Omega^{\sigma}(\hat{P})=\Lambda^\sigma(\hat{P})\gamma^5\mathcal{C}\otimes 
\Lambda^+(\hat{P})
\end{equation}
with $\sigma=\pm$, $\
\mathcal{C}=\gamma^4\gamma^2$ is  the charge conjugation-matrix, 
$\Lambda^\pm(\hat{P})=\left(\mathbb{1}\pm\Slash{\hat{P}}\right)/2$ 
is  the positive- and negative-energy  projector.
These are 128 elements of which one can check that only 64 are linearly independent. Its structure and partial-wave decomposition can be found in \cite{Eichmann2010a}.

Using the elements in (\ref{eq:basis_dirac}) we can construct the building blocks of a mixed 
Dirac-Poincar\'e basis in the following way
\begin{eqnarray}\label{eq:basis_mixed}
\left(\begin{array}{c}
\left[M^g_{ij}\right]^\sigma\\~\\
\left[M^p_{ij}\right]^\sigma\\~\\
\left[M^q_{ij}\right]^\sigma
\end{array}\right)=
\left(
\begin{array}{c}
\gamma^\mu_T\otimes\mathbb{1}\\~\\
\widehat{p_T}^\mu\gamma^5\otimes\mathbb{1}\\~\\
\widehat{q_t}^\mu\gamma^5\otimes\mathbb{1}
\end{array}\right)(M_{ij}^\sigma)(\mathbb{1}\otimes\mathbb{P}^{\mu\nu})\,,
\quad
\end{eqnarray}
where $M\in\{\textnormal{S},\textnormal{P},\textnormal{V},\textnormal{A}\}$ 
are the basis elements defined in (\ref{eq:basis_dirac}) and 
$\mathbb{P}^{\mu\nu}$ is the Rarita-Schwinger projector for 
positive-energy particles
\begin{equation}
 \mathbb{P}_+^{\mu\nu}(\hat{\textnormal{P}})=\Lambda_+(\hat{\textnormal{P}})
 \left(T_P^{\mu\nu}-\frac{1}{3}\gamma^\mu_T\gamma^\nu_T\right)\,,
\end{equation}
with $\gamma^\mu_T=T_P^{\mu\nu}\gamma^\nu$. The set (\ref{eq:basis_mixed}) 
contains 384 elements, but it can be checked that only 128 of them are 
linearly independent. Of course it is physically irrelevant which 128 elements are chosen to form a basis. Nevertheless, from a naive quark-model point of view, the s-wave components (i.e. the relative-momentum independent basis elements, such as $\textnormal{S}_{11}^g$) will play a dominant role and therefore is convenient to include them explicitly in the basis.

To this end, the basis elements can be
classified with respect to their quark-spin and relative orbital angular
momentum in the baryon's rest frame, as explained in the following subsection and summarized in tables \ref{table_basis3_2}
and \ref{table_basis1_2}. For the purpose of this classification, we found the following choice of linearly independent 
elements convenient:
\begin{equation}\label{eq:LI_set}
\begin{array}{c}
\left\{
      \begin{array}{c}
       \textnormal{S}^p_{1j}\,,\textnormal{S}^q_{1j}\\
       \textnormal{P}^p_{1j}\,,\textnormal{P}^q_{1j}\\
       \textnormal{V}^p_{1j}\,,\textnormal{V}^q_{1j}\\
       \textnormal{A}^p_{1j}\,,\textnormal{A}^q_{1j}
      \end{array}
\right\}\,,\quad
\left\{
      \begin{array}{c}
       \textnormal{S}^p_{43}\,,\textnormal{S}^p_{41}\\
       \textnormal{S}^q_{32}\,,\textnormal{S}^q_{34}\\
       \textnormal{P}^p_{43}\,,\textnormal{P}^p_{41}\\
       \textnormal{P}^q_{32}\,,\textnormal{P}^q_{34}
      \end{array}
\right\}~,\\
\quad\\
\left\{
      \begin{array}{c}
       \textnormal{S}^g_{1j}\\
       \textnormal{P}^g_{1j}
      \end{array}
\right\}\,,\quad
\left\{
      \begin{array}{c}
       \textnormal{S}^p_{3j}\,,\textnormal{S}^q_{4j}\\
       \textnormal{P}^p_{3j}\,,\textnormal{P}^q_{4j}
      \end{array}
\right\}~,
\end{array}
\end{equation}
with $j=1\dots4$, and we have omitted the index $\sigma$ for better readability.

\subsection{Partial-wave decomposition of the spin-$\nicefrac{3}{2}$ basis}

    \renewcommand{\arraystretch}{1.4}

\begin{table}[htbp]
\begin{center}
\begin{tabular}{ | c | c || c |} \hline

$s$ &  $\ell$  &   $\sqrt{5}\,\tau^{\sigma,1}_{1j}$ \\ \hline\hline

$\quad\nicefrac{3}{2}\quad$   &  $\quad 0\quad$    & $\sqrt{5}\,\textnormal{S}^g_{11}$ \\

$\nicefrac{3}{2}$   &  $1$     &   $3\,\textnormal{S}^g_{12}+2\,(\textnormal{S}^p_{14}-\textnormal{V}^q_{13})$ \\ 

$\nicefrac{3}{2}$   &  $1$     & $3\,\textnormal{S}^g_{13}+2\,\textnormal{V}^p_{11}$\\

$\nicefrac{3}{2}$   &  $1$     & $3\,\textnormal{S}^g_{14}+2\,\textnormal{V}^q_{11}$ \\ 
\hline\hline 

$s$           &  $\ell$  &   $\frac{1}{\sqrt{3}}\,\tau^{\sigma,1}_{2j}$   \\   \hline\hline
$\quad\nicefrac{3}{2}\quad$   &  $2$     &   $\textnormal{S}^g_{11}+\textnormal{S}^p_{31}+2\,\textnormal{S}^q_{41}-\frac{1}{3}\,(\textnormal{V}^p_{13}+2\,\textnormal{V}^q_{14})$ \\

$\nicefrac{3}{2}$   &  $\quad 2\quad$     &    $\textnormal{S}^g_{12}-2\textnormal{S}^p_{41}-\frac{2}{3}(\textnormal{V}^q_{13}-2\textnormal{V}^p_{14})$  \\ 

$\nicefrac{3}{2}$   &  $2$     &    $\textnormal{S}^g_{13}+2\,(\textnormal{S}^q_{43}-\textnormal{S}^q_{34})+\frac{2}{3}\,(\textnormal{V}^p_{11}+2\,\textnormal{V}^q_{12})$   \\

$\nicefrac{3}{2}$   &  $2$     &    $\textnormal{S}^g_{14}-2\,(\textnormal{S}^p_{43}-\textnormal{S}^p_{34})+\frac{2}{3}\,(\textnormal{V}^q_{11}-2\,\textnormal{V}^p_{12})$  \\

\hline\hline 

$s$           &  $\ell$  &   $\sqrt{5}\,\tau^{\sigma,1}_{3j}$   \\   \hline\hline
$\quad\nicefrac{3}{2}\quad$   &  $2$     &   $\sqrt{5}(\textnormal{S}^g_{11}+3\,\textnormal{S}^p_{31}-\textnormal{V}^p_{13})$ \\

$\nicefrac{3}{2}$   &  $\quad 3\quad$     &    $4\,\textnormal{S}^g_{12}+5\,(\textnormal{S}^p_{32}+\textnormal{S}^q_{42})+\textnormal{V}^p_{14}-\textnormal{V}^q_{13})$  \\ 

$\nicefrac{3}{2}$   &  $3$     &    $\textnormal{S}^g_{13}+5\,\textnormal{S}^p_{33}-\textnormal{V}^p_{11}$   \\

$\nicefrac{3}{2}$   &  $3$     &    $\textnormal{S}^g_{14}+5\,\textnormal{S}^q_{44}-\textnormal{V}^q_{11}$  \\

\hline\hline 

$s$           &  $\ell$  &   $\frac{1}{\sqrt{5}}\,\tau^{\sigma,1}_{4j}$   \\   \hline\hline
$\quad\nicefrac{3}{2}\quad$   &  $3$     &   $\, \textnormal{S}^g_{11}+2\,(\textnormal{S}^q_{32}+\textnormal{S}^q_{41}+\textnormal{S}^p_{31})-\frac{2}{3}\,(\textnormal{V}^p_{13}+2\,\textnormal{V}^q_{14}) \,$ \\

$\nicefrac{3}{2}$   &  $\quad 3\quad$     &    $\textnormal{S}^p_{32}-\textnormal{S}^q_{42}-\frac{1}{3}\,(\textnormal{V}^q_{13}+\textnormal{V}^p_{14})$  \\ 

$\nicefrac{3}{2}$   &  $3$     &    $\textnormal{S}^g_{13}+\textnormal{S}^p_{33}+2\textnormal{S}^q_{43}+\frac{1}{3}\,(\textnormal{V}^p_{11}+2\,\textnormal{V}^q_{12})$   \\

$\nicefrac{3}{2}$   &  $3$     &    $\textnormal{S}^g_{14}+\textnormal{S}^q_{44}+2\textnormal{S}^p_{34}+\frac{1}{3}\,(\textnormal{V}^q_{11}-2\,\textnormal{V}^p_{12})$  \\
\hline
\end{tabular}
\end{center}
\caption{Orthonormal Dirac basis $\mathrm{\tau}_{ij}^{\sigma,k}$  with $s=\nicefrac{3}{2}$ and for $k=1$. We omit the Dirac and Lorentz indices as well as $\sigma$ for better readability. }\label{table_basis3_2}
\end{table}

\begin{table}[htbp]
\begin{center}
\begin{tabular}{ | c | c || c  |  c |} \hline

$s$    &  $\ell$  &   $\frac{1}{\sqrt{3}}\,\tau^{\sigma,1}_{5j}$ &  $\tau^{\sigma,1}_{6j}$ \\ \hline\hline

$\quad\nicefrac{1}{2}\quad$   &  $\quad 1\quad$    &  $\quad \textnormal{S}^p_{14}-\textnormal{S}^q_{13} \quad$ & $\quad \textnormal{V}^p_{14}-\textnormal{V}^q_{13} \quad$ \\

$\nicefrac{1}{2}$   &  $1$    &  $\textnormal{S}^p_{11}$ & $\textnormal{V}^p_{11}$ \\ 

$\nicefrac{1}{2}$   &  $1$    &  $\textnormal{S}^q_{11}$ & $\textnormal{V}^q_{11}$ \\

$\nicefrac{1}{2}$   &  $2$    &  $\textnormal{S}^p_{13}+\textnormal{S}^q_{14}$ & $\textnormal{V}^p_{13}+\textnormal{V}^q_{14}$ \\  
\hline\hline 

$s$    &  $\ell$  &   $\tau^{\sigma,1}_{7j}$ &  $\sqrt{3}\,\tau^{\sigma,1}_{8j}$ \\ \hline\hline

$\nicefrac{1}{2}$   &  $2$    &  $\textnormal{S}^p_{13}-\textnormal{S}^q_{14}$ & $\textnormal{V}^p_{13}-\textnormal{V}^q_{14}$ \\

$\nicefrac{1}{2}$   &  $2$    &  $\textnormal{S}^q_{13}+\textnormal{S}^p_{14}$ & $\textnormal{V}^q_{13}+\textnormal{V}^p_{14}$ \\ 

$\nicefrac{1}{2}$   &  $2$    &  $\textnormal{S}^p_{11}+2\,\textnormal{S}^q_{12}$ & $\textnormal{V}^p_{11}+2\,\textnormal{V}^q_{12}$ \\

$\nicefrac{1}{2}$   &  $2$    &  $\textnormal{S}^q_{11}-2\,\textnormal{S}^p_{12}$ & $\textnormal{V}^q_{11}-2\,\textnormal{V}^p_{12}$ \\  
\hline
\end{tabular}
\end{center}
\caption{Orthonormal Dirac basis $\mathrm{\tau}_{ij}^{\sigma,k}$  
with $s=\nicefrac{1}{2}$ and for $k=1$. We omit the Dirac and 
Lorentz indices as well as $\sigma$ for better readability.}
\label{table_basis1_2}
\end{table}

In the baryon rest frame, the total spin and 
relative angular momentum operators are
\begin{equation}\label{eq:SL_operators}
\begin{array}{l}
\textnormal{S}^2=\frac{9}{4}\left(\mathbb{1}\otimes\mathbb{1}
\otimes\mathbb{1}\right)+\frac{1}{2}\left(\sigma^{\mu\nu}
\otimes\sigma^{\mu\nu}\otimes\mathbb{1}+\textnormal{perm.}\right)~,\quad\\
L^2=L^2_{(p)}+L^2_{(q)}+2\,L_{(p)}\cdot L_{(q)}~,
\end{array}
\end{equation}
with
\begin{equation}\label{eq:L_operators}
\begin{array}{rcl}
L^2_{(p)}&=&2\,\textbf{p}\cdot\nabla_\textbf{p}
+p^i(\textbf{p}\cdot\nabla_\textbf{p})\nabla^i_\textbf{p}
-\textbf{p}^2\Delta_\textbf{p}~,\quad\\
L^2_{(q)}&=&2\,\textbf{q}\cdot\nabla_\textbf{q}
+q^i(\textbf{q}\cdot\nabla_\textbf{q})\nabla^i_\textbf{q}
-\textbf{q}^2\Delta_\textbf{q}~,\quad \\
L_{(p)}\cdot L_{(q)}&=&
p^i(\textbf{q}\cdot\nabla_\textbf{p})\nabla^i_\textbf{q}
-(\textbf{p}\cdot\textbf{q})(\nabla_\textbf{p}\cdot\nabla_\textbf{q})~,\quad
\end{array}
\end{equation}
where $\textbf{p}$ and $\textbf{q}$ are the spatial parts of $p_T$ and $q_t$, 
respectively.

It is useful to realize that the basis elements 
containing $\{\textnormal{S},\textnormal{V}\}$ and 
$\{\textnormal{P},\textnormal{A}\}$, which differ by a 
$\gamma^5\otimes\gamma^5$, and those with a different value for the 
index $\sigma=\pm$, do not mix under the action of 
$\textnormal{S}^2$ or $L^2$ and then can be analyzed independently. 
On the other hand, from Equations\ (\ref{eq:SL_operators}) and 
(\ref{eq:L_operators}) one can infer that the set (\ref{eq:LI_set}) 
can be further subdivided into four subsets which (due to their different 
momentum dependence) do not mix under the action of 
$\textnormal{S}^2$ or $L^2$:
\begin{equation}
\begin{array}{l}
1,p^2,q^2,p^2q^2:\quad \textnormal{S}^g_{11},\textnormal{S}^p_{13},
\textnormal{S}^p_{31},\textnormal{S}^q_{14},\textnormal{S}^q_{41},
\textnormal{S}^q_{32},\textnormal{V}^q_{14},\textnormal{V}^p_{13}~,\\
pq,p^3q,pq^3:\qquad \textnormal{S}^g_{12},\textnormal{S}^p_{14},
\textnormal{S}^p_{41},\textnormal{S}^p_{32},\textnormal{S}^q_{13},
\textnormal{S}^q_{42},\textnormal{V}^p_{14},\textnormal{V}^q_{13}~,\\
p,pq^2,p^3:\qquad \,\,\,\,\,\, \textnormal{S}^g_{13},\textnormal{S}^p_{11},
\textnormal{S}^p_{33},\textnormal{S}^q_{12},\textnormal{S}^q_{34},
\textnormal{S}^q_{43},\textnormal{V}^p_{11},\textnormal{V}^q_{12}~,\\
q,p^2q,q^3:\qquad \,\,\,\,\,\,  \textnormal{S}^g_{14},\textnormal{S}^q_{11},
\textnormal{S}^q_{44},\textnormal{S}^p_{12},\textnormal{S}^p_{34},
\textnormal{S}^p_{43},\textnormal{V}^q_{11},\textnormal{V}^p_{12}~,
\end{array}
\end{equation}
where the left column indicates symbolically the different momentum  dependence
of the basis elements, in powers of  $\widehat{p_T}$ and $\widehat{q_t}$,
denoted as $p$ and $q$, respectively.  This allows to simplify the partial-wave
decomposition by looking for $S^2$ and $L^2$ eigenfunctions only within the
above subsets.

The operator $S^2$ is independent of the momentum content of the basis
elements. Therefore it is sufficient to find the eigenstates at fixed values
of $\widehat{p_T}$ and $\widehat{q_t}$. As we mentioned at the beginning, the problem can
be easily implemented and solved using a symbolic programming language.

It is instructive to study how the decomposition for the orbital angular momentum is performed using some simple 
examples (again, we use Mathematica\texttrademark~ for the full calculation). The $\ell=0$ elements can be found immediately; they are the 
momentum independent elements in (\ref{eq:LI_set}), 
\textit{i.e.}, 
$[\textnormal{S}^g_{11}]^\sigma$ and $[\textnormal{P}^g_{11}]^\sigma$. 
For the remaining basis elements, let us note that they can be written as 
contractions of
\begin{equation}
 p^\alpha,\,\, q^\alpha,\,\, p^\alpha q^\beta,\,\, p^\alpha p^\beta,\,\, 
 q^\alpha q^\beta,\,\, p^\alpha p^\beta q^\delta,\,\,\dots
\end{equation}
with appropriate Dirac-Lorentz momentum-independent tensors. 
For the $\ell=1$ elements it is enough to consider the first three elements 
in the list above. Applying the orbital angular momentum operator one gets
\begin{equation}
\begin{array}{c}
  L^2 p^\alpha=2\,p^\alpha\,,\\
  L^2 q^\alpha=2\,q^\alpha\,,\\
  L^2 p^\alpha q^\beta=4\,p^\alpha q^\beta+2\,q^\alpha p^\beta\,,
\end{array}
\end{equation}
and from here it is clear that the $L^2$ eigenfunctions will come from the 
combinations
\begin{equation}
  p^\alpha,\,\,q^\alpha,\,\,p^\alpha q^\beta-q^\alpha p^\beta\,,
\end{equation}
again contracted with the corresponding Dirac-Lorentz structures. 
For other $\ell$ values the calculation proceeds along the same lines but 
the details are more involved.

The above analysis focused on the subset $\{\textnormal{S},\textnormal{V}\}$, 
which we will denote by $k=1$. Similar results hold for the set 
$\{\textnormal{P},\textnormal{A}\}$, denoted by k=2. The final result of 
the partial-wave decomposition is given in tables \ref{table_basis3_2} 
and \ref{table_basis1_2} for $k=1$. The case $k=2$ is obtained from the 
previous elements by exchanging $\textnormal{S}\rightarrow\textnormal{P}$, 
$\textnormal{V}\rightarrow\textnormal{A}$ and adding an extra minus sign 
to the elements $\textnormal{P}^g_{1j}$. 
This basis fulfills the following orthonormality relation
\begin{equation}
\begin{array}{l}
 \frac{1}{8}\textnormal{Tr}\left(\bar{\tau}^{\sigma,k}_{ij}
 \tau^{\sigma',k'}_{i'j'}\right)=\\
\qquad\qquad\frac{1}{8}
\left(\bar{\tau}^{\sigma,k}_{ij}\right)_{\beta\alpha\delta\gamma}^\mu
\left(\tau^{\sigma',k'}_{i'j'}\right)_{\alpha\beta\gamma\delta}^\mu=
\delta_{ii'}\delta_{jj'}\delta_{kk'}\delta_{\sigma\sigma'}\,,
\end{array}
\end{equation}
where the conjugation of the basis elements is defined as
\begin{equation}
 \bar{\tau}^\mu_{\alpha\beta\gamma\delta}(p,q,P)=-C_{\alpha b}C_{\gamma d}
 \left[\tau^\mu_{abcd}(-p,-q,-P)\right]^TC_{a\beta}^TC_{c\delta}^T\,.
\end{equation}

%%%%%%%%%%%%%%%%%%%%%%%%%%%%%%%%%%%%%%%%%%%%%%%%%%%%%%%%%%
\section{Color traces}\label{sec:appendix_color}
%%%%%%%%%%%%%%%%%%%%%%%%%%%%%%%%%%%%%%%%%%%%%%%%%%%%%%%%%%

To calculate the color factors of the quark DSE and the Faddeev equation we only need the following identities for the Gell-Mann color matrices $t^m_{AB}$
\begin{flalign}
 \sum_{m,C}t^{m}_{AC}t^{m}_{CB}&=-\frac{N_c^2-1}{2N_c}\delta_{AB} \\
 \textnormal{Tr}~t^m&=0 
\end{flalign}
where lower-case and capital indices run over the fundamental and the adjoint representations, respectively, of the color $SU(N_c)$ group.
\begin{figure}[hbtp]
 \begin{center}
  \includegraphics[width=0.7\textwidth,clip]{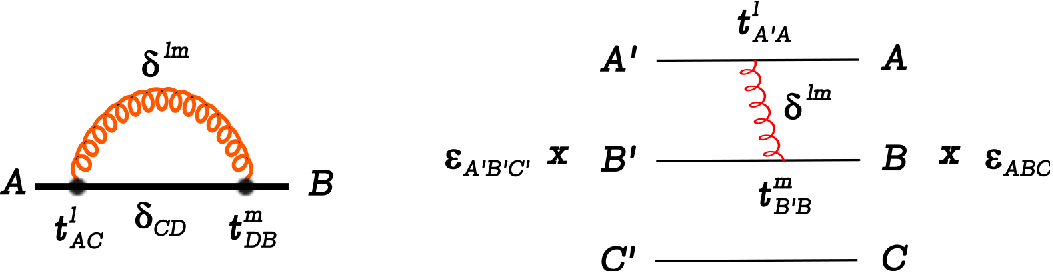}
 \end{center}
 \caption{Diagrams to calculate the color factors in the quark DSE (left) and in the Faddeev equation (right).}\label{fig:color}
\end{figure}

For the quark propagator DSE we have (see Figure \ref{fig:color})
\begin{equation}
 \sum_{l,m,C,D}t^{l}_{AC}\delta^{lm}\delta_{CD}t^m_{DB}=\sum_{m,C}t^{m}_{AC}t^m_{CB}=\frac{N_c^2-1}{2N_c}\delta_{AB}
\end{equation}
and therefore the color factor is $(N_c^2-1)/(2N_c)$, which equals $4/3$ for $N_c=3$.

For the Faddeev equation we consider the color factor is the same for all three diagrams. As shown in Section \ref{sec:basis}, the color part of the Faddeev amplitudes is 
\begin{equation}
 \frac{\epsilon_{ABC}}{\sqrt{6}}
\end{equation}
Therefore, as depicted in Figure \ref{fig:color}, we have
\begin{flalign}
 \sum_{\textnormal{all indices}}&\frac{\epsilon_{A'B'C'}}{\sqrt{6}}\left(t^{l}_{A'A}\delta^{lm}t^m_{B'B}\delta_{C'C}\right)\frac{\epsilon_{ABC}}{\sqrt{6}} \nonumber\\
 &=\sum_{\textnormal{all indices}}\frac{1}{6}~\epsilon_{A'B'C}\epsilon_{ABC}\left(t^{m}_{A'A}t^m_{B'B}\right)
 = \sum_{\textnormal{all indices}}\frac{1}{6}~\left(\delta_{A'A}\delta_{B'B}-\delta_{A'B}\delta_{B'A}\right)\left(t^{m}_{A'A}t^m_{B'B}\right) \nonumber\\
 &=\sum_{m,A,B}\frac{1}{6}~\left(t^{m}_{AA}t^m_{BB}-t^{m}_{BA}t^m_{AB}\right)=-\frac{1}{6}~\sum_{m,A,B}t^{m}_{BA}t^m_{AB}=-\frac{1}{6}~\frac{N_c^2-1}{2N_c}\sum_{B}\delta_{BB}\nonumber\\
 &= -\frac{N_c^2-1}{12}
\end{flalign}
where we used $\epsilon_{A'B'C}\epsilon_{ABC}=\delta_{A'A}\delta_{B'B}-\delta_{A'B}\delta_{B'A}$. The color factor thus gives $-2/3$ for $N_c=3$.

%%%%%%%%%%%%%%%%%%%%%%%%%%%%%%%%%%%%%%%%%%%%%%%%%%%%%%%%%%
\section{Numerical details}\label{sec:numdetails}
%%%%%%%%%%%%%%%%%%%%%%%%%%%%%%%%%%%%%%%%%%%%%%%%%%%%%%%%%%

\subsection{Resolution of the Faddeev equation}

The numerical techniques used in this work are an extension of those explained
in \cite{Eichmann2011a} to the case of the nucleon. We
summarize the main ideas in this appendix and extend them to the case of the $\Delta$-baryon. Using the symmetries of the Faddeev amplitudes, and requiring certain symmetry properties for the interaction kernel, we will be able to relate the three terms in the Faddeev equation (\ref{eq:faddeev_eq}) and, therefore, simplify its resolution.

The full Faddeev amplitude is the product of color, flavor and spin parts. Since it describes baryons in terms of their valence quarks, it must be antisymmetric under the exchange of any two of the three quarks, as
required by the Pauli principle. Furthermore, a baryon must be a color singlet and hence, as explained in Section \ref{sec:basis}, the
color part of the Faddeev amplitude is $\epsilon_{ABC}$ and therefore is always antisymmetric. Thus, the
product of flavor and spin parts must be symmetric. As we show below, these symmetry properties
allow to relate the three terms in the covariant Faddeev equation (see Figure \ref{fig:FaddeevRLeq}).

Although in Equation (\ref{eq:faddeev_eq}) (and throughout this work) only the Rainbow-Ladder interaction kernel is considered, for the derivation in this section we relax this constraint a bit and require the 2-body interaction kernel to be only flavor- and color-independent. In this way the color and flavor parts of the Faddeev amplitude factor out and we are left with an equation for the spin part
  %\begin{flalign}
  % \Gamma_{ABCD}(p,q,P)=&\int \frac{d^4k}{(2\pi)^4}~\Bigl[~K_{AA',BB'}~\delta_{CC''}~S_{A'A''}(-q-\frac{p}{2}+\frac{P}{3}-k)~S_{B'B''}(q-\frac{p}{2}+\frac{P}{3}+k)\times \nonumber\\
   %                  & ~~~~~~~~~~~~~~~~\Gamma_{A''B''C''D}(p,q+k,P)~\Bigr]~+\nonumber\\
   %                   &\int \frac{d^4k}{(2\pi)^4}~\Bigl[~K_{AA',CC'}~\delta_{BB''}~S_{A'A''}(-q-\frac{p}{2}+\frac{P}{3}+k)~S_{C'C''}(p+\frac{P}{3}+k)\times \nonumber\\
   %                  & ~~~~~~~~~~~~~~~~\Gamma_{A''B''C''D}(p-k,q-\frac{k}{2},P)~\Bigr]~+\nonumber\\
   %                   &\int \frac{d^4k}{(2\pi)^4}~\Bigl[~K_{BB',CC'}~\delta_{AA''}~S_{B'B''}(q-\frac{p}{2}+\frac{P}{3}+k)~S_{C'C''}(p+\frac{P}{3}+k)\times \nonumber\\
%                     & ~~~~~~~~~~~~~~~~\Gamma_{A''B''C''D}(p+k,q-\frac{k}{2},P)~\Bigr]
%  \end{flalign}

  \begin{flalign}\label{eq:simbolic_faddeev_original}
   \Psi_{\alpha\beta\gamma \mathcal{I}}(p,q,P)={}&\int \frac{d^4k}{(2\pi)^4}~\Bigl[~K_{\beta\beta',\gamma\gamma'}(k_2,\widetilde{k}_3;k)~\delta_{\alpha\alpha''}~S_{\beta'\beta''}(k_2)~S_{\gamma'\gamma''}(\widetilde{k}_3)\nonumber\\
&~~~~~~~~~~~~~~~~~~~~~~~~~~~~~~~~~~~~~~~~~~~~\times\Psi_{\alpha''\beta''\gamma'' \mathcal{I}}(p^{(1)},q^{(1)},P)~\Bigr]~+ \nonumber \\
                      &\int \frac{d^4k}{(2\pi)^4}~\Bigl[~K_{\alpha\alpha',\gamma\gamma'}(k_3,\widetilde{k}_1;k)~\delta_{\beta\beta''}~S_{\alpha'\alpha''}(\widetilde{k}_1)~S_{\gamma'\gamma''}(k_3)\nonumber\\
&~~~~~~~~~~~~~~~~~~~~~~~~~~~~~~~~~~~~~~~~~~~~\times\Psi_{\alpha''\beta''\gamma'' \mathcal{I}}(p^{(2)},q^{(2)},P)~\Bigr]~+\nonumber\\
                      &\int \frac{d^4k}{(2\pi)^4}~\Bigl[~K_{\alpha\alpha',\beta\beta'}(k_1,\widetilde{k}_2;k)~\delta_{\gamma\gamma''}~S_{\alpha'\alpha''}(k_1)~S_{\beta'\beta''}(\widetilde{k}_2)\nonumber\\
&~~~~~~~~~~~~~~~~~~~~~~~~~~~~~~~~~~~~~~~~~~~~\times\Psi_{\alpha''\beta''\gamma''\mathcal{I}}(p^{(3)},q^{(3)},P)~\Bigr]~,                   
  \end{flalign}
where $k$ here only means the exchanged momentum between the two interacting quarks.

The strategy to relate all terms in this equation is to perform permutations on the quark momenta $\{p_1,p_2,p_3\}$ and quark indices $\{\alpha,\beta,\gamma\}$ of the Faddeev amplitudes, such that the three integrals (which in the following we call $I1$, $I2$ and $I3$) look formally the same. In particular, for reasons we explain at the end of the section, we want all of them to look formally like $I3$. For the first integral the permutation we are interested in is $(123)\rightarrow (231)$ and for the second is $(123)\rightarrow (312)$. 

Using now  $\{ABC\}$ as generic quark indices for Dirac, flavor and color indices (e.g. $A\rightarrow\{\alpha,a,r\}$ where $r$ would be a color index) and using for clarity $\{p_1,p_2,p_3\}$ as the arguments of the Faddeev amplitudes instead of the (equivalent) set $\{p,q,P\}$, it reads
\begin{equation}
 \Gamma_{ABCD}(p_1,p_2,p_3)=\left(\sum_\rho \Psi^\rho_{\alpha\beta\gamma\mathcal{I}}(p_1,p_2,p_3) \otimes F^\rho_{abcd}\right)\otimes \frac{\epsilon_{rst}}{\sqrt{6}}~.
\end{equation}
Here the index $\rho$ denotes the representation of the isospin group to which the baryon belongs and $F$ is the flavor part of the amplitude. For example, in the case of baryons in the octet $SU(3)$ representation (spin-$\nicefrac{1}{2}$ baryons like the nucleon), they can belong to a mixed-symmetric or a mixed-antisymmetric representation, and the physical state is a quantum superposition of both. As explained above, the symmetry properties of the spin part can be deduced from the fact that the Faddeev amplitude must be antisymmetric in the first three indices. The color part is automatically antisymmetric and this forces the product of spin and flavor part to be symmetric. The symmetry properties of the flavor parts are known, since they are obtained from the quark model for baryons, and from them the transformation properties of the spin part can be inferred. 

The application of this ideas to the nucleon and the Delta have been described in \cite{Eichmann2011a} and \cite{SanchisAlepuz:2011jn}, respectively. The nucleon is a member of the octet representation of the SU(3) flavor group. There are two such octets, corresponding to mixed-symmetric and mixed-antisymmetric flavor states (see e.g. \cite{Griffiths:1987tj}). Under permutations of the quark indices, these two representations mix
\begin{align}
 F^\rho_{abcd}&=M_1^{\rho\rho'}F^{\rho'}_{bcad}~, \nonumber\\
 F^\rho_{abcd}&=M_2^{\rho\rho'}F^{\rho'}_{cabd}~,
\end{align}
with $\rho=1,2$ and
\begin{equation}
 M_1=\frac{1}{2}\left(
           \begin{array}{cc}
            -1 & -\sqrt{3} \\
            \sqrt{3} & -1
           \end{array}
\right)~,\qquad
 M_2=\frac{1}{2}\left(
            \begin{array}{cc}
            -1 & \sqrt{3} \\
            -\sqrt{3} & -1
           \end{array}
\right)~,
\end{equation}
and therefore the spin parts must transform in the same way
\begin{align}
 \Psi^\rho_{\alpha\beta\gamma \mathcal{I}}(p_1,p_2,p_3)&=M_1^{\rho\rho'}\Psi^{\rho'}_{\beta\gamma\alpha \mathcal{I}}(p_2,p_3,p_1)~, \nonumber\\
 \Psi^\rho_{\alpha\beta\gamma \mathcal{I}}(p_1,p_2,p_3)&=M_2^{\rho\rho'}\Psi^{\rho'}_{\gamma\alpha\beta \mathcal{I}}(p_3,p_1,p_2)~,
\end{align}
to get a singlet under permutations for the product of spin and flavor parts.
The Delta and Omega baryons belong to the decuplet flavor representation, which is a symmetric one. In this case, $\rho=1$, $M_{1,2}=\mathbb{1}$ and the spin parts are therefore simply symmetric under permutations
\begin{align}\label{eq:permutation_delta1}
 \Psi_{\alpha\beta\gamma \mathcal{I}}(p_1,p_2,p_3)&=\Psi_{\beta\gamma\alpha \mathcal{I}}(p_2,p_3,p_1)~, \nonumber\\
 \Psi_{\alpha\beta\gamma \mathcal{I}}(p_1,p_2,p_3)&=\Psi_{\gamma\alpha\beta \mathcal{I}}(p_3,p_1,p_2)~.
\end{align}

In both cases, we show now that the permutation of the quark momenta is equivalent to evaluate the amplitude with non-permuted quark momenta (i.e. $I3$) at different relative momenta $\{p',q'\}$ and $\{p'',q''\}$, respectively. First, let us recall the definitions of the relative and total momenta
\begin{equation}\label{eq:defpqAppendix}
\begin{array}{rl@{\quad}rl}
        p &= (1-\zeta)\,p_3 - \zeta (p_1+p_2)\,, &  p_1 &=  -q -\dfrac{p}{2} + \dfrac{1-\zeta}{2} P\,, \\[0.25cm]
        q &= \dfrac{p_2-p_1}{2}\,,         &  p_2 &=   q -\dfrac{p}{2} + \dfrac{1-\zeta}{2} P\,, \\[0.25cm]
        P &= p_1+p_2+p_3\,,                &  p_3 &=   p + \zeta  P~,
\end{array}
\end{equation}
as well as the definitions for the internal relative momenta
\begin{equation}\label{internal-relative-momentaAppendix}
\begin{array}{l@{\quad}l@{\quad}l}
p^{(1)} = p+k~,& p^{(2)} = p-k~,& p^{(3)} = p~,\\
q^{(1)} = q-k/2~,& q^{(2)} = q-k/2~, & q^{(3)} = q+k~,
\end{array}
\end{equation}
with $k_i=p_i-k$ and $\tilde{k}_i=p_i+k$ the internal quark momenta. Now, let us show that the quark and internal relative momenta in the terms $I1$ and $I2$ of the Faddeev equation can be written as those in $I3$ by introducing new external relative momenta $\{p',q'\}$ and $\{p'',q''\}$,
\begin{flalign}
 \textnormal{I}3:& \left\{\begin{array}{l}
       k_1=-q-\frac{p}{2}+\frac{P}{3}-k \\
       \widetilde{k}_2=q-\frac{p}{2}+\frac{P}{3}+k \\
       p_3=p+\frac{P}{3}
      \end{array}\right. \nonumber \\
 \textnormal{I}1:& \left\{\begin{array}{l}
       (p_1=-q-\frac{p}{2}+\frac{P}{3})\equiv(p'_3=p'+\frac{P}{3}) \\
       (k_2=q-\frac{p}{2}+\frac{P}{3}-k)\equiv (k'_1=-q'-\frac{p'}{2}+\frac{P}{3}-k) \\
       (\widetilde{k}_3=p+\frac{P}{3}+k)\equiv (\widetilde{k}'_2=q'-\frac{p'}{2}+\frac{P}{3}+k)
      \end{array}\right.\quad\Rightarrow \nonumber \\
    &\qquad\qquad\Rightarrow\boxed{p'=-q-\frac{p}{2}~,\quad q'=-\frac{q}{2}+\frac{3p}{4}}\label{eq:permutation_delta2} \\
 \textnormal{I}2:& \left\{\begin{array}{l}
       (\widetilde{k}_1=-q-\frac{p}{2}+\frac{P}{3}+k)\equiv(\widetilde{k}''_2=q''-\frac{p''}{2}+\frac{P}{3}+k) \\
       (p_2=q-\frac{p}{2}+\frac{P}{3})\equiv (p''_3=p''+\frac{P}{3}) \\
       (k_3=p+\frac{P}{3}-k)\equiv (k''_1=-q''-\frac{p''}{2}+\frac{P}{3}-k)
      \end{array}\right.\quad\Rightarrow \nonumber \\
    &\qquad\qquad\Rightarrow\boxed{p''=q-\frac{p}{2}~,\quad q''=-\frac{q}{2}-\frac{3p}{4}}\label{eq:permutation_delta3}
\end{flalign}
This identification is possible if one has $(1-\zeta)/2=\zeta$ in (\ref{eq:defpqAppendix}). To this end, we need to set $\zeta=1/3$. 

Making these substitutions in (\ref{eq:simbolic_faddeev_original}), we get
  \begin{flalign}\label{eq:simbolic_faddeev_primes}
   \Psi^\rho_{\alpha\beta\gamma\mathcal{I}}(p,q,P)={}&M_1^{\rho\rho'}\int \frac{d^4k}{(2\pi)^4}~\Bigl[~K_{\beta\beta',\gamma\gamma'}(k'_1,\widetilde{k}'_2,k)~\delta_{\alpha\alpha''}~S_{\beta'\beta''}                            (k'_1)~S_{\gamma'\gamma''}(\widetilde{k}'_2)\nonumber\\
&~~~~~~~~~~~~~~~~~~~~~~~~~~~~~~~~~~~~~~~~~~~~\times\Psi^{\rho'}_{\beta''\gamma''\alpha''\mathcal{I}}(p'^{(3)},q'^{(3)},P)~\Bigr]~+ \nonumber \\
                      &M_2^{\rho\rho'}\int \frac{d^4k}{(2\pi)^4}~\Bigl[~K_{\alpha\alpha',\gamma\gamma'}(k''_1,\widetilde{k}''_2,k)~\delta_{\beta\beta''}~S_{\gamma'\gamma''}(k''_1)~S_{\alpha'\alpha''}(\widetilde{k}''_2)\nonumber\\
&~~~~~~~~~~~~~~~~~~~~~~~~~~~~~~~~~~~~~~~~~~~~\times\Psi^{\rho'}_{\gamma''\alpha''\beta''\mathcal{I}}(p''^{(3)},q''^{(3)},P)~\Bigr]~+\nonumber\\
                      &\delta^{\rho\rho'}\int \frac{d^4k}{(2\pi)^4}~\Bigl[~K_{\alpha\alpha',\beta\beta'}(k_1,\widetilde{k}_2,k)~\delta_{\gamma\gamma''}~S_{\alpha'\alpha''}(k_1)~S_{\beta'\beta''}(\widetilde{k}_2)\nonumber\\
&~~~~~~~~~~~~~~~~~~~~~~~~~~~~~~~~~~~~~~~~~~~~\times\Psi^{\rho'}_{\alpha''\beta''\gamma''\mathcal{I}}(p^{(3)},q^{(3)},P)~\Bigr]~,                      
  \end{flalign}
with the internal relative momenta defined, as in (\ref{internal-relative-momenta}),
\begin{equation}\label{internal-relative-momenta2}
\begin{array}{l@{\quad}l@{\quad}l}
p^{(1)} = p+k,& p^{(2)} = p-k,& p^{(3)} = p,\\
q^{(1)} = q-k/2,& q^{(2)} = q-k/2, & q^{(3)} = q+k.
\end{array}
\end{equation}
 Renaming dummy indices it becomes clear that, if the interaction kernel is such that
\begin{equation}\label{eq:condition_on_kernel}
 K_{\alpha\alpha',\gamma\gamma'}(k''_1,\widetilde{k}''_2,k)=K_{\gamma\gamma',\alpha\alpha'}(k''_1,\widetilde{k}''_2,k)~,
\end{equation}
then the three integrals are formally the same and, denoting $I3$ by $\Psi^{(3)}_{\alpha\beta\gamma\mathcal{I}}(p,q,P)$, we have
\begin{equation}\label{eq:simbolic_faddeev_short}
 \Psi^\rho_{\alpha\beta\gamma\mathcal{I}}(p,q,P)=\Psi^{(3),\rho}_{\alpha\beta\gamma\mathcal{I}}(p,q,P)+M_1^{\rho\rho'}\Psi^{(3),\rho'}_{\beta\gamma\alpha\mathcal{I}}(p',q',P)+M_2^{\rho\rho'}\Psi^{(3),\rho'}_{\gamma\alpha\beta\mathcal{I}}(p'',q'',P)
\end{equation}
so that in practice we need to calculate only $\Psi^{(3)}$ and by further evaluating it at the points $\{p',q'\}$ and $\{p'',q''\}$ we obtain the full Faddeev amplitude. 

The requirement (\ref{eq:condition_on_kernel}) is obviously satisfied by the Rainbow-Ladder kernel (\ref{eq:ladder})
\begin{equation}
	K\sim\frac{\alpha_{eff}(k^2)}{k^2}~
	T_{\mu\nu}(k)~\gamma^\mu_{\alpha\alpha'}\gamma^\nu_{\gamma\gamma'}~.
\end{equation}
Nevertheless, the condition (\ref{eq:condition_on_kernel}) does not seem to be a very stringent one and it may be fulfilled by many 'beyond Rainbow-Ladder' kernels.

In practical calculations one solves for the scalar coefficients in the expansion of the Faddeev amplitudes
\begin{equation}
 \Psi^\rho_{\alpha\beta\gamma\mathcal{I}}(p,q,P)=f^{(\rho),(i)}(p^2,q^2,z_0,z_1,z_2)~\tau^{(\rho),(i)}_{\alpha\beta\gamma\mathcal{I}}(p,q,P)~,
\end{equation}
with $z_0=\widehat{p_T}\cdot\widehat{q_T}$,  $z_1=\widehat{p}\cdot\widehat{P}$ and  $z_2=\widehat{q}\cdot\widehat{P}$.
Then, the Faddeev equation for the coefficients $f^{(\rho),(i)}$ is
\begin{flalign}\label{eq:Faddeev_coeff}
 f^{(\rho),(i)}(p^2,q^2,z_0,z_1,z_2)=&f^{(\rho),(i),(3)}(p^2,q^2,z_0,z_1,z_2)+ \nonumber\\
       & M_1^{\rho\rho'}H_1^{ij}f^{(\rho'),(j),(3)}(p'^2,q'^2,z'_0,z'_1,z'_2)+ \nonumber\\
       & M_2^{\rho\rho'}H_2^{ij}f^{(\rho'),(j),(3)}(p''^2,q''^2,z''_0,z''_1,z''_2)~,
\end{flalign}
with
\begin{equation}
\begin{aligned}
 H_1^{ij}&=\left[\bar{\tau}^i_{\beta\alpha\mathcal{I}\gamma}(p,q,P)\tau^j_{\beta\gamma\alpha\mathcal{I}}(p',q',P)\right]~,\\
 H_2^{ij}&=\left[\bar{\tau}^i_{\beta\alpha\mathcal{I}\gamma}(p,q,P)\tau^j_{\gamma\alpha\beta\mathcal{I}}(p'',q'',P)\right]~,
\end{aligned}
\end{equation}
and
\begin{multline}\label{eq:Faddeev_coeff3}
f^{(\rho),(i),(3)}(p^2,q^2,z_0,z_1,z_2)=\int \frac{d^4k}{(2\pi)^4}~\textnormal{Tr}\Bigl[~\bar{\tau}^{(\rho),(i)}_{\beta\alpha\mathcal{I}\gamma}(p,q,P)K_{\alpha\alpha',\beta\beta'}(k_1,\widetilde{k}_2,k)~\delta_{\gamma\gamma''}\\
~~~~~~~~~~\times S_{\alpha'\alpha''}(k_1)~S_{\beta'\beta''}(\widetilde{k}_2)~\tau^{(\rho),(j)}_{\alpha''\beta''\gamma''\mathcal{I}}(p^{(3)},q^{(3)},P)~\Bigr]f^{(\rho),(j),(3)}((p^{(3)})^2,(q^{(3)})^2,z^{(3)}_0,z^{(3)}_1,z^{(3)}_2)~.
\end{multline}

Let us show why it is convenient to represent all diagrams in terms of $I3$. According to (\ref{internal-relative-momenta2}), in this case the internal and external relative momentum $p$ is the same ($p^{(3)}=p$), and can be chosen as in (\ref{eq:def_external_momenta}). On the other hand, if we change the integration variable\footnote{This is, in principle, only possible if we use a translation-invariant regularization of the integrals.} from the gluon momentum $k$ to the internal relative momentum $q^{(3)}$ (the Jacobian of this transformation is, according to (\ref{internal-relative-momenta2}), unity), then the transverse momenta which appear in the definition of the Faddeev basis (see Section \ref{sec:basis}) simplify to
\begin{equation}
 \widehat{p_T}= \widehat{p^{(3)}_T}=\left(
 \begin{array}{c}
  0 \\
  0 \\
  1 \\
  0
 \end{array}
\right)~,\qquad 
 \widehat{q_t}=\left(
 \begin{array}{c}
  0 \\
  1 \\
  0 \\
  0
 \end{array}
\right)~,\qquad 
 \widehat{q^{(3)}_t}=\left(
 \begin{array}{c}
  \sin\varphi \\
  \cos\varphi \\
  0 \\
  0
 \end{array}
\right)
\end{equation}
with $\varphi$ an angular integration variable. Now, defining the Faddeev wave-function as
\begin{equation}\label{eq:def_wavefunction}
\Phi_{\alpha\beta\gamma\mathcal{I}}(p,q,P)\equiv S_{\alpha\alpha'}(p_1)~S_{\beta\beta'}(p_2)~\Psi_{\alpha'\beta'\gamma'\mathcal{I}}(p,q,P)
\end{equation}
it can be expanded in terms of the same Faddeev basis. Then, the interaction kernel matrix becomes
\begin{equation}
\mathcal{K}^{(3)}_{ij}~^{\mu\nu}(\varphi)\sim\bar{\tau}^{i}_{\beta\alpha\mathcal{I}\gamma}(p,q,P)\left(\gamma^{\mu}_{\alpha\alpha'}\gamma^{\nu}_{\beta\beta'}\mathbb{1}_{\gamma\gamma'}\right)\tau^{j}_{\alpha'\beta'\gamma'\mathcal{I}}(p^{(3)},q^{(3)},P)
\end{equation}
and depends only on the angular variable $\varphi$. In the numerical resolution of the Faddeev equation, this kernel is treated as a matrix, which is now small enough to be calculated in advance, stored in memory and reused during the iteration process. This reduces considerably the computation time.

Finally, let us briefly describe how the Faddeev equation (\ref{eq:Faddeev_coeff}) is solved numerically. The first step is to modify (\ref{eq:Faddeev_coeff3}) by a multiplicative factor $\lambda$, thus transforming it into an eigenvalue equation. The physical solution corresponds to the case $\lambda=1$. The simplest method to solve an eigenvalue problem is by iteration. A test baryon-mass as well as a starting function for the coefficients $f^{(\rho),(i)}$ in the integral (\ref{eq:Faddeev_coeff3}) must be chosen for the first iteration. The resulting functions are used as an input for the next iteration and so on. If the converged eigenvalue is not $1$, then one chooses a different value for the test mass and repeats the procedure.

\begin{figure}[ht!]
 \begin{center}
  \includegraphics[width=0.7\textwidth,clip]{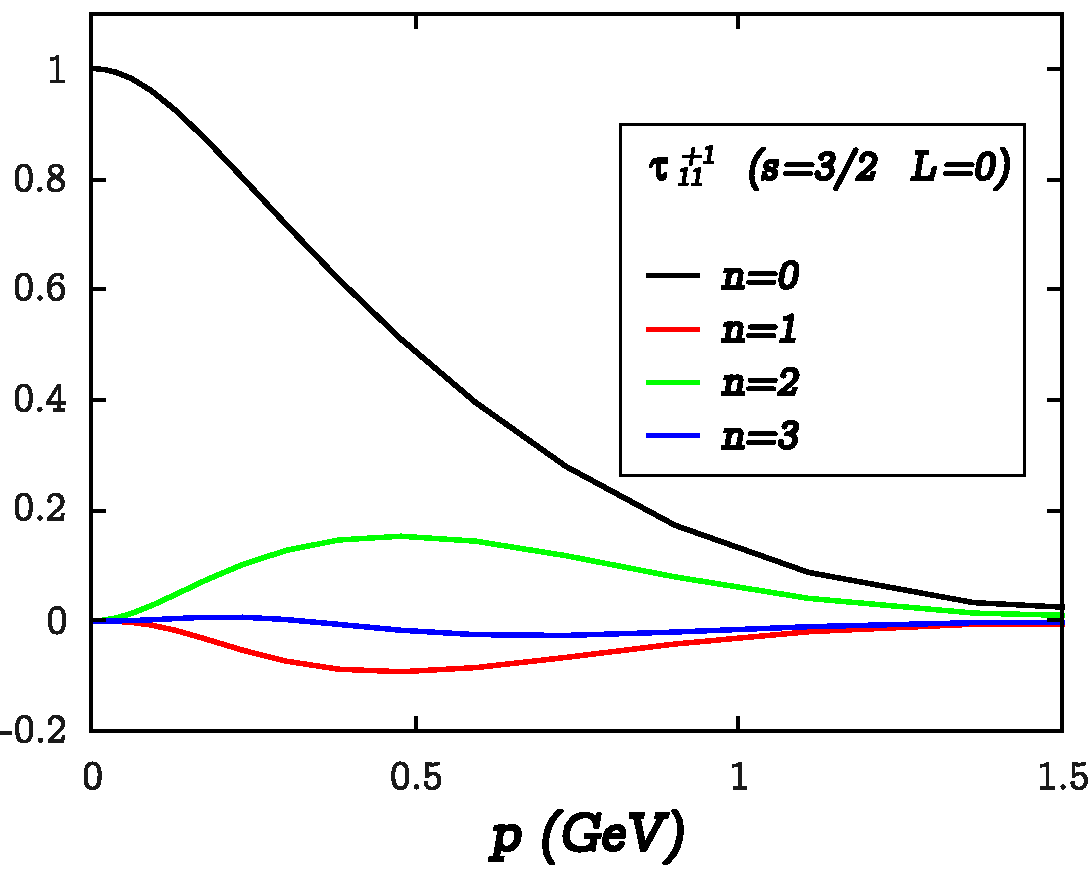}
 \end{center}
 \caption{Chebyshev moments of the $z_1$ variable as a function of $p$, for $q=0$, $z_0=0$ and $z_2=0$.}\label{fig:chebys-dominant}
\end{figure}

To optimize the numerics, we expand the angular dependence of the coefficients in Chebyshev polynomials
\begin{equation}\label{eq:cheby_expansion}
 f^{(\rho),(i)}(p^2,q^2,z_0,z_1,z_2)=f^{(\rho),(i)}_{lnm}(p^2,q^2)C_l(z_0)C_n(z_1)C_m(z_2)
\end{equation}
with $C_n(z)\equiv i^nU_n(z)$ and $U_n(z)$ the Chebyshev polynomials of the second kind
\begin{equation}
 U_n(z)=2^n\prod_{k=1}^{n}\left(z-\cos\frac{\pi k}{n+1}\right)~,\quad(n\ge 0)~.
\end{equation}
The main advantage of this expansion is that the angular dependence of the Faddeev amplitudes is typically weak, and therefore a very small number of Chebyshev moments $f^{(\rho),(i)}_{lnm}(p^2,q^2)$ are needed (for an illustration of a typical case, see Figure \ref{fig:chebys-dominant}).

Although the resulting baryon masses are not very sensitive to the precision used for the numerical integration in the Faddeev equation, for the calculation of form factors one needs a very precise determination of the Faddeev amplitudes. For the calculation in this work we used $30$~ Gauss-Legendre quadrature points for $\{p^2,q^2\}$, $8$~ Gauss-Chebyshev points for $\{z,z_0,z_1,z_2\}$ and $8$~ Gauss-Legendre points for $\{y,\varphi\}$ (using the notation of Section \ref{sec:conventions}). We use four Chebyshev moments in the expansion (\ref{eq:cheby_expansion}). For the evaluation of the amplitudes $f^{(\rho),(i)}_{lnm}(p^2,q^2)$ at $\{p^2,q^2\}$-values different from the Gaussian quadrature points, we use cubic-spline interpolation.

\subsection{Form factor calculation}

Using the same transformations we introduced in previous section, the three terms in the equation for the electromagnetic current in Rainbow-Ladder truncation (\ref{eq:FFeqRL}) can be expressed in terms of one of them. In this section, we limit the discussion to the case of spin-$\nicefrac{3}{2}$ baryons (the spin-$\nicefrac{1}{2}$ case has been treated in \cite{Eichmann2011a}).

The equation for the current is
\begin{flalign}\label{eq:FFsRL_appendix}
 J_{\mathcal{I}'\mathcal{I}}^\mu=\mathcal{Q}_1\int_p\int_q\bar{\Psi}_{\beta'\alpha'\mathcal{I}'\gamma'}(p_f^{\{1\}},q^{\{1\}}_f,P_f)\left[\left(S(p_1^f)\Gamma^\mu(p_1,Q)S(p_1^i)\right)_{\alpha'\alpha}S_{\beta'\beta}(p_2)S_{\gamma'\gamma}(p_3)\right]\times\nonumber\\
~~~~~~~~~~~~~~~~~~~~~~~~~~~~~~\left(\Psi_{\alpha\beta\gamma\mathcal{I}}(p^{\{1\}}_i,q^{\{1\}}_i,P_i)-\Psi^{\{1\}}_{\alpha\beta\gamma\mathcal{I}}(p^{\{1\}}_i,q^{\{1\}}_i,P_i)\right)\nonumber\\
+\mathcal{Q}_2\int_p\int_q\bar{\Psi}_{\beta'\alpha'\mathcal{I}'\gamma'}(p^{\{2\}}_f,q^{\{2\}}_f,P_f)\left[S_{\alpha'\alpha}(p_1)\left(S(p_2^f)\Gamma^\mu(p_2,Q)S(p_2^i)\right)_{\beta'\beta}S_{\gamma'\gamma}(p_3)\right]\times\nonumber\\
~~~~~~~~~~~~~~~~~~~~~~~~~~~~~~\left(\Psi_{\alpha\beta\gamma\mathcal{I}}(p^{\{2\}}_i,q^{\{2\}}_i,P_i)-\Psi^{\{2\}}_{\alpha\beta\gamma\mathcal{I}}(p^{\{2\}}_i,q^{\{2\}}_i,P_i)\right)\nonumber\\
+\mathcal{Q}_3\int_p\int_q\bar{\Psi}_{\beta'\alpha'\mathcal{I}'\gamma'}(p^{\{3\}}_f,q^{\{3\}}_f,P_f)\left[S_{\alpha'\alpha}(p_1)S_{\beta'\beta}(p_2)\left(S(p_3^f)\Gamma^\mu(p_3,Q)S(p_3^i)\right)_{\gamma'\gamma}\right]\times\nonumber\\
~~~~~~~~~~~~~~~~~~~~~~~~~~~~~~\left(\Psi_{\alpha\beta\gamma\mathcal{I}}(p^{\{3\}}_i,q^{\{3\}}_i,P_i)-\Psi^{\{3\}}_{\alpha\beta\gamma\mathcal{I}}(p^{\{3\}}_i,q^{\{3\}}_i,P_i)\right)~,
\end{flalign}
where we now wrote explicitly the charge of the quark $i$, $\mathcal{Q}_i$, that is part of the quark-photon vertex. Let us apply the transformations (\ref{eq:permutation_delta1}-\ref{eq:permutation_delta3}) to this equation. For example, the first term becomes
\begin{flalign}
\mathcal{Q}_1\int_p\int_q\bar{\Psi}_{\gamma'\beta'\mathcal{I}'\alpha'}(p_f'^{\{3\}},q'^{\{3\}}_f,P_f)\left[\left(S(p_3'^f)\Gamma^\mu(p'_3,Q)S(p_3'^i)\right)_{\alpha'\alpha}S_{\beta'\beta}(p'_1)S_{\gamma'\gamma}(p'_2)\right]\times\nonumber\\
~~~~~~~~~~~~~~~~~~~~~~~~~~~~~~\left(\Psi_{\beta\gamma\alpha\mathcal{I}}(p'^{\{3\}}_i,q'^{\{3\}}_i,P_i)-\Psi^{\{3\}}_{\beta\gamma\alpha\mathcal{I}}(p'^{\{3\}}_i,q'^{\{3\}}_i,P_i)\right)
\end{flalign}
where we used the definitions (\ref{eq:permutation_delta2}) and (\ref{eq:relative_momenta_withQ}), with $\zeta=1/3$, and we also introduced
\begin{equation}
\begin{aligned}
 p'_1&=-q'-\frac{p'}{2}+\frac{P}{3}=q-\frac{p}{2}+\frac{P}{3}=p_2~,\\
 p'_2&=q'-\frac{p'}{2}+\frac{P}{3}=p+\frac{P}{3}=p_3~,\\
 p'_3&=p'+\frac{P}{3}=-q-\frac{p}{2}+\frac{P}{3}=p_1~.
\end{aligned}
\end{equation}
Performing the change of variables $\{p,q\}\rightarrow\{p',q'\}$ in the integral (which is a transformation of Jacobian unity), this term is exactly equal to the third term in (\ref{eq:FFsRL_appendix}). In a similar way, the second term in (\ref{eq:FFsRL_appendix}) can be written formally like the third one. Therefore, introducing the total baryon charge $\mathcal{Q}_B=\mathcal{Q}_1+\mathcal{Q}_2+\mathcal{Q}_3$, the calculation of the electromagnetic current simplifies to
\begin{flalign}\label{eq:FFsRL_simplified}
 J_{\mathcal{I}'\mathcal{I}}^\mu=\mathcal{Q}_B\int_p\int_q\bar{\Psi}_{\beta'\alpha'\mathcal{I}'\gamma'}(p^{\{3\}}_f,q^{\{3\}}_f,P_f)\left[S_{\alpha'\alpha}(p_1)S_{\beta'\beta}(p_2)\left(S(p_3^f)\Gamma^\mu(p_3,Q)S(p_3^i)\right)_{\gamma'\gamma}\right]\times\nonumber\\
~~~~~~~~~~~~~~~~~~~~~~~~~~~~~~\left(\Psi_{\alpha\beta\gamma\mathcal{I}}(p^{\{3\}}_i,q^{\{3\}}_i,P_i)-\Psi^{\{3\}}_{\alpha\beta\gamma\mathcal{I}}(p^{\{3\}}_i,q^{\{3\}}_i,P_i)\right)~,
\end{flalign}
and it is clear that for a neutral spin-$\nicefrac{3}{2}$ baryon, all electromagnetic form factors will vanish identically. This as a drawback of assuming isospin symmetry in a covariant Bethe-Salpeter approach. Finally, this simplification is valid only in Rainbow-Ladder or, more generally, whenever the interaction kernel does not couple to the external photon and fulfills (\ref{eq:condition_on_kernel}).

As already mentioned in previous section, a reliable calculation of (\ref{eq:FFsRL_simplified}) requires a very precise determination of $\Psi$. A naive approach to solve (\ref{eq:FFsRL_simplified}) with the number of integration points used in this work, implies an evaluation of $(128\times4\times4\times4)\sim 6\cdot10^7$ functions (each of the coefficients in the expansion (\ref{eq:cheby_expansion})) at $(30\times8\times8\times8)^2\sim 8\cdot 10^8$ quadrature points. It is clear that one needs to arrange the integration in a way such that the number of evaluations is reduced. In this respect, if we use the Faddeev wave functions (\ref{eq:def_wavefunction}) we can write Equation (\ref{eq:FFsRL_simplified}) as 
\begin{multline}\label{eq:FFsRL_simplifiedwaves}
 J_{\mathcal{I}'\mathcal{I}}^\mu=\mathcal{Q}_B\int_p\int_q\bar{\Psi}_{\beta'\alpha'\mathcal{I}'\gamma'}(p^{\{3\}}_f,q^{\{3\}}_f,P_f)\left(S(p_3^f)\Gamma^\mu(p_3,Q)\right)_{\gamma'\gamma}\\
\times\left(\Phi_{\alpha\beta\gamma\mathcal{I}}(p^{\{3\}}_i,q^{\{3\}}_i,P_i)-\Phi^{\{3\}}_{\alpha\beta\gamma\mathcal{I}}(p^{\{3\}}_i,q^{\{3\}}_i,P_i)\right)~,
\end{multline}
and since $p_3$ and $p_3^f$ are independent of $q$, we can distribute the integration
\begin{multline}\label{eq:FFsRL_simplifiedwavesfactored}
 J_{\mathcal{I}'\mathcal{I}}^\mu=\mathcal{Q}_B\int_p\left(S(p_3^f)\Gamma^\mu(p_3,Q)\right)_{\gamma'\gamma}\\
\times\int_q\bar{\Psi}_{\beta'\alpha'\mathcal{I}'\gamma'}(p^{\{3\}}_f,q^{\{3\}}_f,P_f)\left(\Phi_{\alpha\beta\gamma\mathcal{I}}(p^{\{3\}}_i,q^{\{3\}}_i,P_i)-\Phi^{\{3\}}_{\alpha\beta\gamma\mathcal{I}}(p^{\{3\}}_i,q^{\{3\}}_i,P_i)\right)~.
\end{multline}
Moreover, it is easy to check that the momenta $\widehat{q_{i/f}}$, that appear in the arguments of the Faddeev basis, and the angles $\{z_{0}^{i/f},z_{1}^{i/f},z_{2}^{i/f}\}$ are independent of $q^2$, which simplifies the numerical integration.

\end{appendix}

\bibliography{bibliography}{}
\bibliographystyle{phaip} %{apsrev4-1}

\chapter*{Acknowledgements}

First of all I want to thank my advisor Prof. Reinhard Alkofer for the opportunity to work with him, his continuous support and for his open-mindedness towards all areas of Physics and towards physicists coming from all areas. I am also grateful to Gernot Eichmann, Diana Nicmorus and Richard Williams for many useful discussions and their help during the development of my thesis. I would also like to thank Vicente Vento for the pleasant research stay at the University of Valencia and for his support afterwards.

Life is not life without fun, and to this Elmar, Justine, Joe, Ana, Diana and Valentina have contributed substantially with beer, gossip, discussions and general complaints. Other people who have made these three years an enjoyable time are Selym, Maria, Georg, Tina, Nat\'alia, Matthias and in general all the colleagues and staff of the Doctoral School. In particular, I want to thank Claudia Spidla for helping me and my family solving real-life problems.

Last but most important, I will always be grateful to Helena for letting me try to be a physicist and being determined to come with me all around the world.
\\

\textbf{Financial support:} This thesis was supported by the Austrian Science Fund FWF under Project No. P20592-N16 and the Doctoral Program W1203 (Doctoral Program ``Hadrons in vacuum, nuclei and stars'').

\end{document}